\documentclass[manuscript]{aastex}
\usepackage{amsmath,bm}
\usepackage{morefloats}
\usepackage{natbib}
\usepackage{threeparttable}

\title{Validating Time-Distance Helioseismology With Realistic Quiet Sun Simulations}
\author{
K. DeGrave,\altaffilmark{1}
J. Jackiewicz,\altaffilmark{1}
M. Rempel\altaffilmark{2}
}

\affil{\altaffilmark{1}New Mexico State University, Department of Astronomy, 1320 Frenger Mall, Las Cruces, NM 88003, USA; degravek@nmsu.edu, jasonj@nmsu.edu}
\affil{\altaffilmark{2}National Center for Atmospheric Research, HAO Division, 3080 Center Green Drive, Boulder, CO 80301, USA; rempel@ucar.edu}

\begin{document}

\begin{abstract}
Linear time-distance helioseismic inversions are carried out for vector flow velocities using travel times measured from two $\sim 100^2\,{\rm Mm^2}\times 20\,{\rm Mm}$ realistic magnetohydrodynamic quiet-Sun simulations of about 20~hr. The goal is to test current seismic methods on these state-of-the-art simulations. Using recent three-dimensional inversion schemes, we find that inverted horizontal flow maps correlate well with the simulations in the upper $\sim 3$~Mm of the domains for several filtering schemes, including phase-speed, ridge, and combined phase-speed and ridge measurements. In several cases, however, the velocity amplitudes from the inversions severely underestimate those of the simulations, possibly indicating nonlinearity of the forward problem. We also find that, while near-surface inversions of the vertical velocites are best using phase-speed filters, in almost all other example cases these flows are irretrievable due to noise, suggesting a need for statistical averaging to obtain better inferences.
\end{abstract}

\section{Introduction}
Time-distance helioseismology \citep{duvall1993} is one of the few tools available that allows us to peer beneath the photosphere to better understand the Sun's internal structure and dynamics. Through helioseismic inversions of wave travel-times measured at the solar surface, we can, in principle, better understand and characterize different kinds of subsurface perturbations (i.e. magnetic fields, density anomalies, flows, etc.) that may be present in the upper convection zone. This method has been used to study the near-surface structure of supergranulation \citep{gizon2003, zhao2003, jackiewicz2008, duvall2010, svanda2012} and sunspots \citep{zhao2001, couvidat2006, kosovichev2011a}.

One way to test the accuracy of helioseismology is through comparisons with other helioseismic methods (i.e. comparing time-distance inversion results with those of ring-diagram analysis or helioseismic holography) \citep[e.g.,][]{gizon2009, kosovichev2011b}. However, these types of studies can be misleading as agreement among methods does not necessarily guarantee correctness (which may be especially true around active regions). Perhaps the best way to to test the validity of time-distance inversions is by inverting for specific features present in solar simulations. This approach is advantageous in that the inversion results can easily be compared directly with the known values taken from the simulation. Synthetic data also allow us to carry out additional testing that would otherwise be impossible if only real solar data were available (i.e. testing the effects of data filter choice on results, tests of kernel performance through forward-modeling, etc.).

A small number of studies have been carried out previously to validate time-distance helioseismology by applying the method to simulated quiet-Sun data possessing varying degrees of realism. \citet{zhao2007} performed time-distance inversions of realistic convection simulations \citep{benson2006} in the ray approximation and found that the recovered horizontal flows agree well (correlation coefficients of $\sim$0.5--0.9) with those of the simulation in the upper 5~Mm of the domain, though the vertical velocities were anticorrelated at nearly every depth. \citet{svanda2011} used a snapshot of a convective simulation flow-field to produce forward-modeled travel-time maps, inverting them using the first Born approximation methods \citep{gb02} to validate their time-distance inversions. They found high correlation over the same range of depths for the horizontal flow components, and improved correlation for the vertical component.

In some sense, however, both cases may be too idealized. In the case of \citet{zhao2007}, the work was based upon a simulation which contained no magnetic fields. The work of \citet{svanda2011} did indeed validate the inversion procedure, yet it was based on simplified measurements, as forward-modeled travel times of a frozen flow field may be too idealized in comparison to a full time-distance analysis of a Doppler time series.

\begin{figure*}[t]
\begin{center}$
\begin{array}{c}
\includegraphics[width=1\linewidth,clip=]{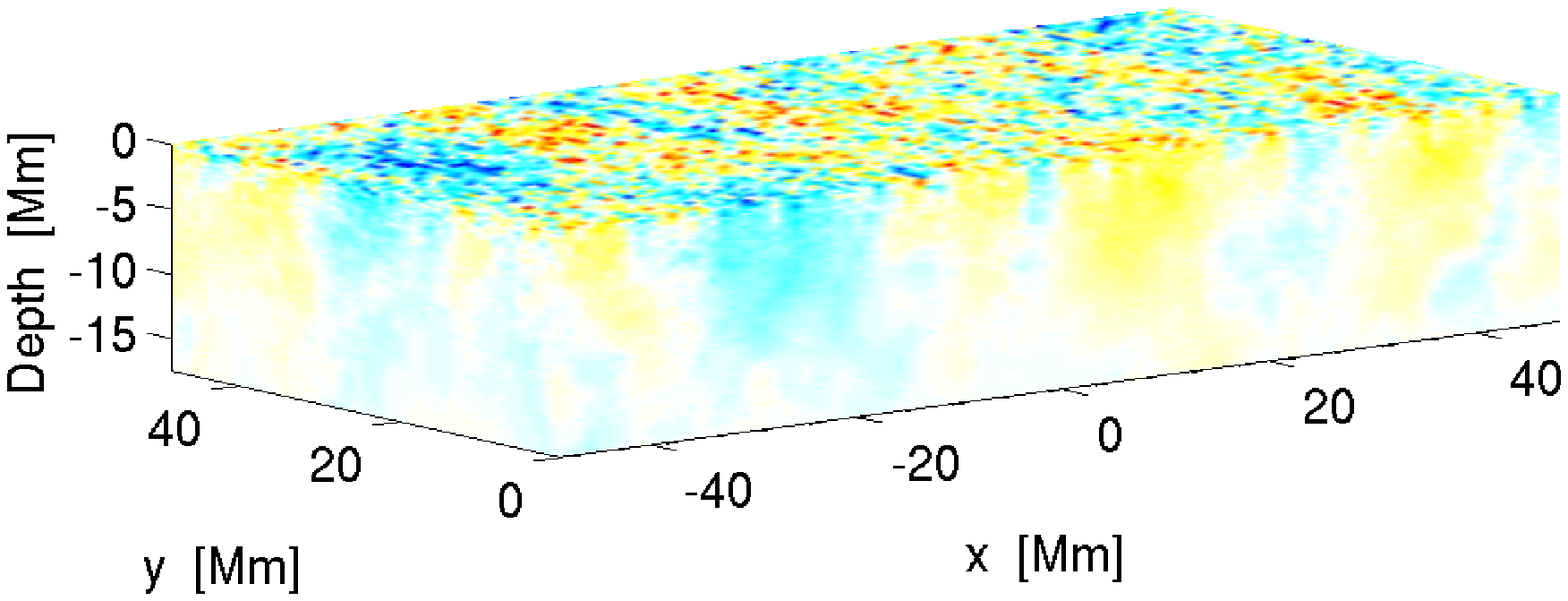} \\
\includegraphics[width=1\linewidth,clip=]{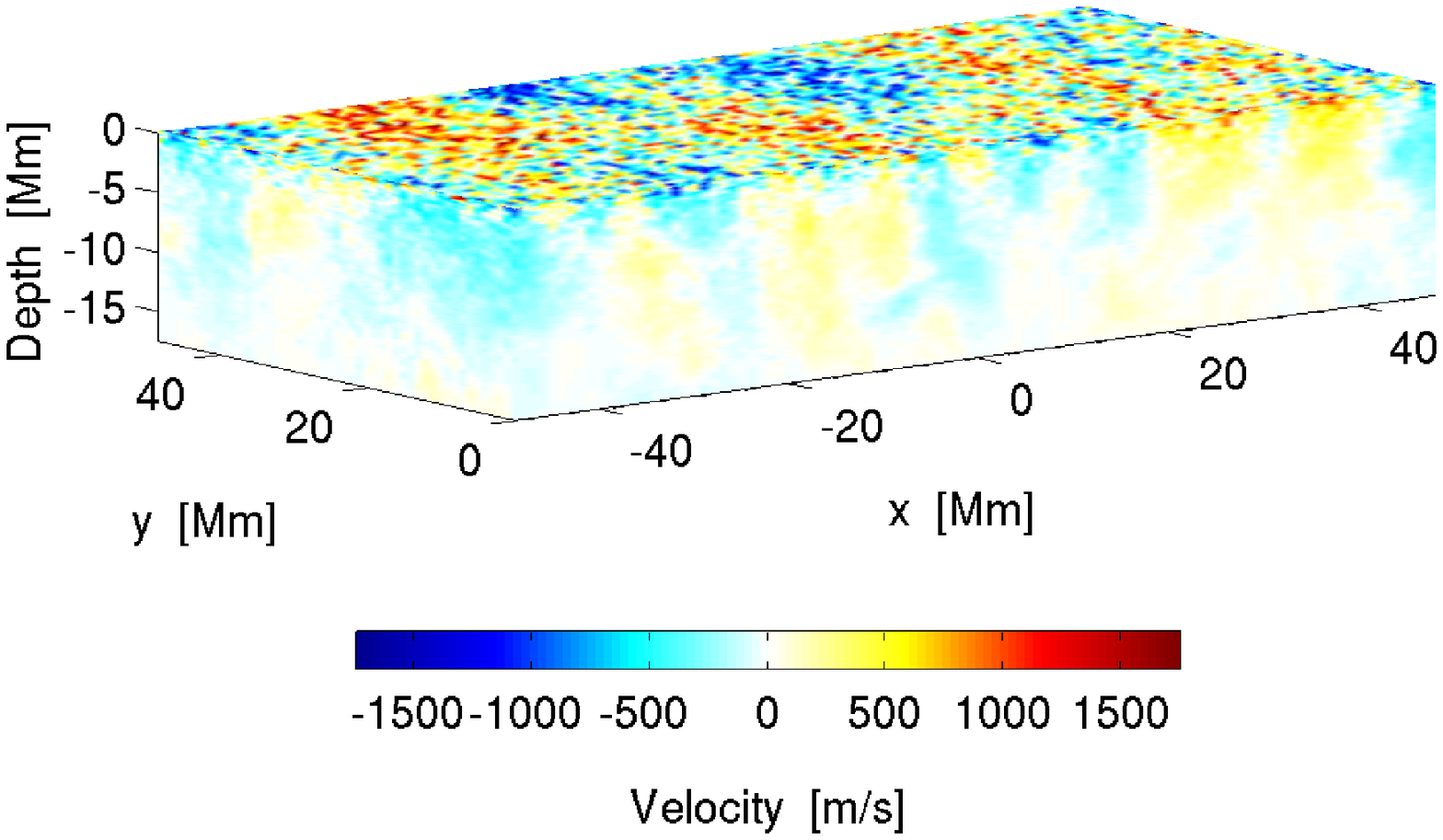}
\end{array}$
\end{center}
\caption{The two simulations used in this study. Shown are time averages (over $\approx 20$~hr) of the $v_x$ velocity for half of the computation domain ($y\ge 0$) for QS1 (top) and QS2 (bottom). The horizontal slice at the top is taken at the $\tau_{500} =0.01$ ($z=0$) level. The color scale is the same in each image.}
\label{fig:QS}
\end{figure*}

The goal of this work is to test current time-distance methods using the most realistic solar simulations available to us today. These magnetohydrodynamic simulations are fully convective and nonlinear in nature and contain large-scale supergranular-type flows. Such data, described in Section~\ref{sec:sim}, will allow us to test methods in a regime that is as close as possible to the real Sun. In Section~\ref{sec:filt} the data filtering is described where two methods are considered. The filtered data are used to measure a large set of travel times in two different ways as discussed in Section~\ref{sec:tt}. Sections~\ref{sec:kernels} and \ref{sec:inv} briefly review the forward and inversion problems in the first Born approximation. Also shown are comparisons of measured and forward-modeled travel times. Inversion results for the three vector flow components are given in Section~\ref{sec:results}. We discuss the results in the context of validation of these time-distance methods in Section~\ref{sec:dis}.

This initial work is carried out with the more difficult goal of testing time-distance inversions near active regions in mind. The inversion procedure described here will be applied to a similarly realistic sunspot simulations to assess how capable or limited current linear time-distance inversions may be at recovering subsurface flows in these strong perturbation regimes. As sunspot helioseismology is an active area of research, such work should prove useful.

\section{Quiet-Sun Simulation Data}\label{sec:sim}
The two quiet-Sun simulations (referred to as QS1 and QS2 hereafter) used for this study are based on the work of \citet{rempel2009a, rempel2009b}. They are publicly available for use in testing helioseismic methods\footnote{Pleave visit \url{http://www2.hao.ucar.edu/}, then ``Observations/Data'' and then ``Numerical Sunspot Models.''}. Both Cartesian simulation domains are 98.304 $\times$ 98.304 $\times$ 18.432~Mm in the horizontal and vertical directions. QS1 was computed with horizontal and vertical grid spacings of 0.064~Mm and 0.032~Mm respectively, while QS2 was computed with grid spacings of 0.128~Mm and 0.048~Mm. Boundary conditions for both are periodic in the horizontal directions. The top boundary of the domains is closed and located about $700~\rm{km}$ above the $\tau=1$ level. The bottom boundary is open (i.e. it allows convective flows to cross the boundary). This is implemented by imposing a symmetric boundary condition on all mass flux components (i.e. these quantities are mirrored into the ghost cells across the boundary). In QS1 the mean pressure is extrapolated into the ghost cells such that the value right at the boundary (between first ghost and domain cell) is fixed, while fluctuations are damped ($5\%$ reduction in the first ghost cell, $10\%$ reduction in the second). In QS2, the pressure fluctuations are set to zero at the boundary. Fixing the (mean) pressure at the boundary ensures that the total mass in the simulation domain remains constant apart from small fluctuations. The entropy is specified in inflow regions such that the resulting radiative losses in the photosphere lead to a solar-like energy flux, while down flows use a symmetric boundary condition.

The quiet-Sun simulations presented here are magnetic. We use a setup with zero net flux in which a turbulent magnetic field is maintained through a small-scale dynamo. We use a LES approach, relying only on numerical diffusivities similar to those introduced in \citep{rempel2009b} in order to minimize diffusivities. At the top boundary the magnetic field is vertical in QS2, while it is matched to a potential field extrapolation in QS1. We use a symmetric boundary condition for all three field components at the bottom. For details of the dynamo setup, we refer the reader to a forthcoming paper (Rempel 2014 in preparation).

\begin{figure}[ht]
\begin{center}$
\begin{array}{c}
\includegraphics[width=1.0\linewidth,clip=]{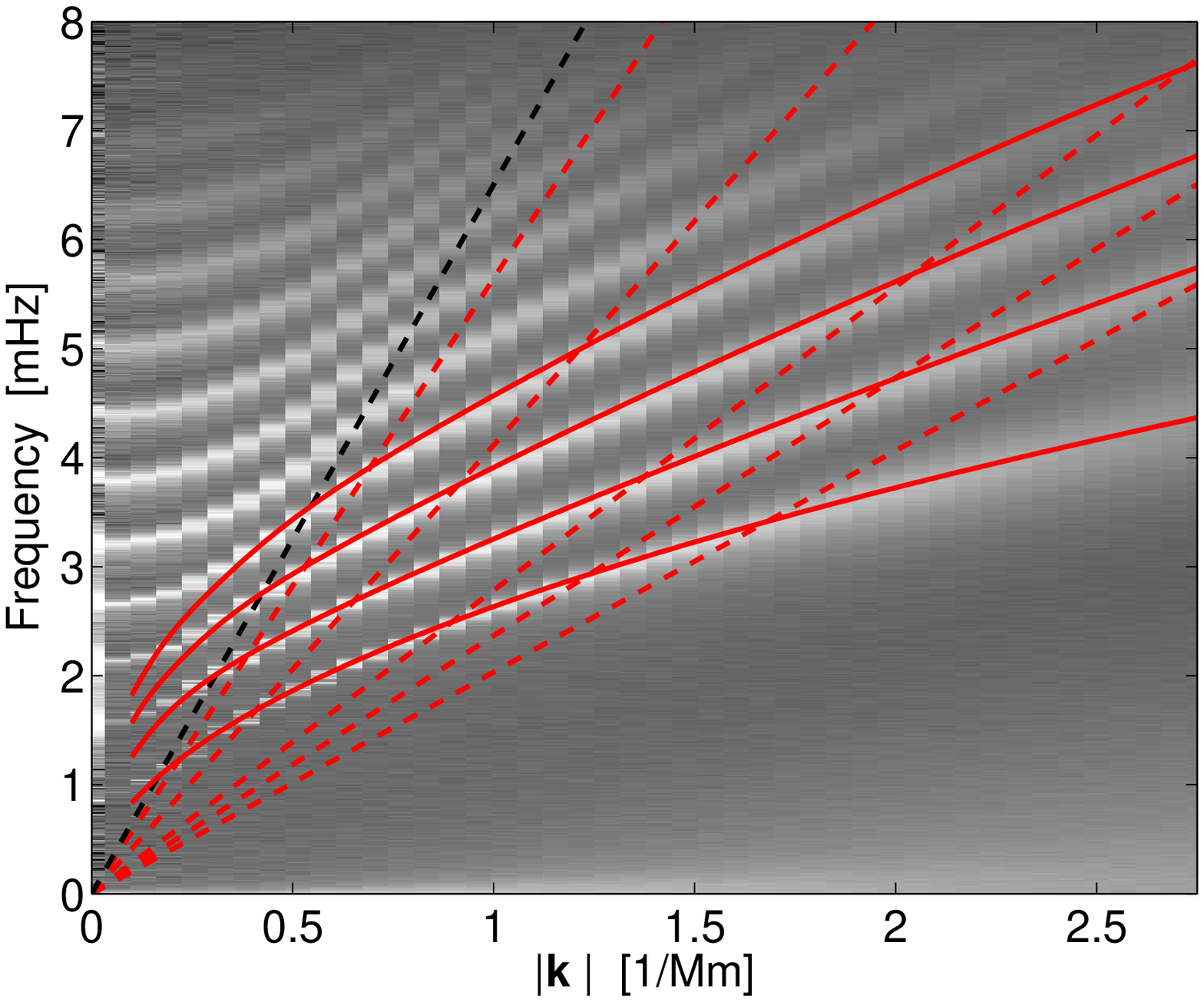}
\end{array}$
\end{center}
\caption{The QS1 power spectrum after azimuthally averaging over wavenumber. Overplotted are the lines marking the centers of each of the phase-speed (dashed red lines) and ridge (solid red lines) filters. The 40~$\rm{kms^{-1}}$ phase-speed line representing a lower turning point depth of 12~Mm is also shown (black dashed line).}
\label{fig:filters}
\end{figure}

Both simulations show large-scale flow structure that extends deeply through their domains (Fig.~\ref{fig:QS}). The near-surface horizontal flow magnitudes exhibit a root-mean-square (RMS) average of the order $350-400~\rm{m\,s^{-1}}$, while the average vertical velocity is $\sim 230~\rm{m\,s^{-1}}$, where we define the root-means-square (RMS) value for $n$ measurements of a quantity, $x$, as

\begin{equation}
 x_{\rm{RMS}} = \sqrt{\frac{1}{n} \sum_i^n x_i^2}.
\end{equation}
In general, the near-surface flows of QS1 are somewhat weaker than those of QS2. The simulations contain surface-gravity and acoustic waves that compare well to solar modes (see Fig.~\ref{fig:filters}), except where the boundaries of the domain have influence.

From each of these simulation runs we extracted a vertical velocity Doppler time series at the $\tau=0.01$ layer. We define this data cube as $v_z (\bm{r}, z=0, t)$, where $\bm{r}=(x,y)$ is the horizontal coordinate and $z$ is the vertical coordinate. The QS1 time series spans 25~hr, while QS2 spans 19.2~hr, and both are sampled with a time cadence $h_t= 45$~s. For the helioseismic analysis, the simulations were interpolated onto grids with $h_x=h_y=1$~Mm horizontal spacing, which is sufficiently small to capture the spatial variations due to acoustic waves.

\section{Filtering}\label{sec:filt}
The first step in the analysis is to filter the data to obtain signal only from those waves whose travel times we would ultimately like to measure. For this work we chose to study the effects on time-distance inversion results of two filtering types -- phase-speed filters, that isolate waves of similar phase speed over a range of radial orders, and ridge filters, that isolate wave packets that have the same radial order.

Phase-speed filters extract the signal from waves that propagate to approximately the same depth in the solar interior. \citep{couvidat2006} showed that there exists an optimal set of such filters that effectively maximizes the signal-to-noise (S/N) of measured travel times. For this work, we selected the first five lowest phase-speed filters defined in \citet{couvidat2006}, referred to as ${\rm td_1}$--${\rm td_5}$ hereafter. Each of these filters is centered upon a line of constant phase-speed and is constructed according to the following analytic form:
\begin{equation}
 F(\omega, k) = e^{[- (\omega/k - v_{\rm ph})^2/ 2\delta v_{\rm ph}^2]}.
 \label{phase}
\end{equation}
Here, $v_{\rm{ph}}$ is the value of the central phase-speed; $\delta v_{\rm{ph}}=\rm{FWHM}/[2(\log 2)^{1/2}]$, where FWHM is the full width at half-maximum of the squared filter; $k = |\bm{k}|$ is the horizontal wavenumer, and $\omega$ is the angular frequency. Figure~\ref{fig:filters} shows the central phase-speed of the five phase-speed filters. These particular filters were chosen to keep only signal from those waves whose lower turning points are above a depth of $\sim$12~Mm (i.e. those waves whose phase-speeds are less than roughly 40~$\rm{km\,s^{-1}}$) to avoid any signal contamination from wave reflections at the bottom boundary of the simulation domain. When phase-speed filtering is used, the $f$-mode is first removed, as it follows a dispersion relation different from that of the acoustic $p$-modes.

As an alternative to phase-speed filtering, ridge filtering has been used to isolate wave signal by selecting those modes possessing the same radial order \citep{duvall2000,jackiewicz2008,gizon2009}. Here we use filters for the modes along ridges $f$, $p_1$--$p_3$ in the analysis. Higher radial order modes were not selected due to the fact that the amount of ridge power between the 40~$\rm{km\,s^{-1}}$ phase-speed line and the acoustic cutoff frequency was quite small, and we have seen some evidence of problems with our travel-time sensitivity kernels for these higher-order modes. The solid red lines in Fig.~\ref{phase} show each of the centers of the filters along the ridges considered here. 

Each ridge filter, centered on its corresponding mode, was designed to keep full wave power extending 40\% of the way in frequency to the next adjacent ridge on either side. This ``flat gap" of full power transmission is then terminated by cosine wings that smoothly decrease to zero at 60\% of the way to the next adjacent ridge. All filters, both ridge and phase-speed, were confined within the frequency range of 2.5--5.3~mHz.

With our initial Doppler time series, $\phi(x, y, t)$, we apply filters by multiplying the Fourier transform of the data cube with the square root of each filter $F_m (\bm{k},\omega)$ as

\begin{equation}
 \phi_m (\bm{k},\omega) = \phi (\bm{k},\omega) \sqrt{F_m (\bm{k},\omega)}.
\end{equation}
Here, $\phi_m (\bm{k},\omega)$ is the filtered cube in Fourier space containing wave signal isolated using filter $m$. The last step is to transform the filtered data back to real space, giving $\phi_m(x,y,t)$.

\section{Travel-Time Measurements}\label{sec:tt}
Temporal cross-covariances are computed from the filtered data $\phi_m (\bm{r},t)$ in the center-to-annulus and quadrant configurations \citep[e.g.,][]{duvall1997}. This is done by cross-correlating the signal at a point on the surface of the simulation domain with the signal averaged over a concentric annulus of some radius $\Delta$, and over 90$^\circ$ quadrants centered at the four discrete north, south, east, and west directions. The cross-covariances are computed for a range of $\Delta$ values depending on filter type. For the ridge-filtered data, $\Delta = 11-27$~Mm in increments of 4~Mm, totaling five radii per ridge. For the phase-speed filtered data, the valid $\Delta$ values over which the cross-correlations can be computed depend on central phase-speed as discussed in \citet{couvidat2006}. For this work, the ranges of $\Delta$ values used are 5--9, 7--11, 9--15, 15--19, 19--29~Mm in increments of 2~Mm for filters $\rm{td_1}$--$\rm{td_5}$ respectively. In total, we have $M=117$ measurements.

\subsection{Comparison of Two Travel-Time Measurement Methods}

\begin{figure}[t]
\begin{center}$
\begin{array}{c}
\includegraphics[width=1.0\linewidth,clip=]{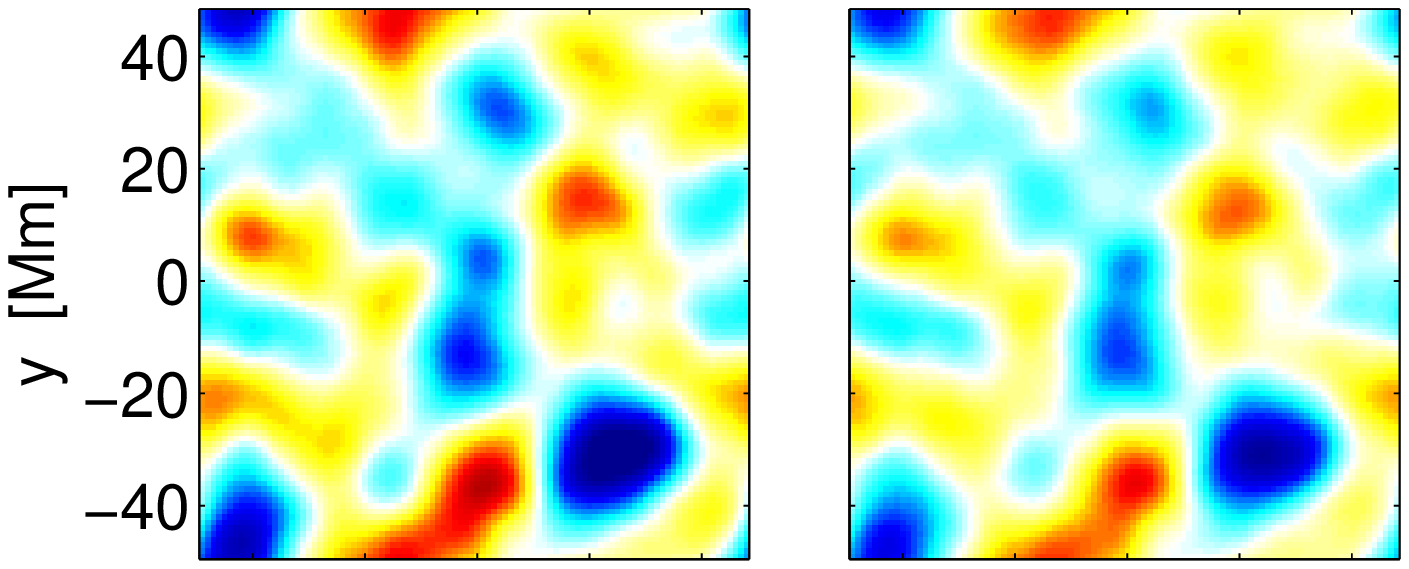}\\
\includegraphics[width=1.0\linewidth,clip=]{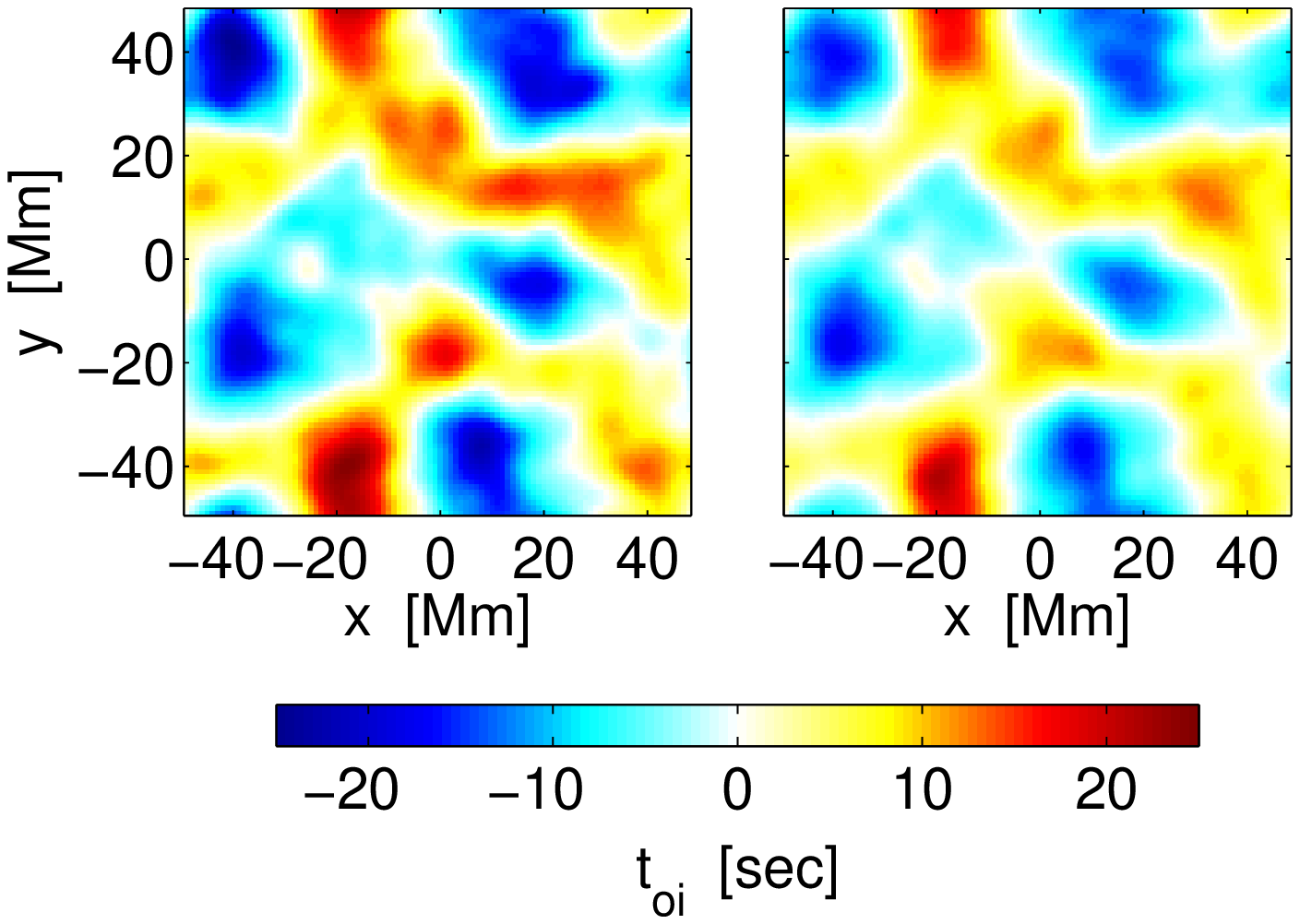}
\end{array}$
\end{center}
\caption{Example oi travel-time difference maps computed from QS1 (top row) and QS2 (bottom row) for $\Delta$ = 15~Mm measured with the GB04 travel-time definition. These examples were computed using $p_2$ (left column) and ${\rm td}_3$ (right column) filtered data.}
\label{ttoi}
\end{figure}

Travel-time differences were computed from each cross-covariance in the `oi' (out minus in), `we' (west - east), and `ns' (north - south) geometries according to the two methods defined in \citet{gb02, gb04} (hereafter referred to as GB02 and GB04, respectively). Both GB02 and GB04 measurement methods rely on calculating the difference between the cross-covariances at each point and a time-symmetric reference cross-covariance function, obtained either from a model or from an appropriate averaging of the data. The GB04 method is a linearization of the least-squares minimization prescribed in GB02. Example $p_1$ and ${\rm td}_3$ oi travel-time difference maps computed under the GB04 definition for both simulations are shown in Fig.~\ref{ttoi} for $\Delta = 15$~Mm. The travel times suggest that both simulations possess large ($\sim 20$~Mm scale) regions of convergent (positive travel-time difference) and divergent (negative travel-time difference) flows.

GB04 assumes that the travel times are linearly related to the small changes in the cross-covariances due to small subsurface perturbations, and will eventually become inaccurate in regions where strong perturbations dominate. GB02 has been shown to be more robust in a larger range of perturbation regimes \citep{couvidat2012}. Inspection of the data in Fig.~\ref{fig:QS} shows that both simulations possess large ($>$~500~$\rm{m\,s^{-1}}$) flows that could potentially cause a problem for linear measurements and inversions \citep{jackiewicz2007a}. Therefore, travel times were computed under both definitions for use in the inversions to see if recovered flows vary in any significant way.

\begin{figure}[t]
\begin{center}$
\begin{array}{c}
\includegraphics[width=1.0\linewidth,clip=]{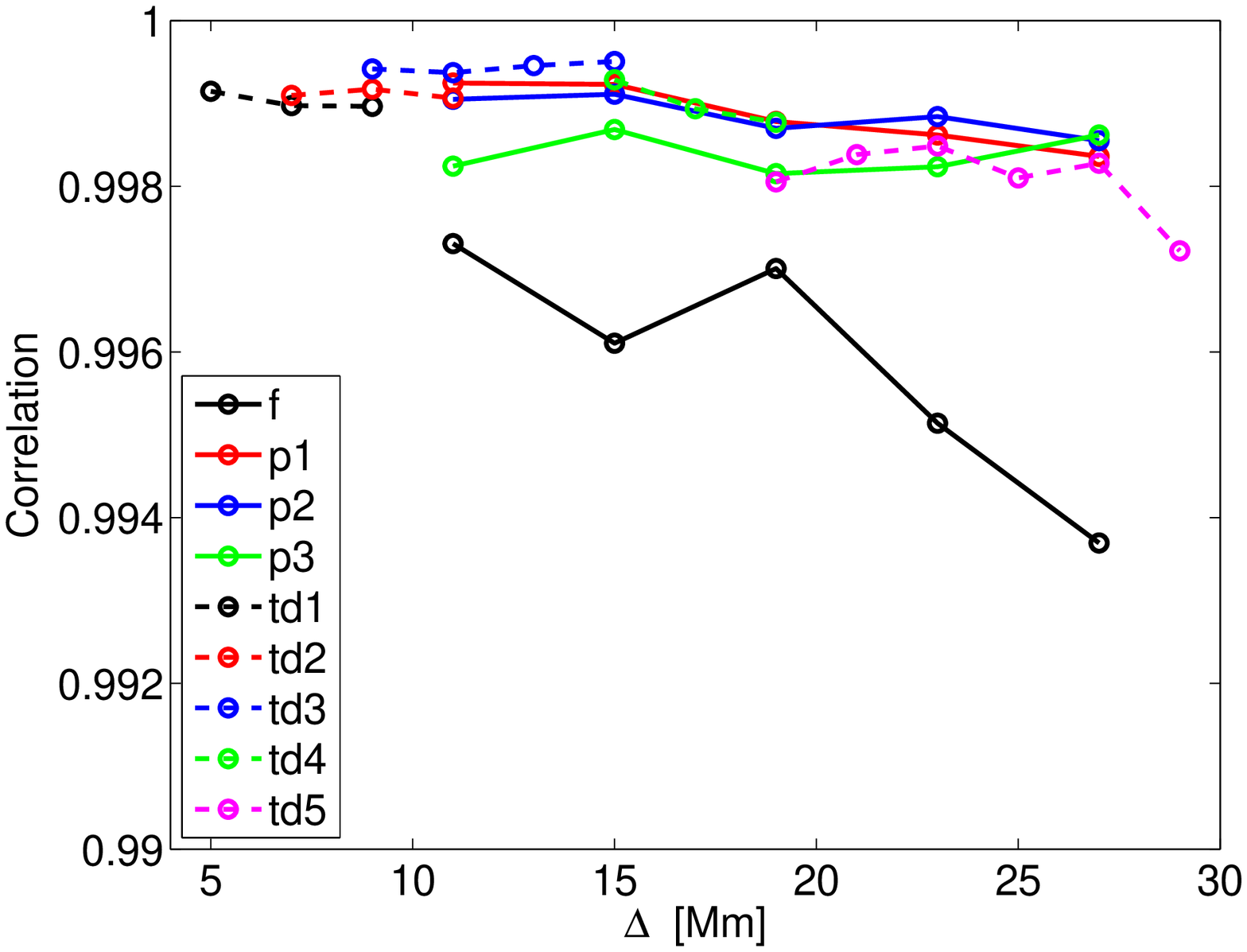}
\end{array}$
\end{center}
\caption{The 2D spatial correlation between the GB02 and GB04 QS1 oi travel-time maps for each filter versus $\Delta$. The correlation between the two methods is very high ($>$ 0.99) in every case, and, with the slight exception of the $f$-mode, shows no trend with filter type or distance.}
\label{gb0204}
\end{figure}

Figure~\ref{gb0204} shows the Pearson correlation coefficient between QS1 GB02 and GB04 travel-time maps computed for each filter and $\Delta$ in the oi geometry. We find a very high correlation between the two and there seems to be no trend with distance or filter type, except in the case of the $f$-mode. Even here, though, the correlation between the methods is greater than 0.99. However, we find that travel times measured with GB04 are on average 10\% larger than those of GB02 in terms of RMS average.

A comparison of MDI travel times carried out by \citet{couvidat2012} found a similarly high correlation between GB02 and GB04 travel time definitions. In their analysis, however, they found that at large ($\sim 30$~Mm) distances, GB02 and GB04 are highly correlated only at small travel-time differences between $\pm 5$ seconds. Outside this range, GB02 travel-times were generally found to be larger than GB04.

\subsection{Travel-Time Noise}
Solar oscillations are driven by stochastic convective motions, and measured travel times contain some level of realization noise depending on the observation length. This noise is very important to characterize as it induces correlations in the travel-time measurements and propagates through the inversion \citep{jensen2003, couvidat2005, couvidat2006, jackiewicz2008}. For this work, we estimate the noise covariances for travel times computed from the simulation data by employing the Monte Carlo technique outlined in \citet{gb04}. The basic idea is to study small segments (90~min in this case) of the Doppler data and compute a series of noise realizations by adding complex Gaussian noise to its average power spectrum. Travel-times are measured from this noisy data, and the covariance between every travel-time map is measured for each available filter-$\Delta$-geometry combination. Rather than simply using the RMS variation in the travel times to estimate noise \citep{zhao2012}, the noise covariance matrices should give a more accurate representation of the resulting error in the flow inversions.

\section{Sensitivity Kernels}\label{sec:kernels}
To infer flows, we relate our travel-time measurements to subsurface conditions present in the simulations. The relationship between travel-time perturbations and flow velocity, $\bm{v}$, developed in the GB02 and GB04 formulation, is given in the form of a linear integral equation

\begin{figure}[t]
\begin{center}$
\begin{array}{c}
\includegraphics[width=.7\linewidth,clip=]{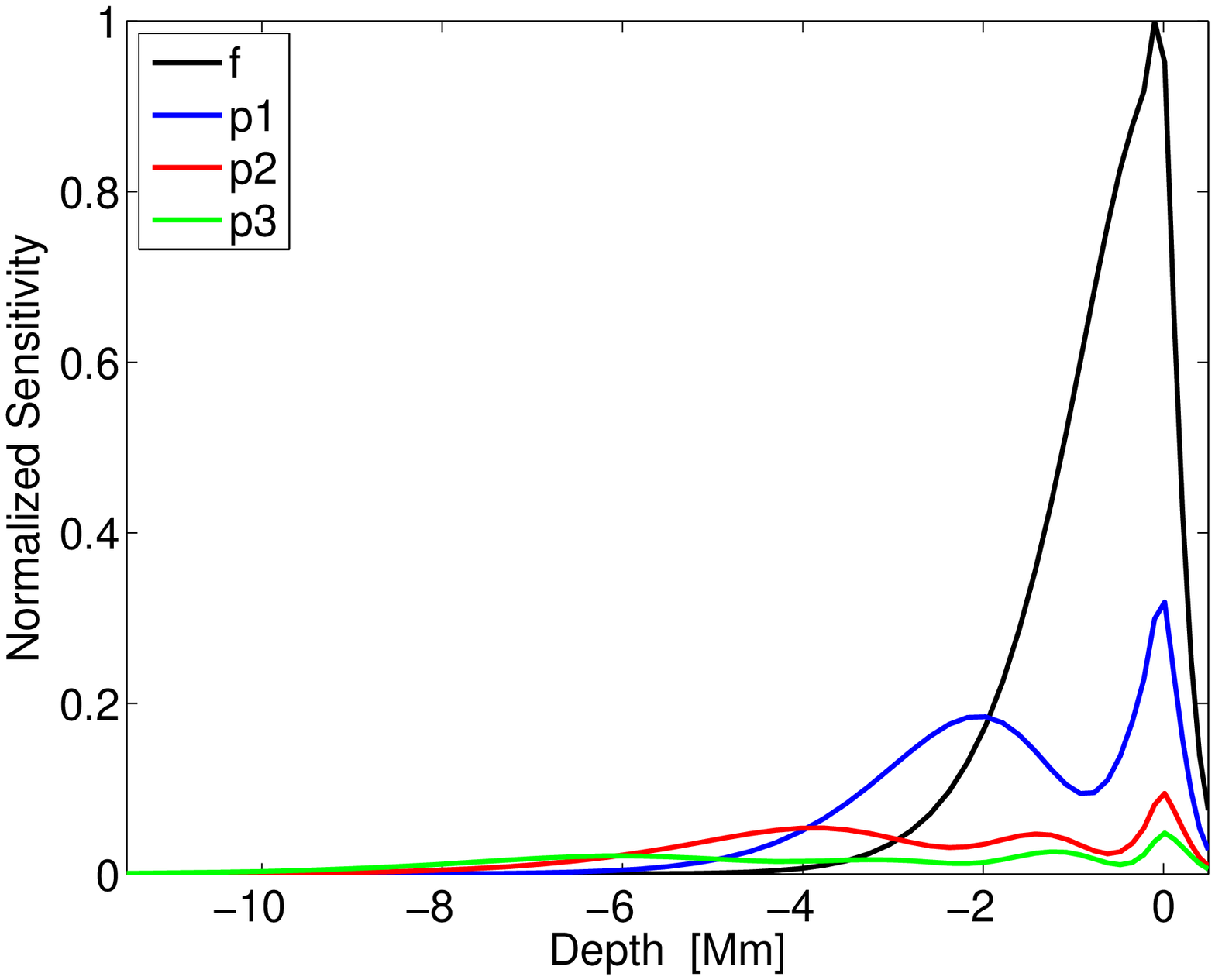} \\
\includegraphics[width=.7\linewidth,clip=]{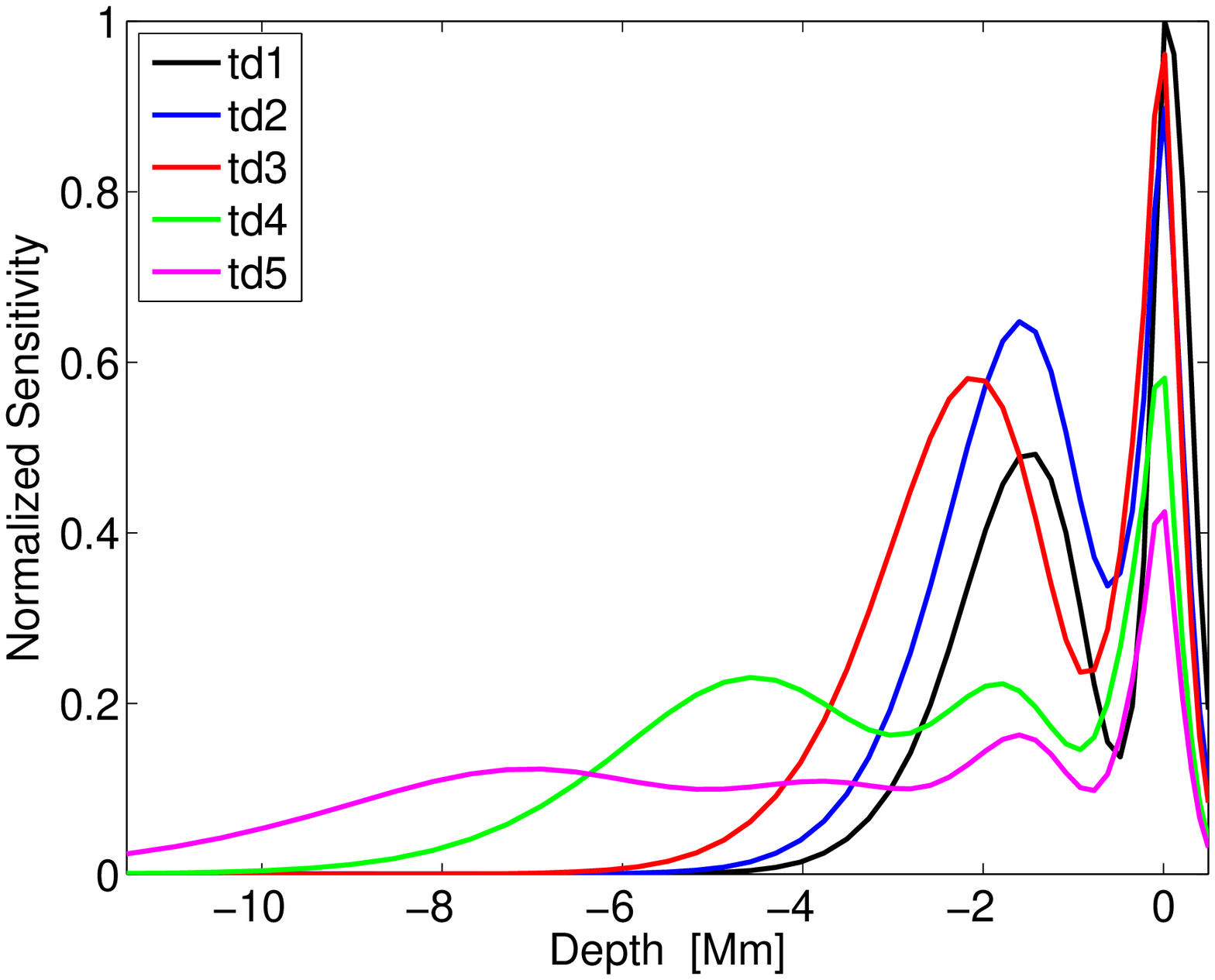}
\end{array}$
\end{center}
\caption{Horizontally integrated `we' sensitivity kernels as a function of depth. The top panel shows examples of kernels for ridge filters and the bottom panel shows those for phase-speed filters. In each case, the absolute value is shown, and each profile has been normalized by the largest sensitivity value in that particular plot. The kernel distances for which these are computed correspond to the mid-range $\Delta$ value for each filter.}
\label{fig:kerns}
\end{figure}

\begin{align}
 \delta\tau^{a}(\bm{r}) = h_rh_z\sum_{ij} \bm{K}^{a} (\bm{r}_i - \bm{r},z_j) \cdot \bm{v} (\bm{r}_i,z_j) + N^{a} (\bm{r})
 \label{eq:dt}
\end{align}

\begin{figure}[ht]
\begin{center}$
\begin{array}{c}
\includegraphics[width=1.0\linewidth,clip=]{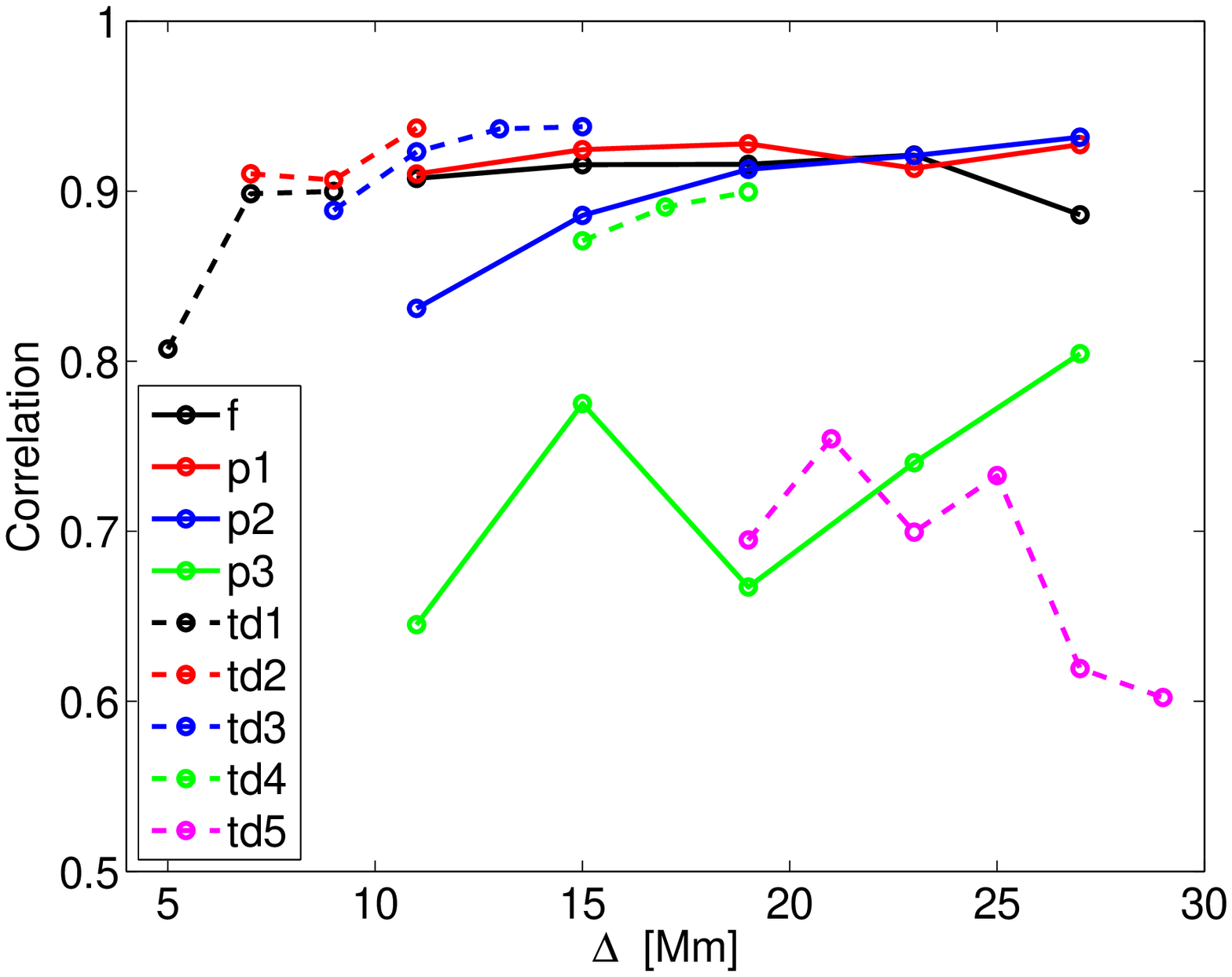}
\end{array}$
\end{center}
\caption{The correlation between the measured and forward-modeled oi travel-time maps of QS1 for each filter and for every $\Delta$.}
\label{fig:forward}
\end{figure}

\noindent where $h_r=h_xh_y$ and $h_z$ is the vertical grid spacing, $\bm{K}^a$ are three-dimensional vector-valued kernels describing the sensitivity of wave travel times to flows for each particular measurement geometry, filter, and $\Delta$, captured in the $a$ index. $N^a$ represents the noise for travel-time measurement $a$. The summation over $i$ represents a horizontal spatial convolution.

A set of kernels for flows $\{K_{v_x}, K_{v_y}, K_{v_z}\}$ in the $+\hat{\bm{x}}$ direction was computed in the single scattering Born approximation \citep{birch2007}. The model power spectra that are needed for the kernel computation (as prescribed in \citet{gb02}) are filtered exactly as the initial data power spectra used to derive the travel times for which we are modeling. These ``point-to-point'' kernels are then combined in an appropriate way to be consistent with the `oi', `we', and `ns' travel-time geometries. Figure~\ref{fig:kerns} shows the approximate depth sensitivity of a selection of kernels computed using ridge and phase-speed filtering. All the kernels are strongly peaked at the top of the box with lower-amplitude peaks slightly below this region. Only one set of kernels and covariance matrices is subsequently used in the inversions, as the power spectra of the simulations are indistinguishable for the modes of interest.

Because we are working with synthetic rather than real solar data, we have an advantage in that we can directly asses the abilities of our kernels through forward modeling. To do this, we computed a set of forward-modeled travel-time maps by convolving our sensitivity kernels with the known flow fields taken directly from the simulations (Eq.~\ref{eq:dt}, without the noise term). In the case of no measurement noise and perfect kernels, each modeled travel-time map would match its measured counterpart exactly. Of course this is not the case, so we expect instead some mismatch. Figure~\ref{fig:forward} shows the correlation between measured and forward-modeled travel times for every filter over the appropriate range of $\Delta$ values for QS1. In the ridge-filter case, we find that the correlation is high for filters $f$, $p_1$, and $p_2$, but drops significantly for $p_3$. This trend of decreasing correlation has also been observed for $p_4$. This was ultimately the basis for using only $f$, $p_1$--$p_3$ in our analysis. The phase-speed filters show something similar with the correlation starting high when phase-speed is low, but dropping sharply when when it becomes large. QS2 shows the same phenomenon, with the $p_3$ and td$_5$ correlations actually being slightly worse than the QS1 case. It is not entirely obvious why such a decrease in correlation should be observed. As the computation of sensitivity kernels is heavily dependent upon accurately modeling the data power spectra, a mismatch between data and model is likely contributing to the poorer agreement in these cases. We do indeed find that it is often more difficult to model the acoustic power of higher-order modes and the power at higher phase-speeds mostly because of the weaker mode power at longer wavelengths due to the finite simulation domain size. Another consideration is that the kernels are computed using a standard solar model, and while the simulation and model stratification generally agree, deviation does occur in near-surfaces layers where the energy transport in the simulation is treated more realistically. The effects of this have not been studied in any quantitative manner.

\section{SOLA Inversion Method}\label{sec:inv}
The goal of the time-distance inversions is to recover subsurface vector flows ($\bm{v}$ in Eq.~\ref{eq:dt}) in the upper layers of QS1 and QS2 given a set of measured travel times, sensitivity kernels, and the appropriate noise-covariance matrices. To do this, we employ the Subtractive Optimally Localized Averaging (SOLA) method \citep{pijpers1992}, modified for computations in the Fourier spatial domain \citep{jackiewicz2012}. The general idea behind a SOLA inversion is to find a set of `inversion weights' that average the set of travel times in a linear fashion. Once obtained, the weights in this case are spatially convolved with the travel times to give an estimate of a chosen component $\alpha=\{x,y,z\}$ of the flow at a targeted location centered at a depth $z_0$ within the simulation domain 
\begin{equation}
 v_{\alpha}^{\rm inv} (\bm{r}; z_0) = \sum_i \sum_{a=1}^M w_a^{\alpha} (\bm{r}_i - \bm{r}; z_0) \delta \tau^a(\bm{r}_i),
 \label{vinv}
 \end{equation}
where we recall that $M$ is the number of travel-time maps considered in a particular inversion. As described in other work \citep{svanda2011,jackiewicz2012}, the inversion weights are used to construct linear combinations of the sensitivity kernels, an averaging kernel, that is localized in 3D space and acceptably matches a pre-defined (usually Gaussian) `target function' $T$. In practice the match is usually far from perfect due to noise and a limited number of travel times. Another consideration is the extent to which the two flow components not being inverted for ``leak'' into the inferred velocity, known as the `cross-talk.' This is a major factor only in inversions for the vertical velocity.

The SOLA method attempts to minimize the misfit between averaging kernel and target function while at the same time minimizing inversion noise, cross-talk, and the spatial localization of the weights \citep{jackiewicz2007b, svanda2011}. This amounts to an optimization problem with the balance of these quantities being determined by a set of regularization parameters, denoted here by $\mu$, $\nu$, and $\epsilon$ respectively . One is free to choose regularization values, though when inverting for horizontal flow components, priority is often placed first on reaching an acceptable noise level rather than focusing on any of the other parameters. For example, when inverting for supergranular flows with a typical RMS velocity $\sim$250~$\rm{m\,s^{-1}}$, one might choose to accept a noise level of $\sim 30 - 40~\rm{m\,s^{-1}}$. On the other hand, inversions for the vertical flow component will likely place a greater importance on minimizing cross-talk and the localization of the weights. In practice, when inversions are performed, a series of regularization values are tested to find which combination gives the best results.

\section{Inversion Results}\label{sec:results}
We carry out inversions using each of the three filtering schemes (i.e. ridge, phase-speed, and ridge+phase-speed) for all three velocity components $(v_x, v_y, v_z)$ at depths of 1, 3, and 5~Mm below the surface using the GB04 travel-time definition. To our knowledge, only one previous study \citep{svanda2013} has employed and tested a combination of ridge+phase-speed filtered data in helioseismic inversions. 

In what follows, inversions for flows, denoted by $v_\alpha^{\rm inv}$, are directly compared to the flows from the numerical simulations. To make direct comparisons on the appropriate spatial scales at each targeted depth, the artificial data are smoothed to the expected resolution of the inverted flows. This is carried out by a convolution of the inversion target function $T$, with the raw simulation flow field to obtain the ``targeted'' answer $v_\alpha^{\rm tgt}$ for flow component $\alpha$
\begin{equation}
 v_\alpha^{\rm tgt}(\bm{r};z_0) = h_r h_z\sum_{ij} T_\alpha(\bm{r}_i - \bm{r},z_j;z_0)v_\alpha(\bm{r}_i,z_j),
 \label{valph}
\end{equation}
similar to the notation in \citet{svanda2011}. The sum over index $i$ represents a horizontal convolution, while the sum of the products over $j$ takes place at the same depth slices. In the case of an ``ideal'' inversion when the averaging kernel matches the target function with minimal noise variance, the quantities $v_\alpha^{\rm tgt}$ and $v_\alpha^{\rm inv}$ will be similar. As is often the case in many of the example inversions presented here, however, the averaging kernel does not sufficiently match the target, particularly for the vertical flow inversions. When this is the case, we will also be interested in studying the quantities
\begin{equation}
 v_\alpha^{(\beta)}(\bm{r};z_0) = h_r h_z\sum_{ij} \mathcal{K}_\alpha^{(\beta)}(\bm{r}_i - \bm{r},z_j;z_0)v_\beta(\bm{r}_i,z_j),
 \label{vbet}
\end{equation}
where $\mathcal{K}_\alpha^\beta$ is the $\beta$ component of the averaging kernel from an inversion for $\alpha$. Again, in the ideal case $v_\alpha^{(\beta)}$ will be zero when $\alpha\neq\beta$. More likely, however, the individual components obtained from this expression are nonzero and can be used to quantify the cross-talk from any inversion. In addition, the component $v_\alpha^{(\alpha)}\equiv v_\alpha$ is the targeted flow, equivalent to Eq.~(\ref{valph}) when the target and averaging kernel are identical.

\subsection{Horizontal Flow Inversions}

\begin{figure*}[t]
 \begin{center}$
  \begin{array}{c c c}
   \includegraphics[width=0.2\linewidth,clip=]{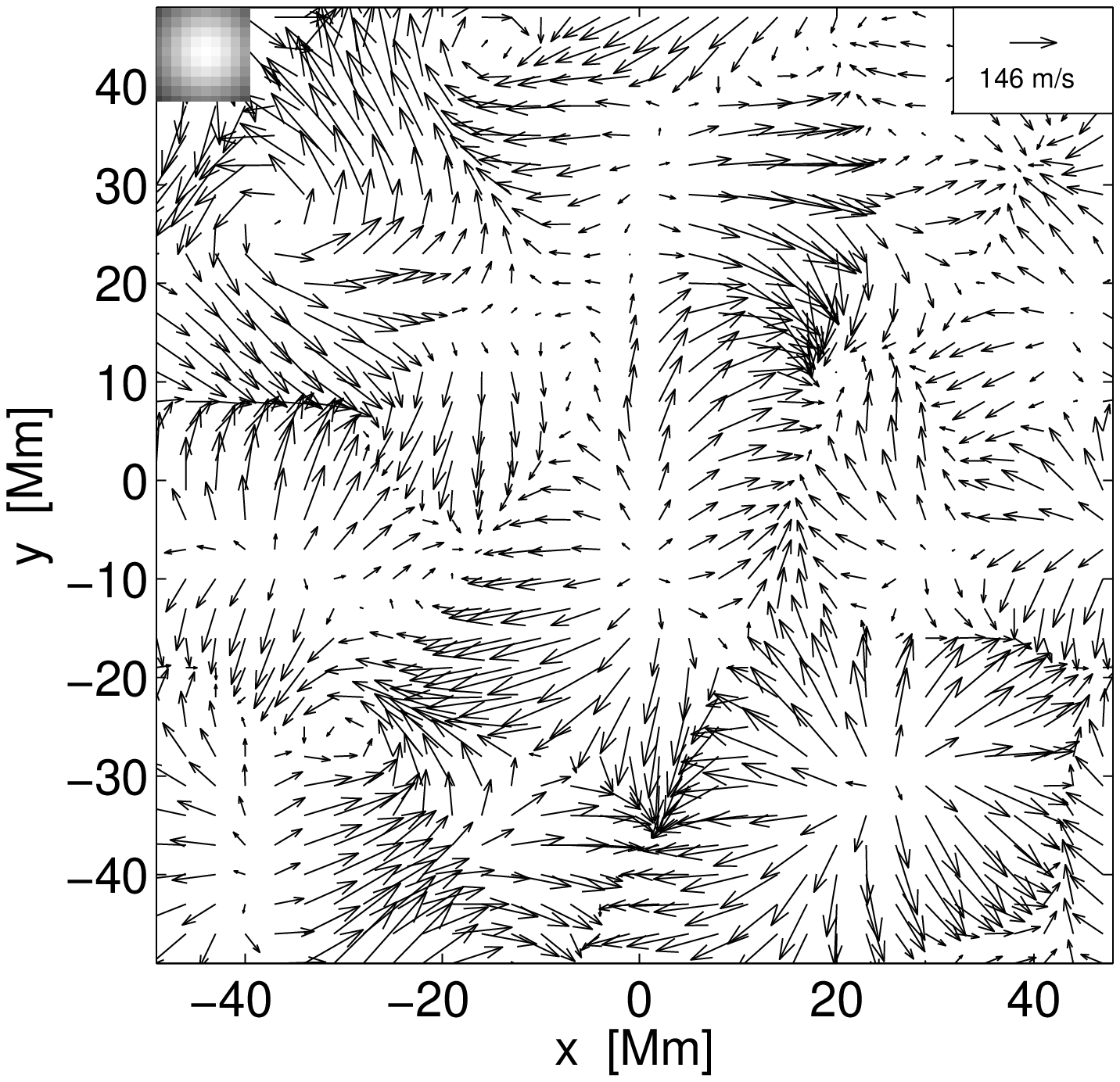} &
   \includegraphics[width=0.2\linewidth,clip=]{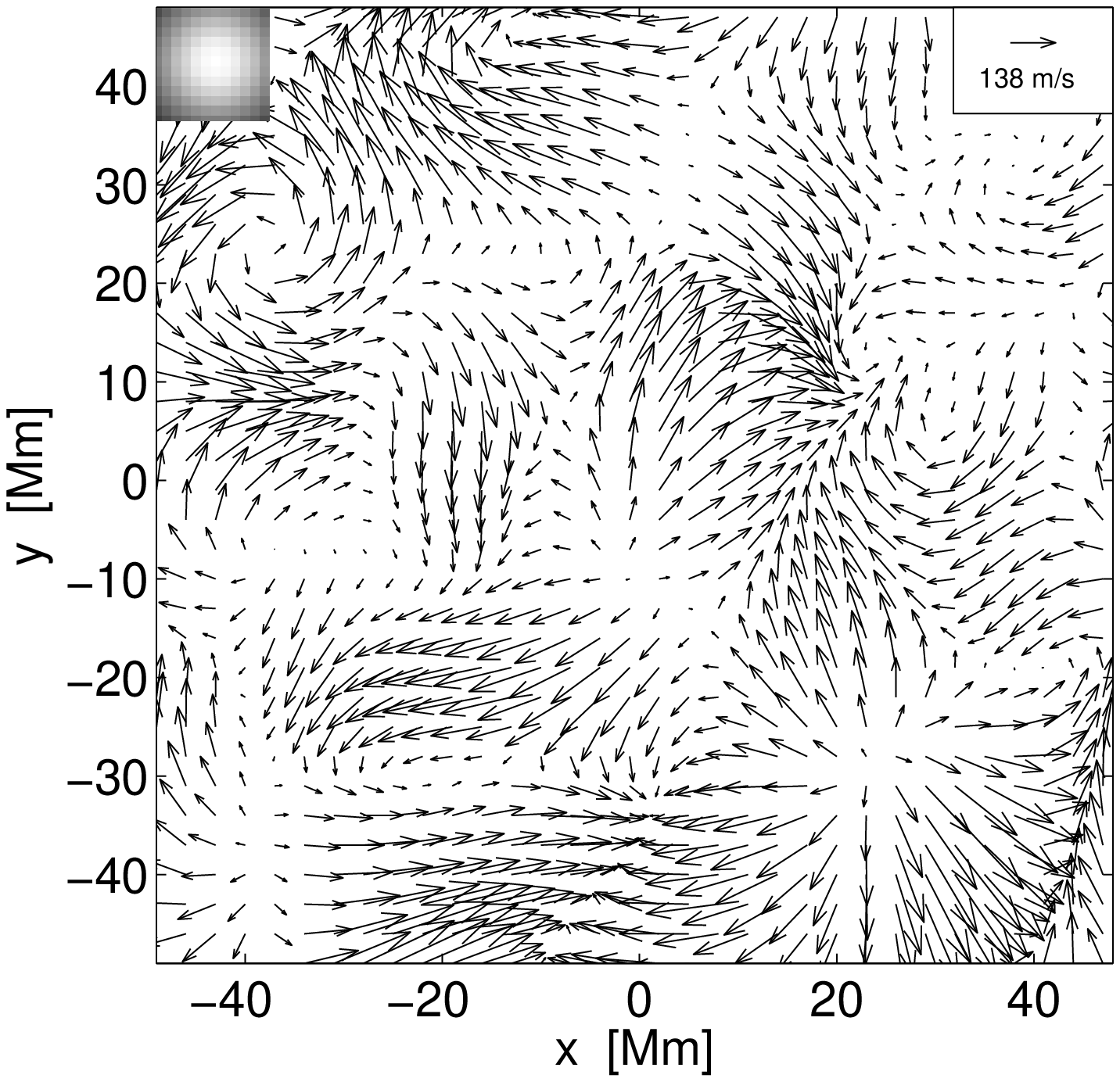} &
   \includegraphics[width=0.2\linewidth,clip=]{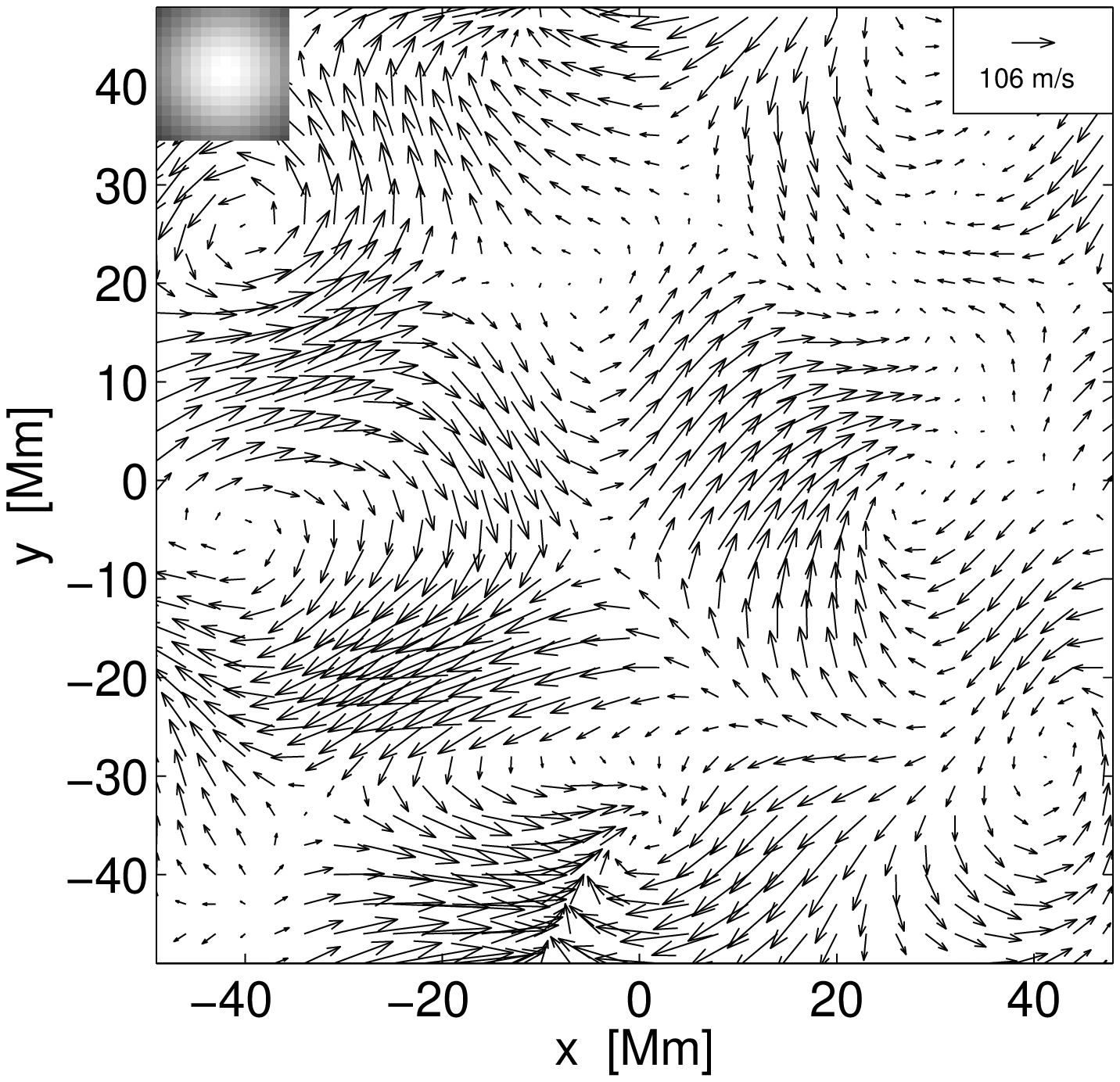} \\
   \includegraphics[width=0.2\linewidth,clip=]{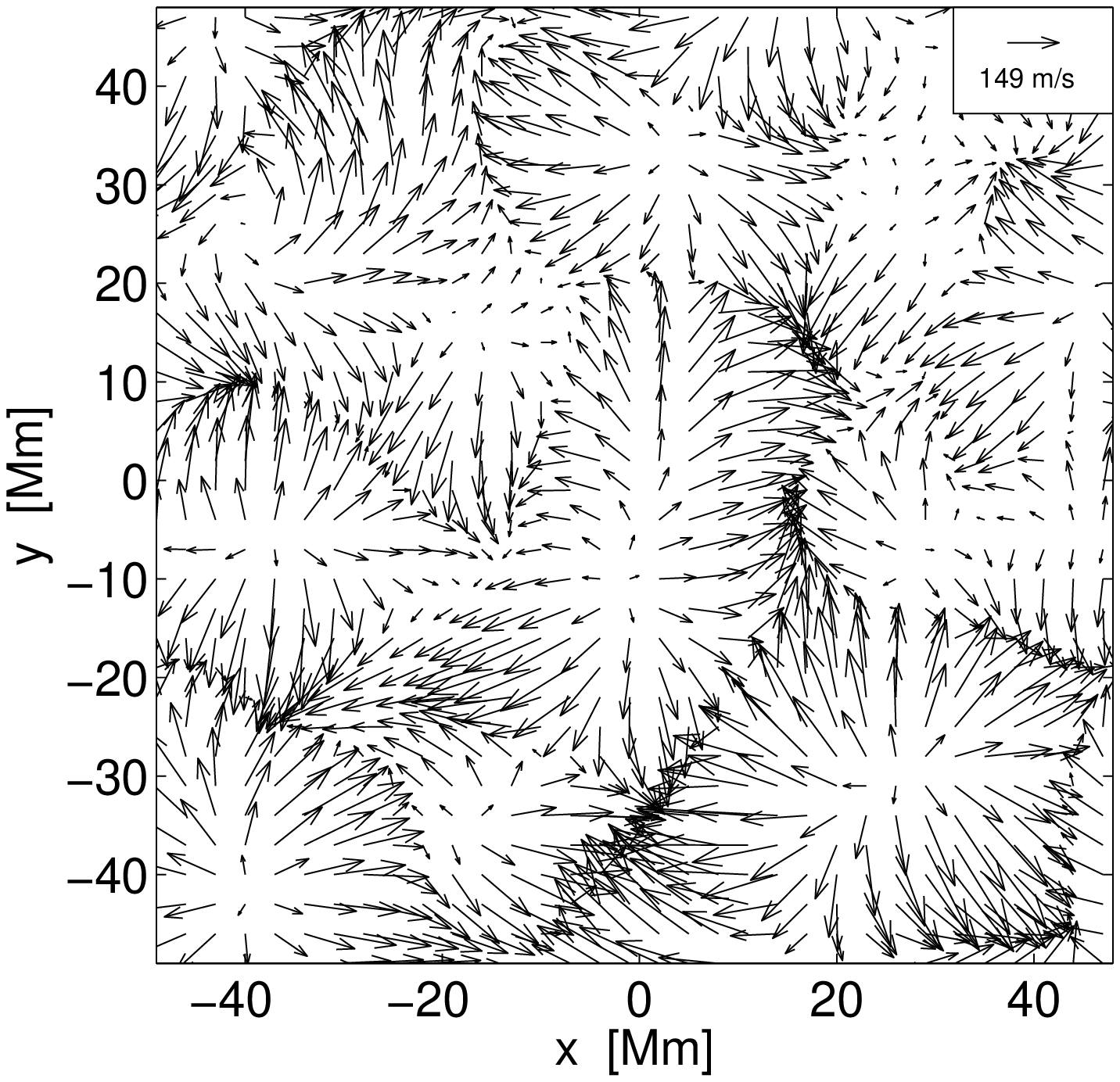} &
   \includegraphics[width=0.2\linewidth,clip=]{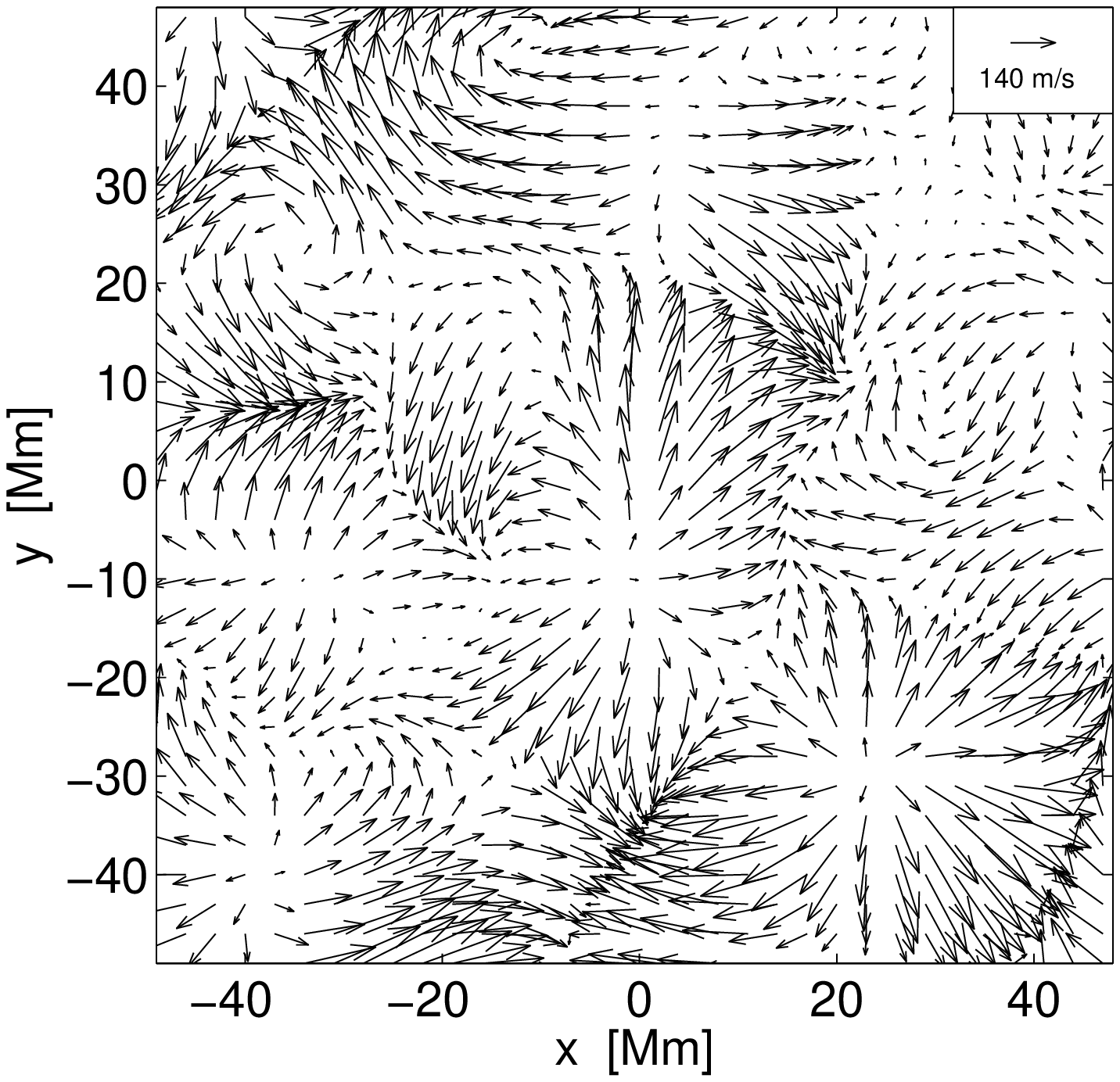} &
   \includegraphics[width=0.2\linewidth,clip=]{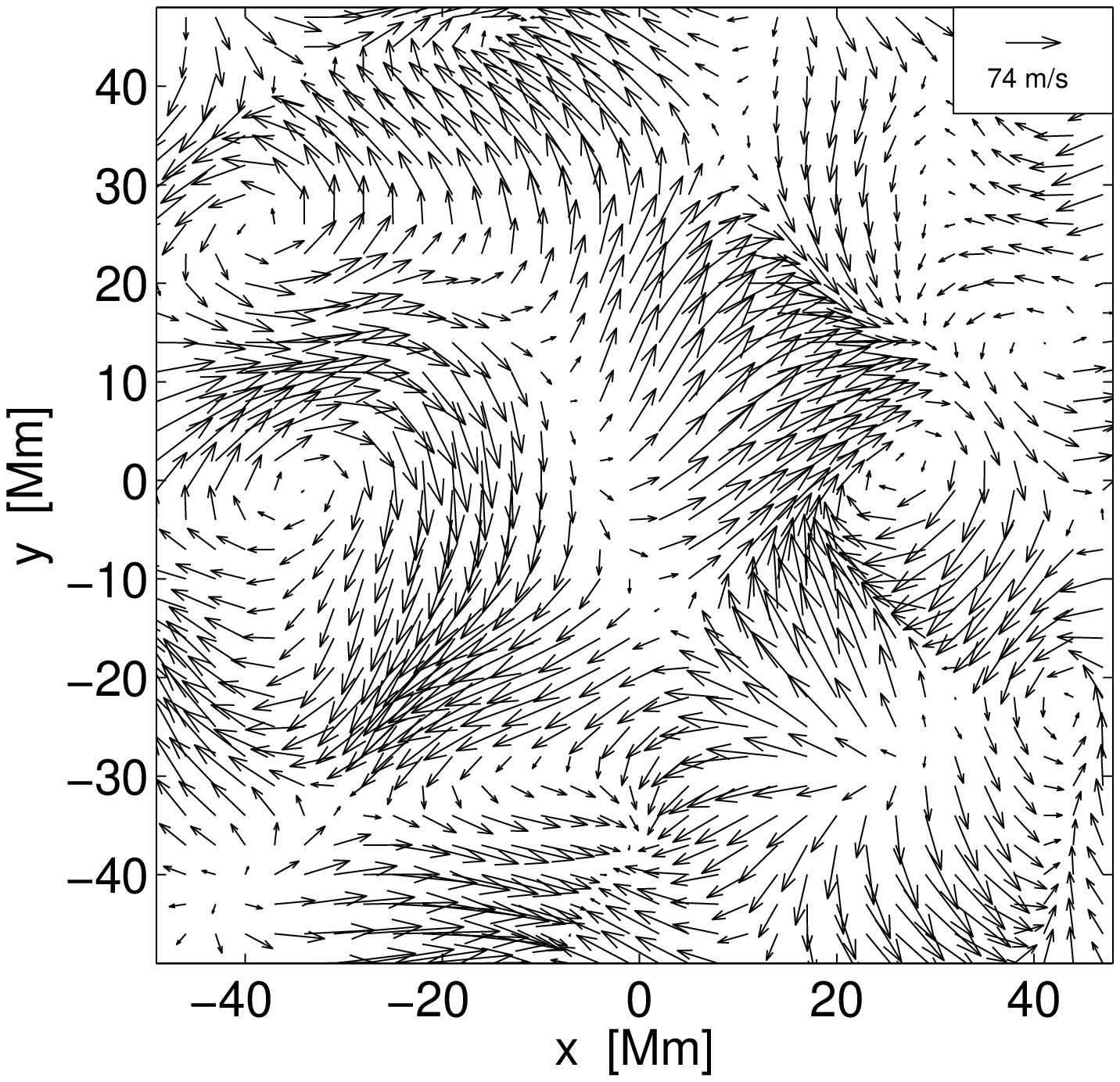} \\
   \includegraphics[width=0.2\linewidth,clip=]{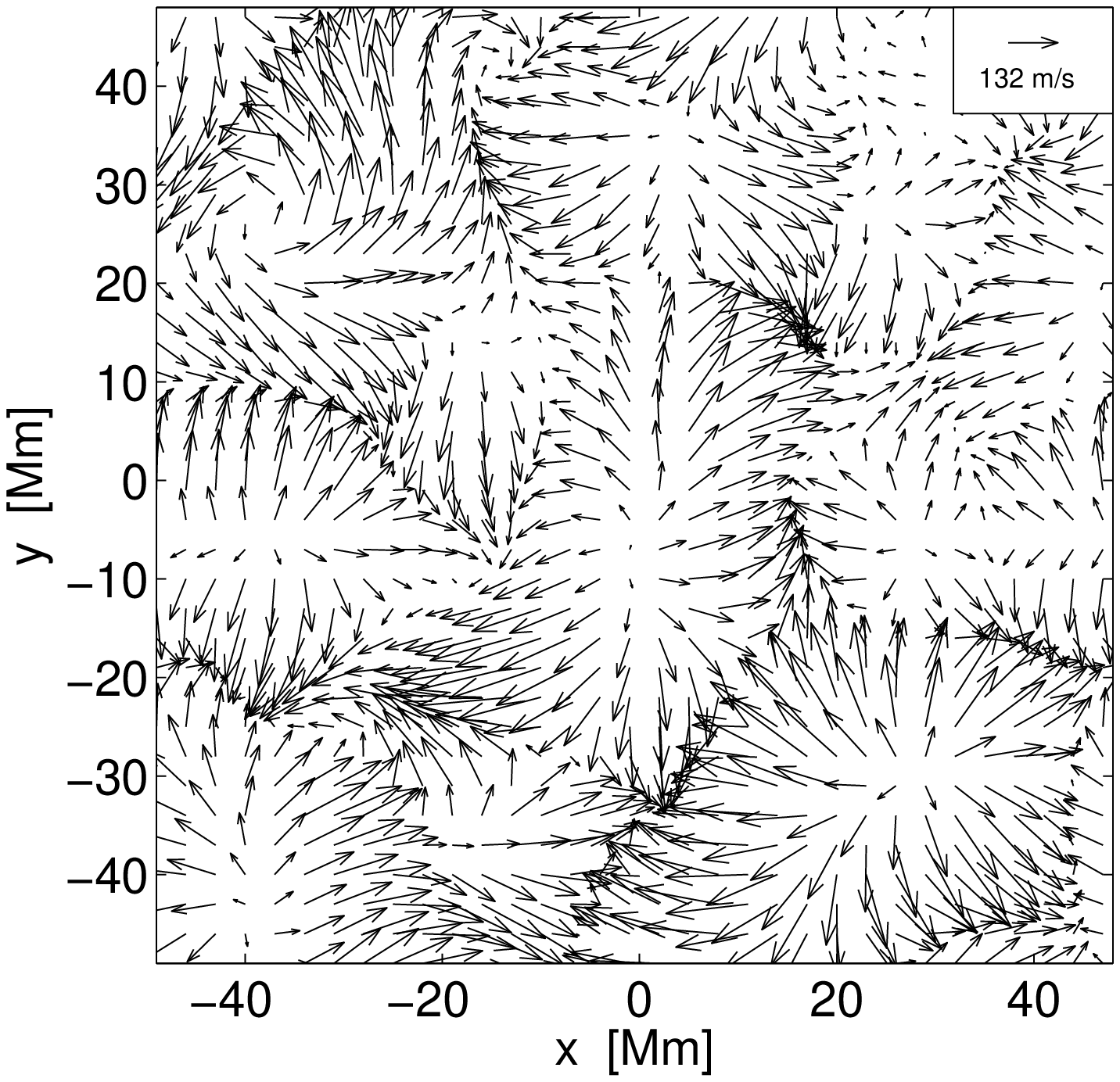} &
   \includegraphics[width=0.2\linewidth,clip=]{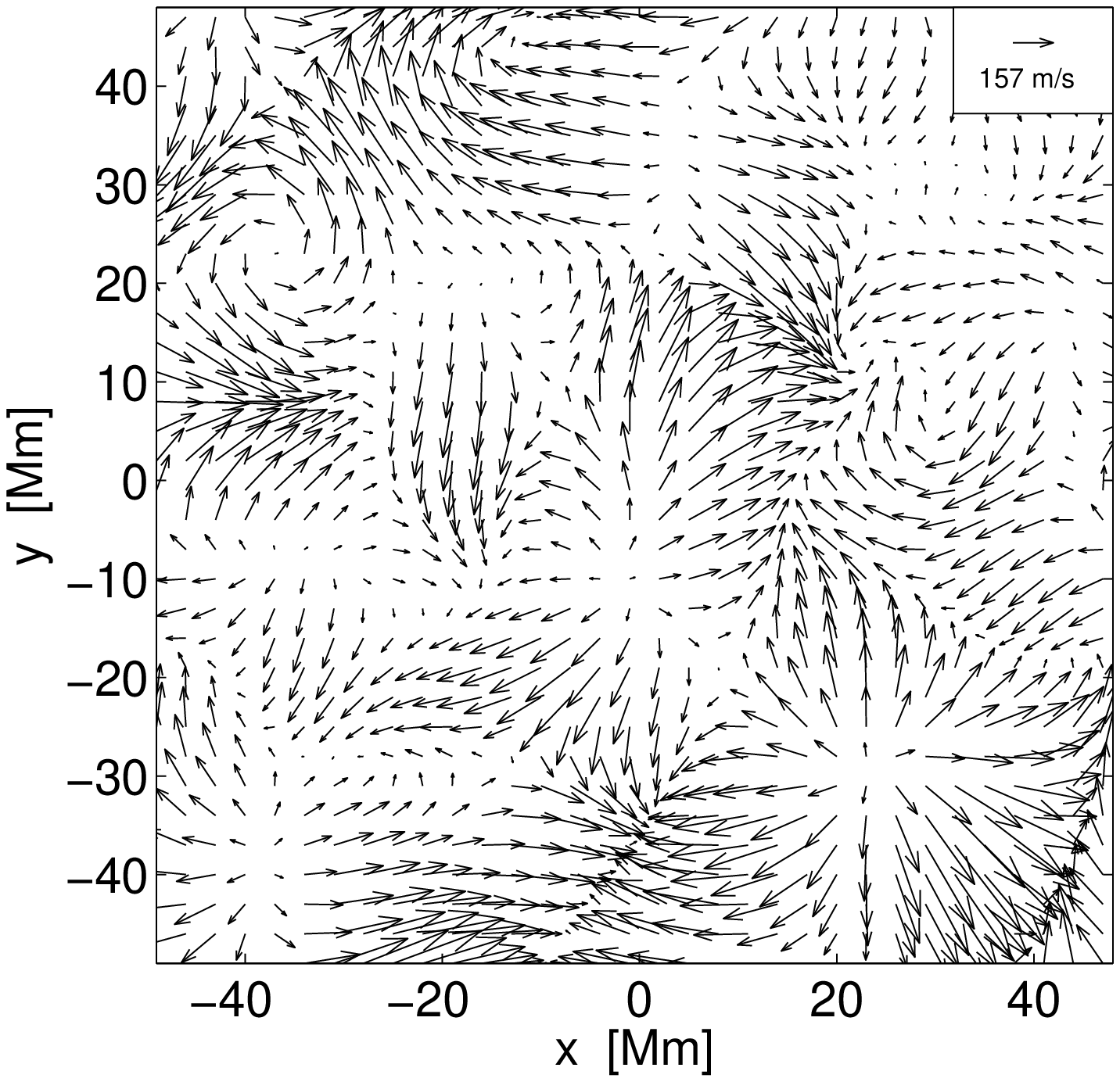} &
   \includegraphics[width=0.2\linewidth,clip=]{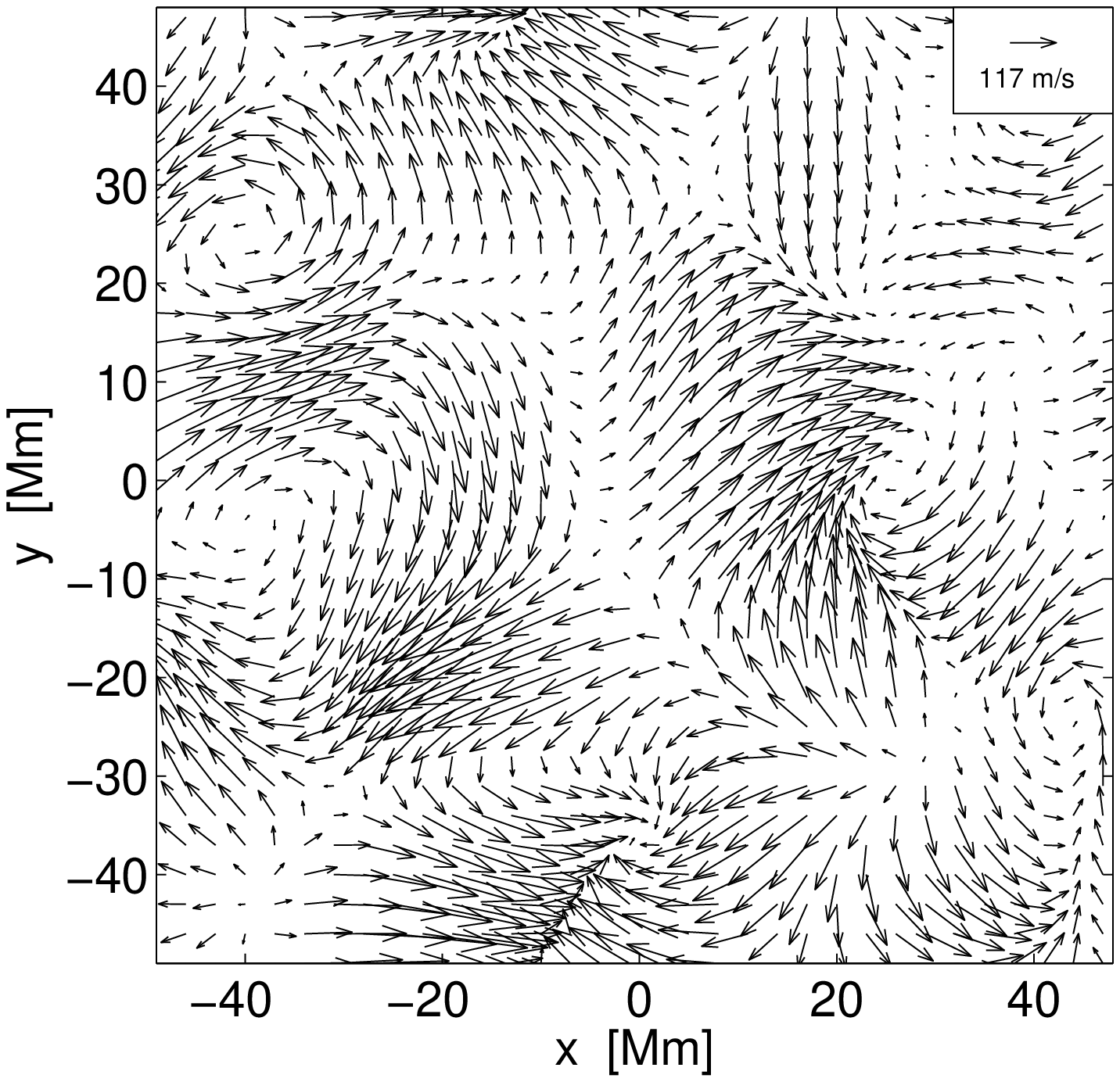} \\
   \includegraphics[width=0.2\linewidth,clip=]{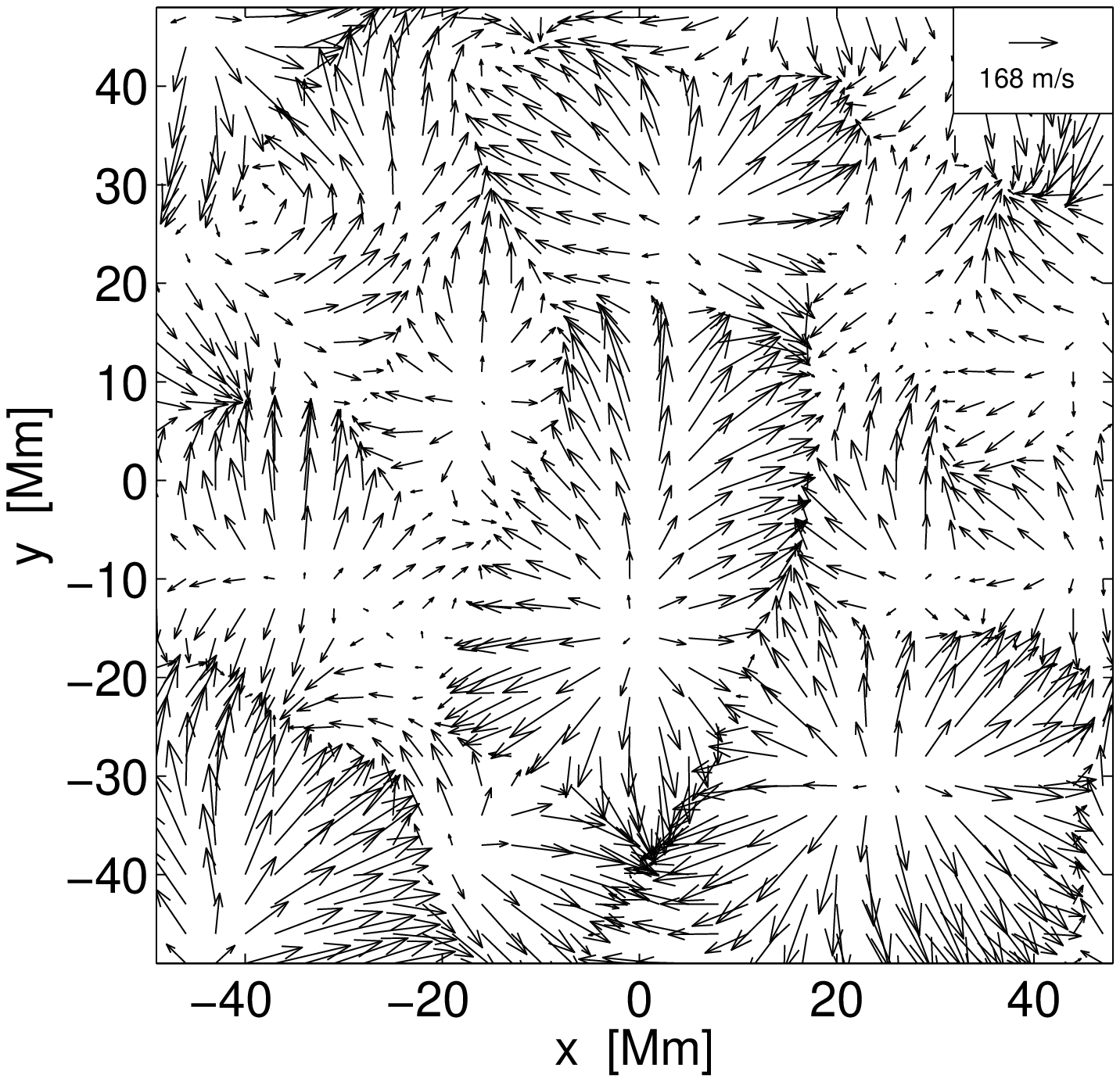} &
   \includegraphics[width=0.2\linewidth,clip=]{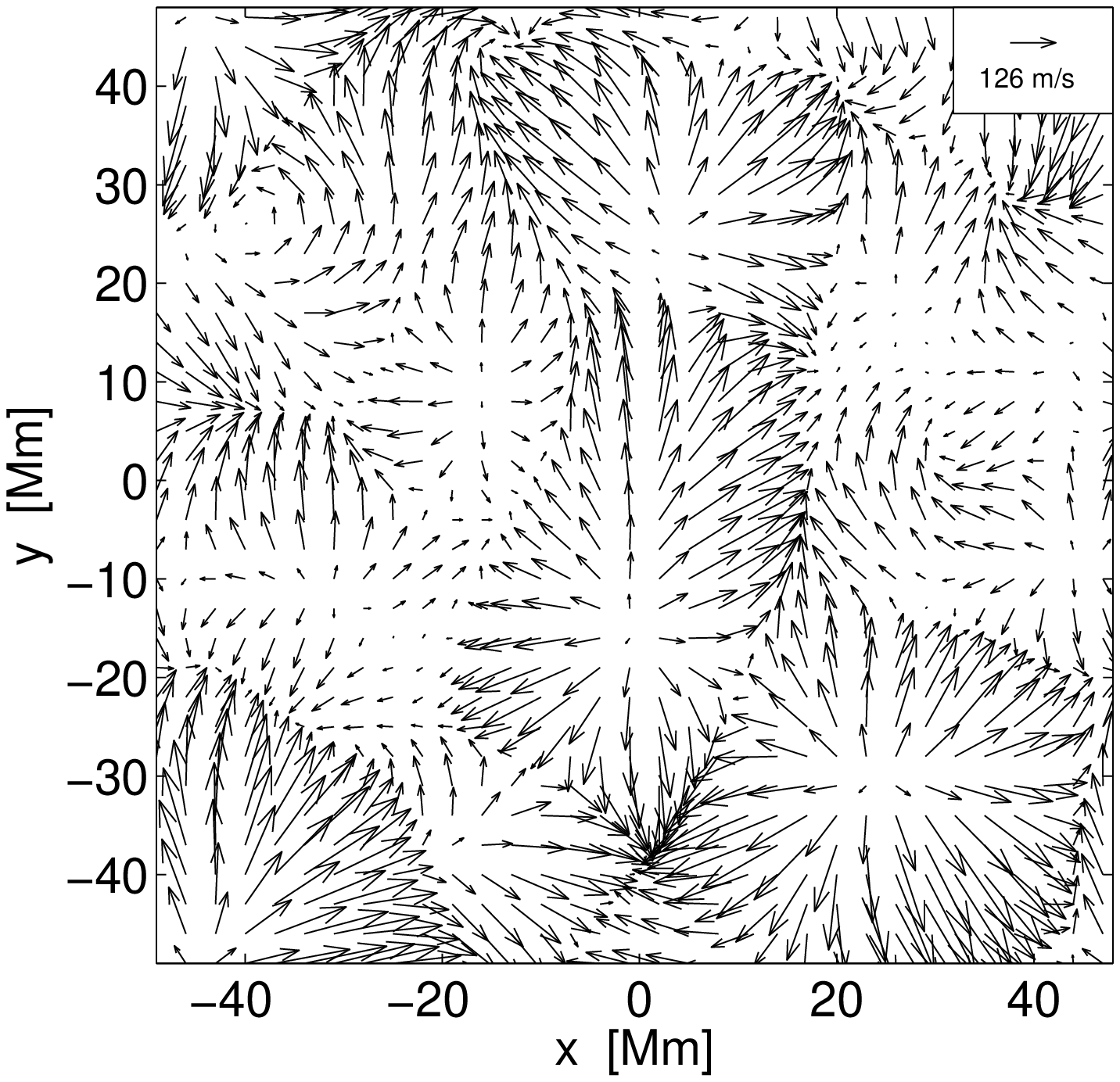} &
   \includegraphics[width=0.2\linewidth,clip=]{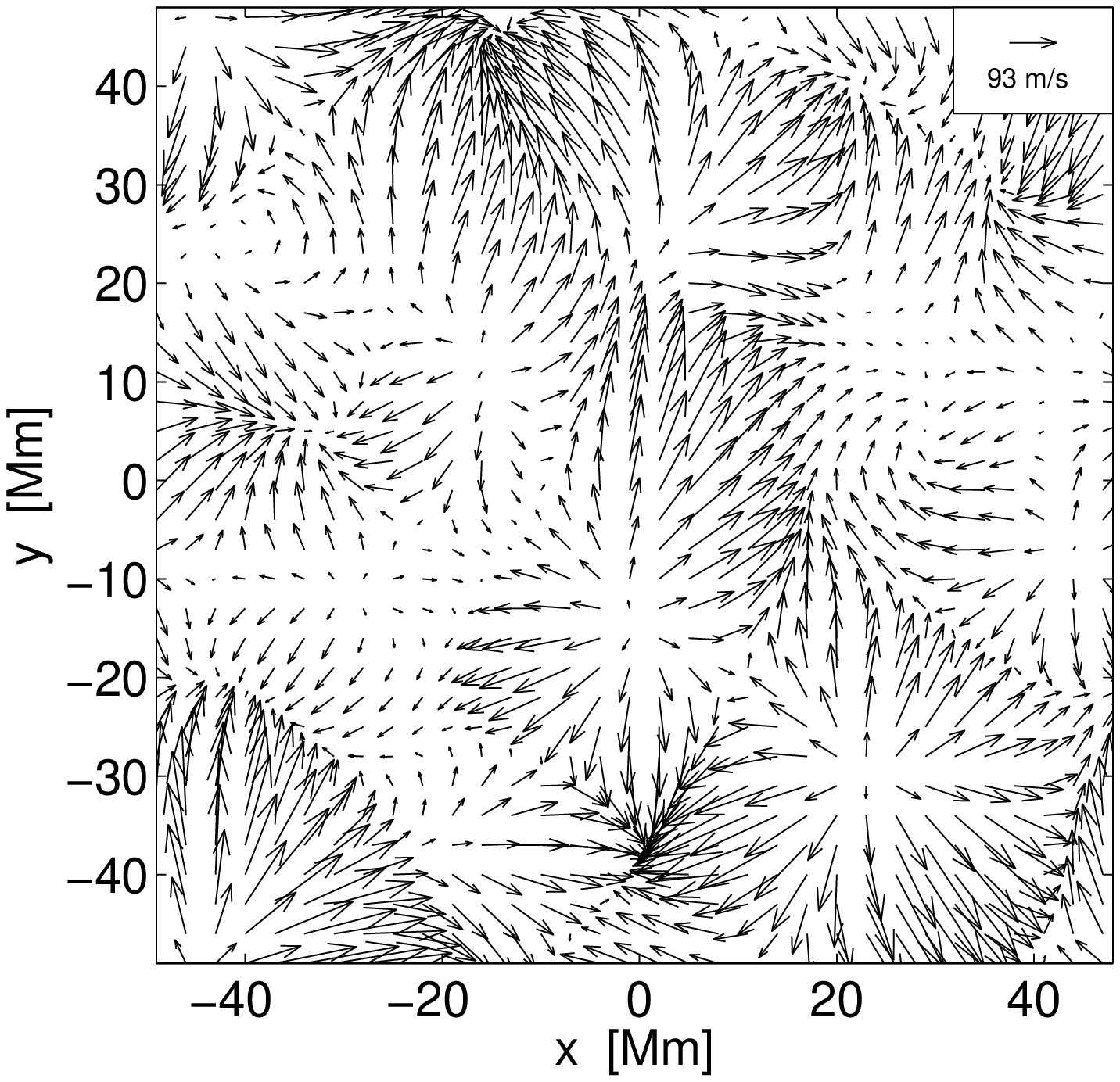}
  \end{array}$
 \end{center}
 \caption{QS1 horizontal $(v_x, v_y)$ inversion flow maps for the ridge (first row), phase-speed (second row), and ridge+phase-speed (third row) travel-time differences for depths (left to right) 1, 3 and 5~Mm. The smoothed simulation flow maps (i.e. $v_{x,y}^{\rm tgt}$) at these depths are shown in the bottom row. The noise for each inversion is $\sim$35~$\rm{ms^{-1}}$ and the reference arrows represent the RMS velocity corresponding to each flow map. The 2D target function at each depth is shown in the upper lefthand corner of the first row figures. The width of the box corresponds to the horizontal FWHM of each target function and represents the approximate spatial resolution of each flow map. All maps in the same column have identical horizontal resolution. Other parameters of each inversion are presented in Table~\ref{tab1}, inversion set~1. Correlation coefficients found between each inversion and the simulation are presented in Table~\ref{tab2}.}
\label{fig:qs1vx}
\end{figure*}

\begin{figure*}[t]
\begin{center}$
\begin{array}{c c c}
\includegraphics[width=0.2\linewidth,clip=]{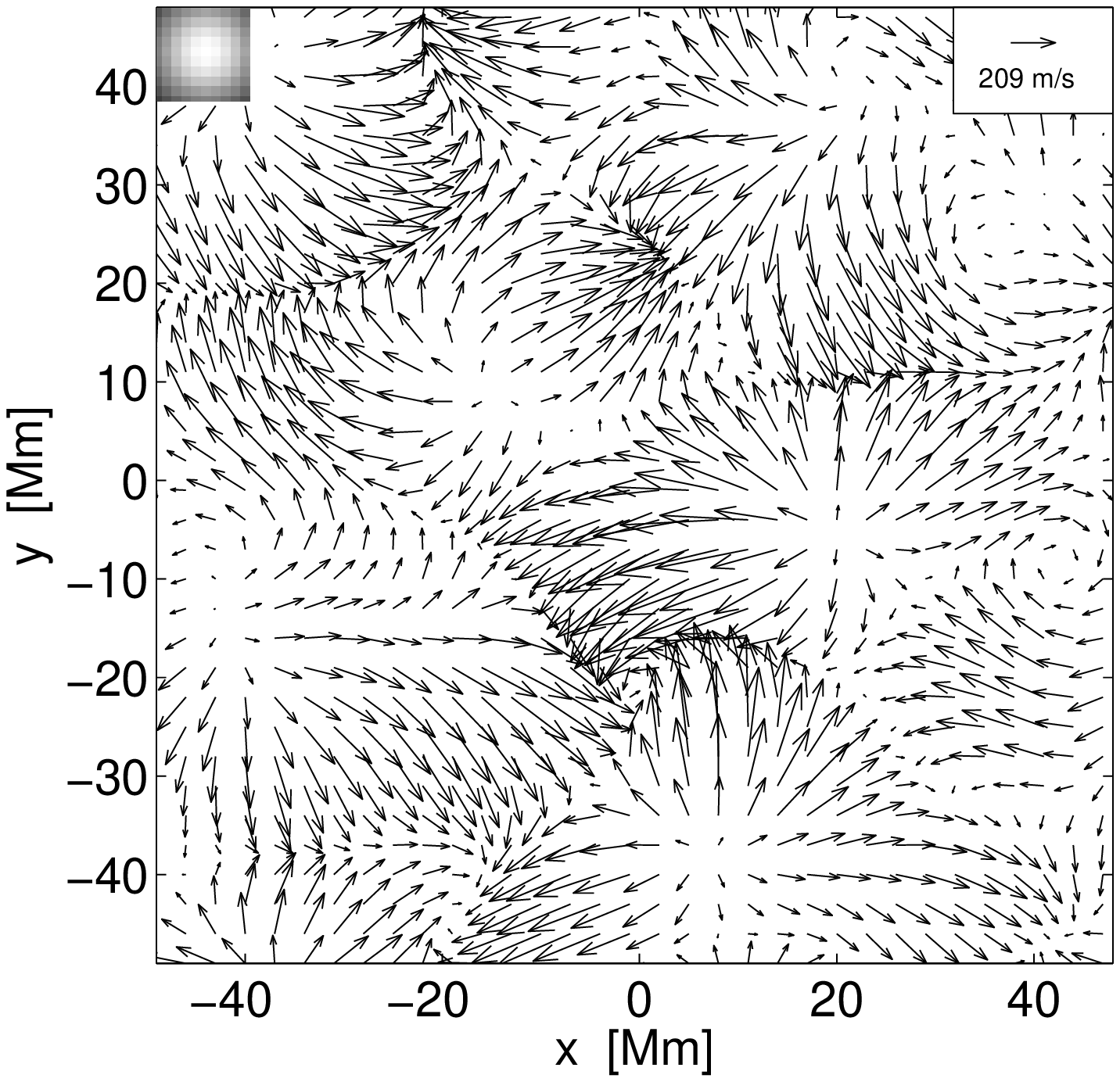} &
\includegraphics[width=0.2\linewidth,clip=]{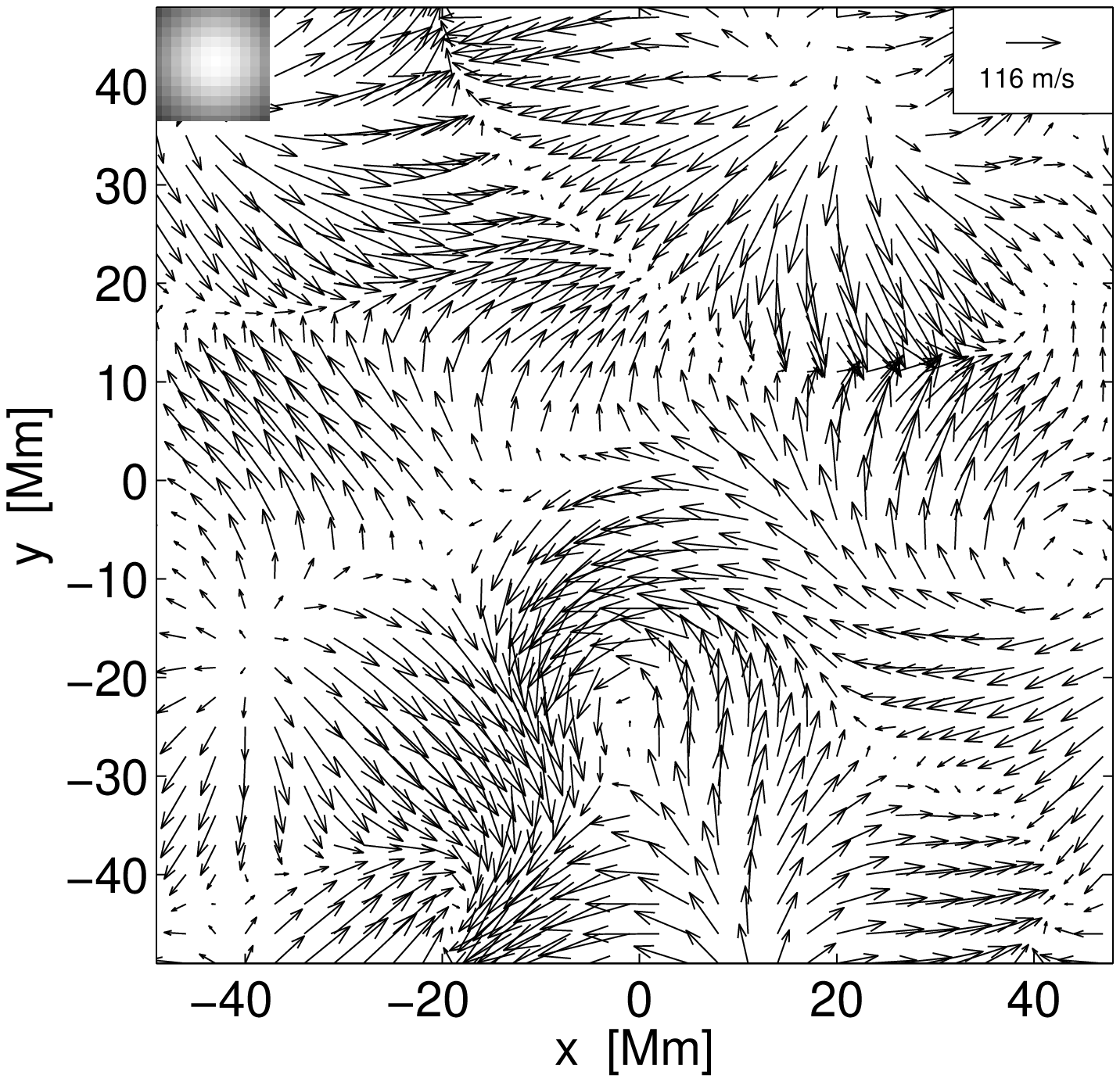} &
\includegraphics[width=0.2\linewidth,clip=]{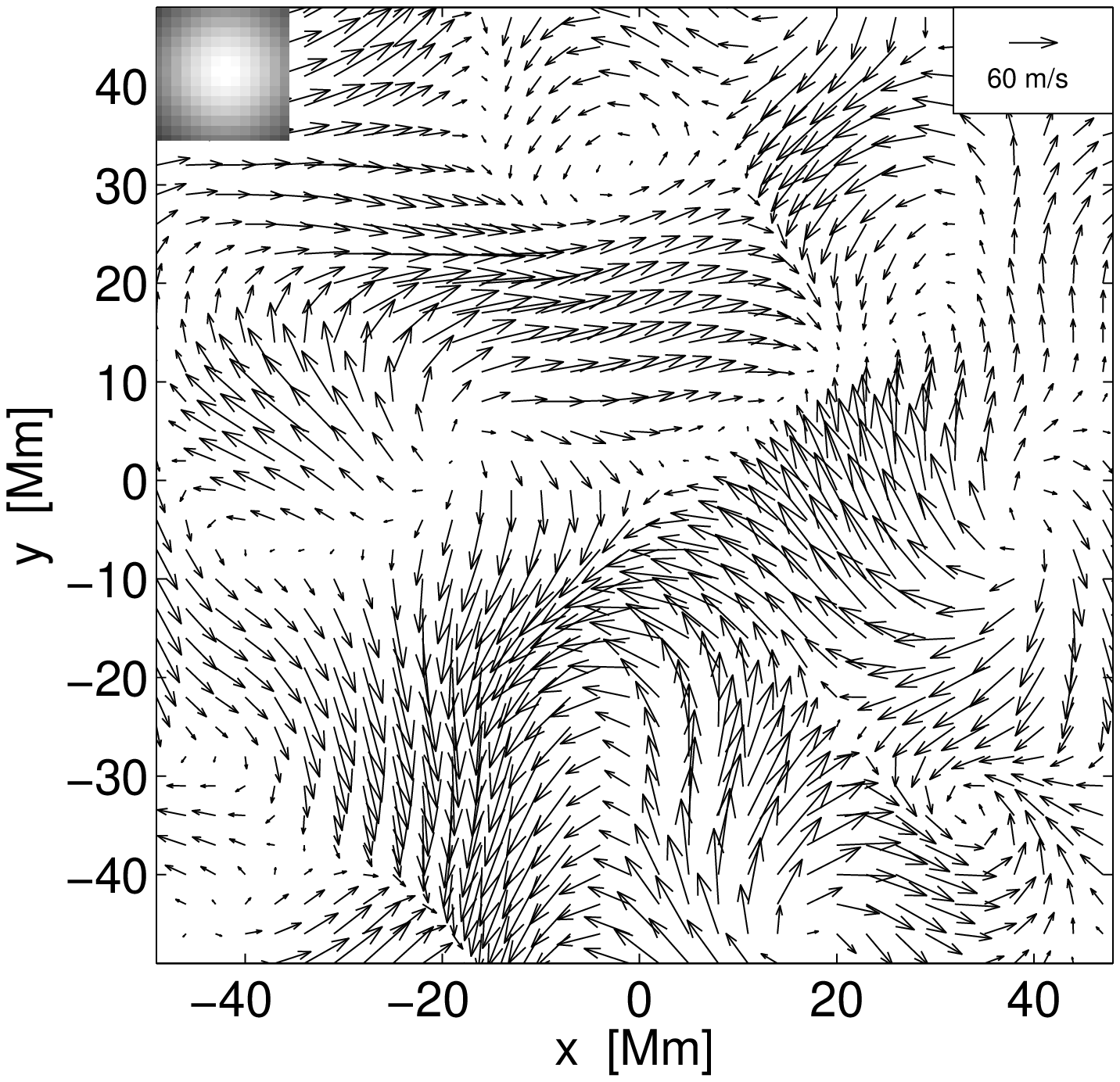} \\
\includegraphics[width=0.2\linewidth,clip=]{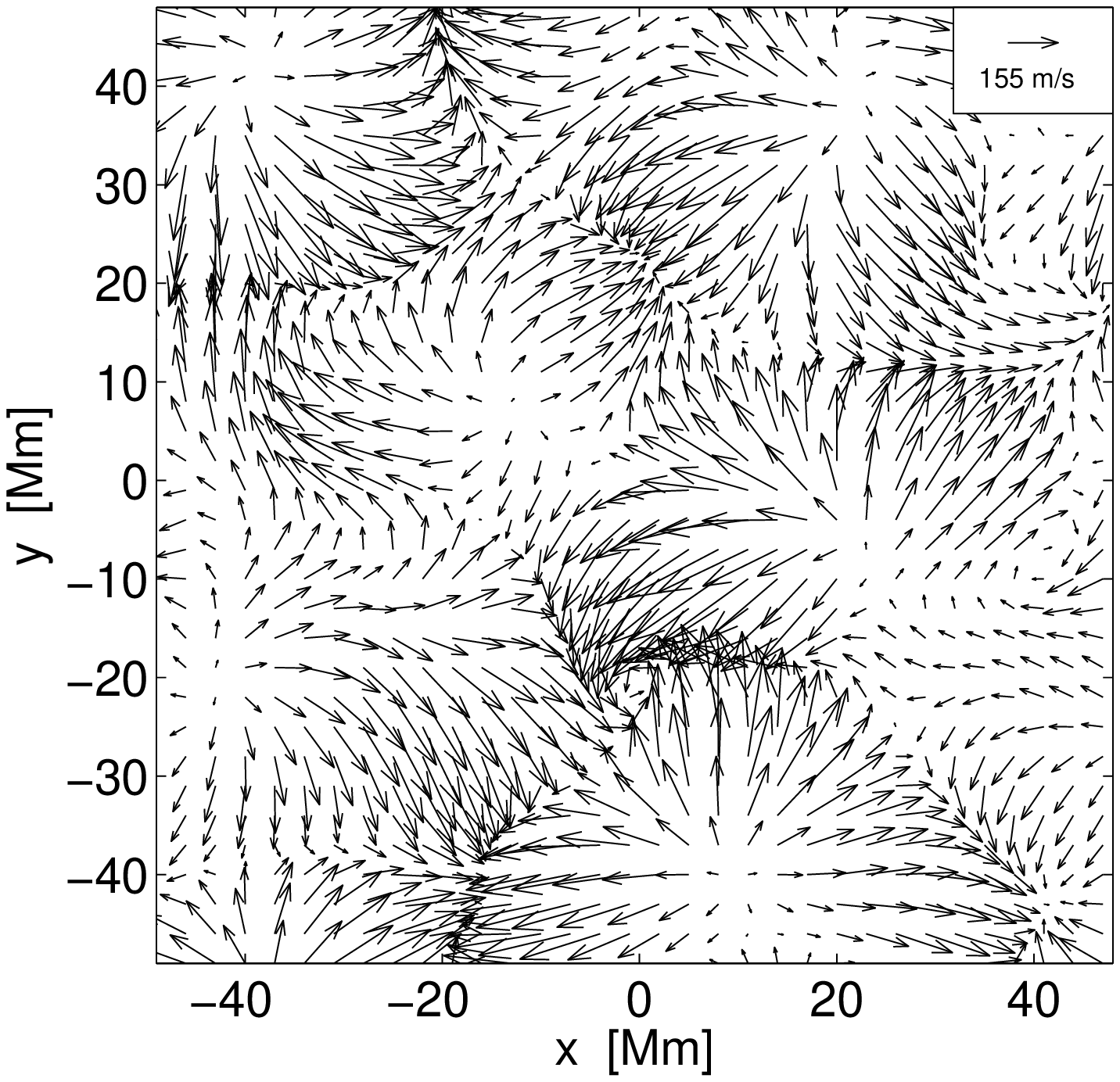} &
\includegraphics[width=0.2\linewidth,clip=]{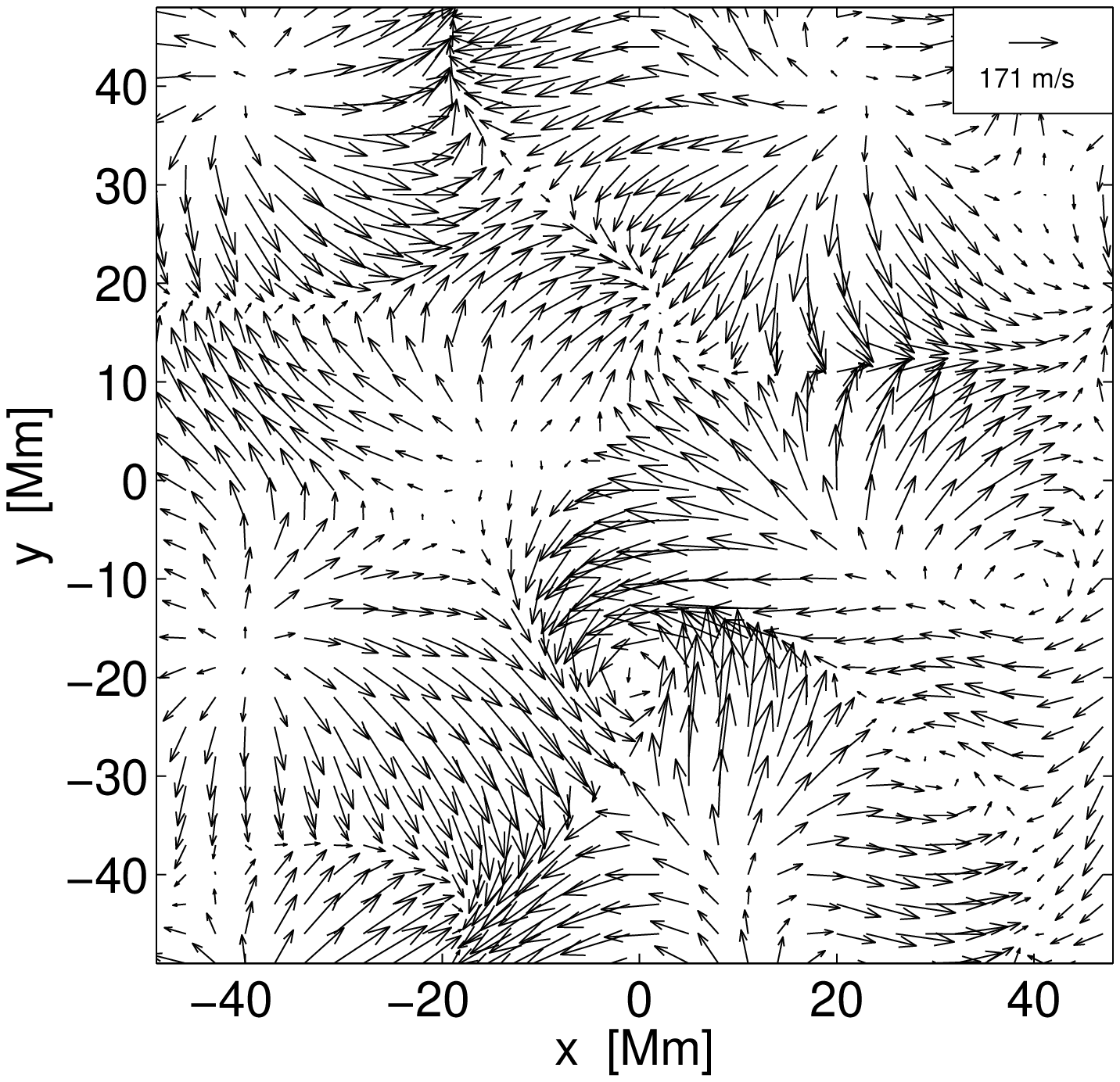} &
\includegraphics[width=0.2\linewidth,clip=]{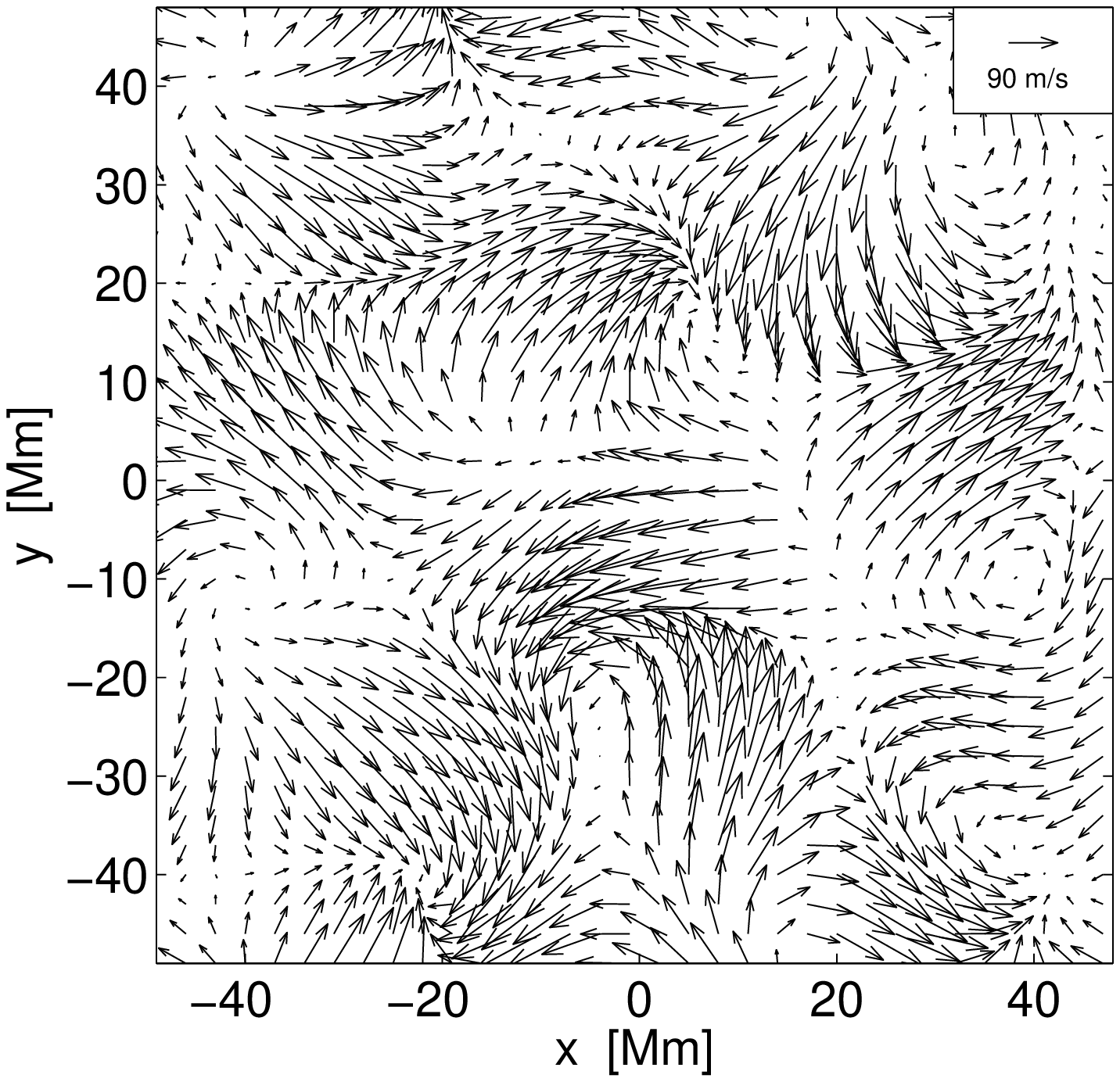} \\
\includegraphics[width=0.2\linewidth,clip=]{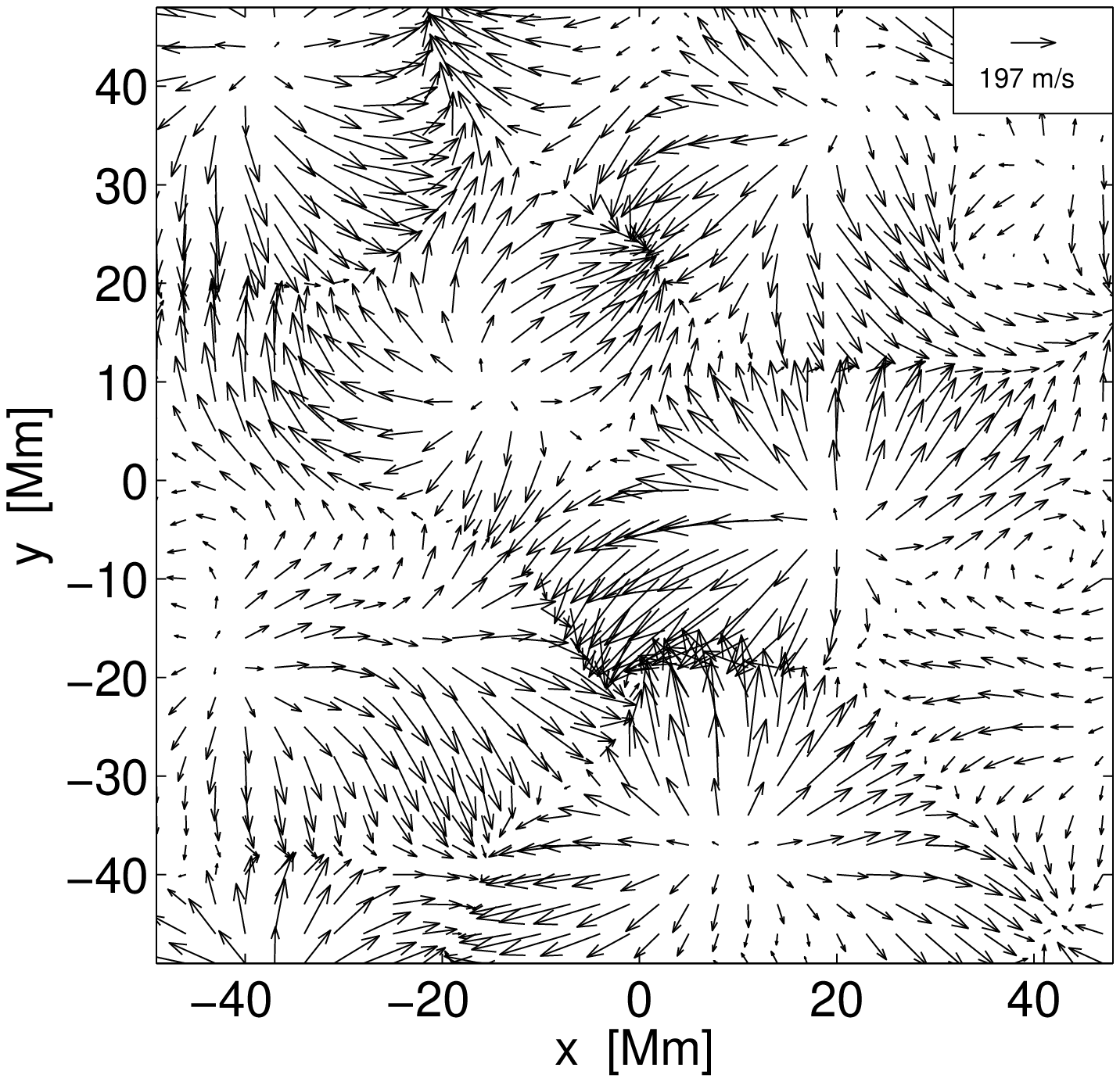} &
\includegraphics[width=0.2\linewidth,clip=]{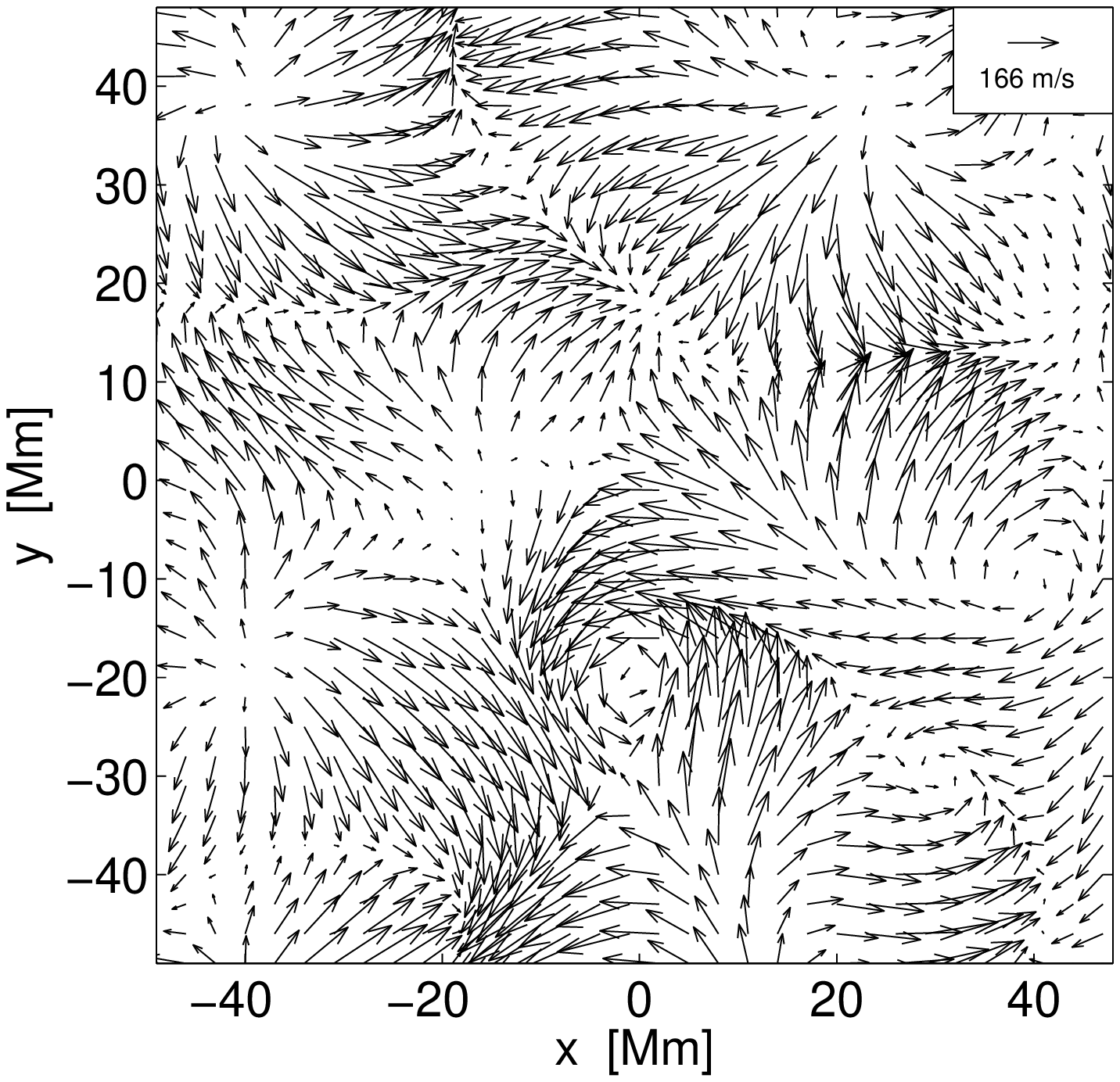} &
\includegraphics[width=0.2\linewidth,clip=]{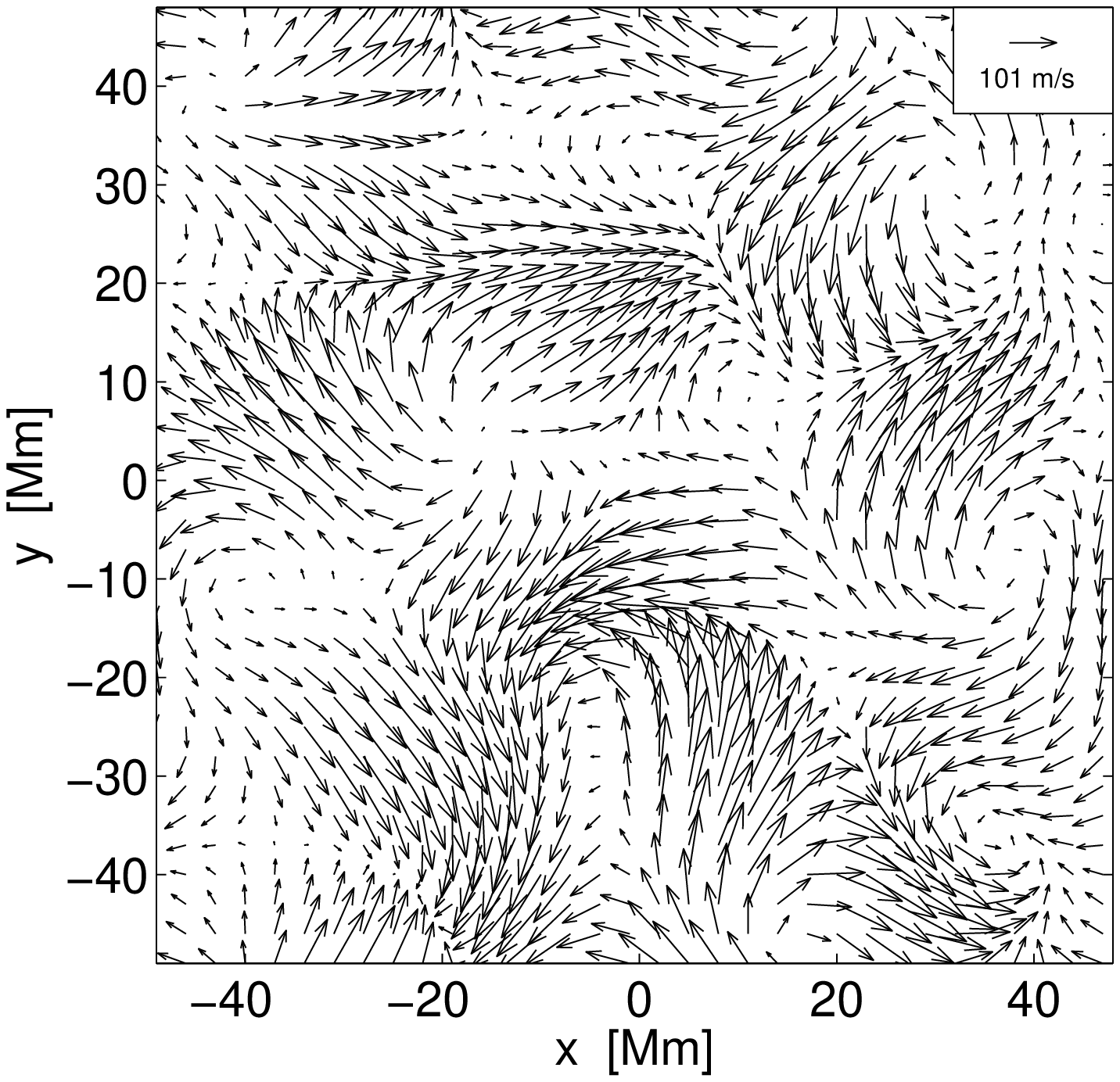} \\
\includegraphics[width=0.2\linewidth,clip=]{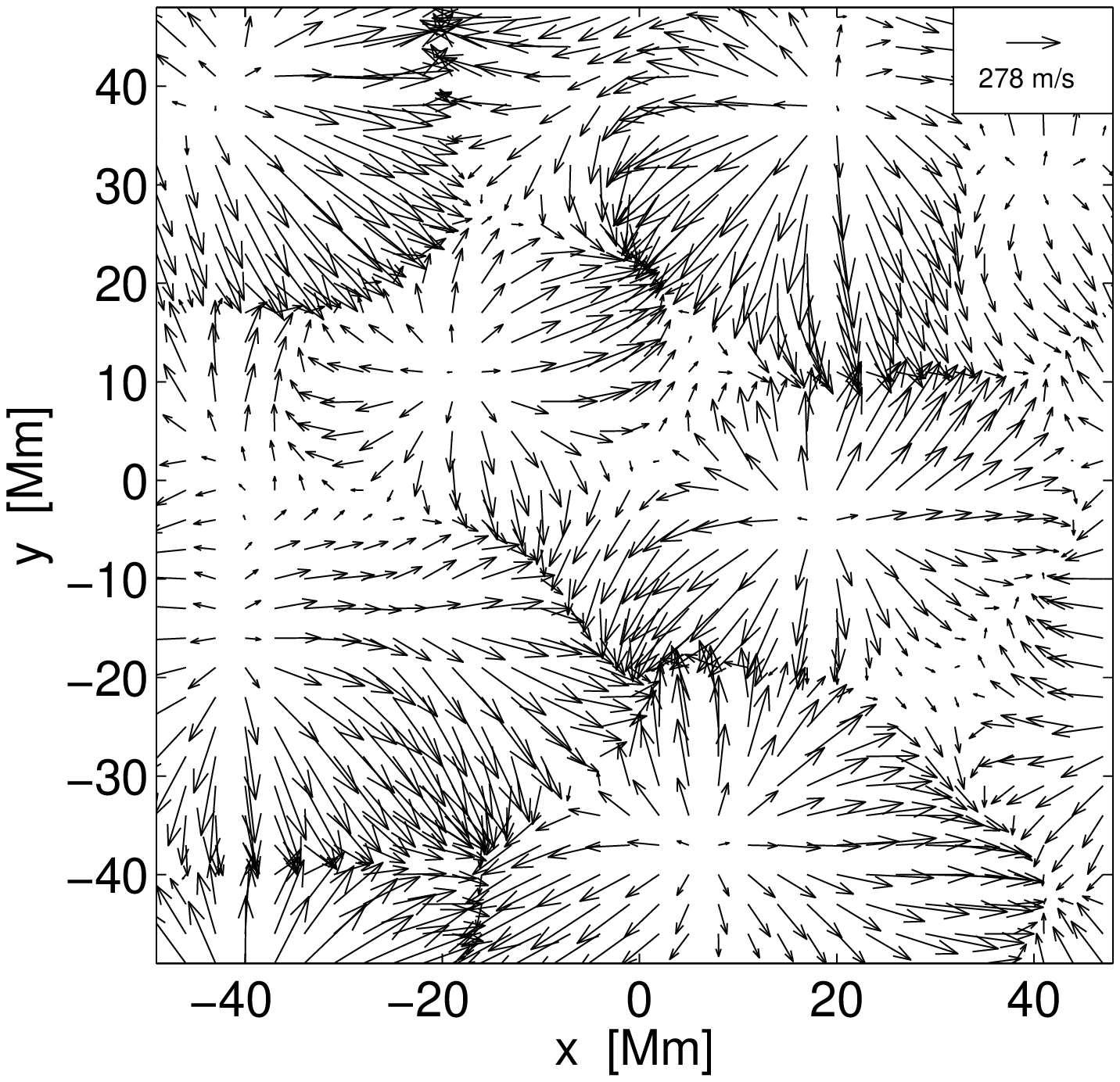} &
\includegraphics[width=0.2\linewidth,clip=]{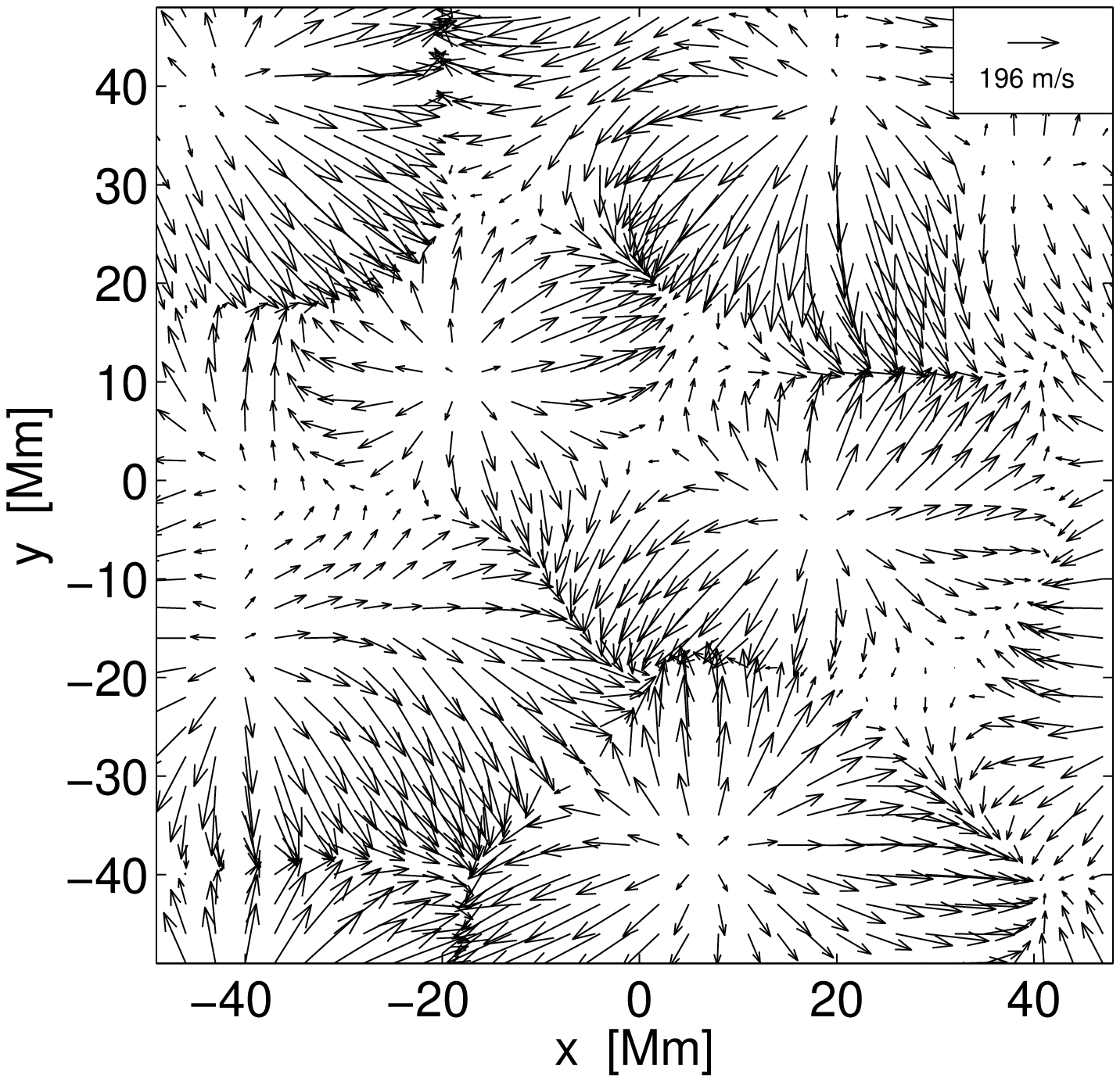} &
\includegraphics[width=0.2\linewidth,clip=]{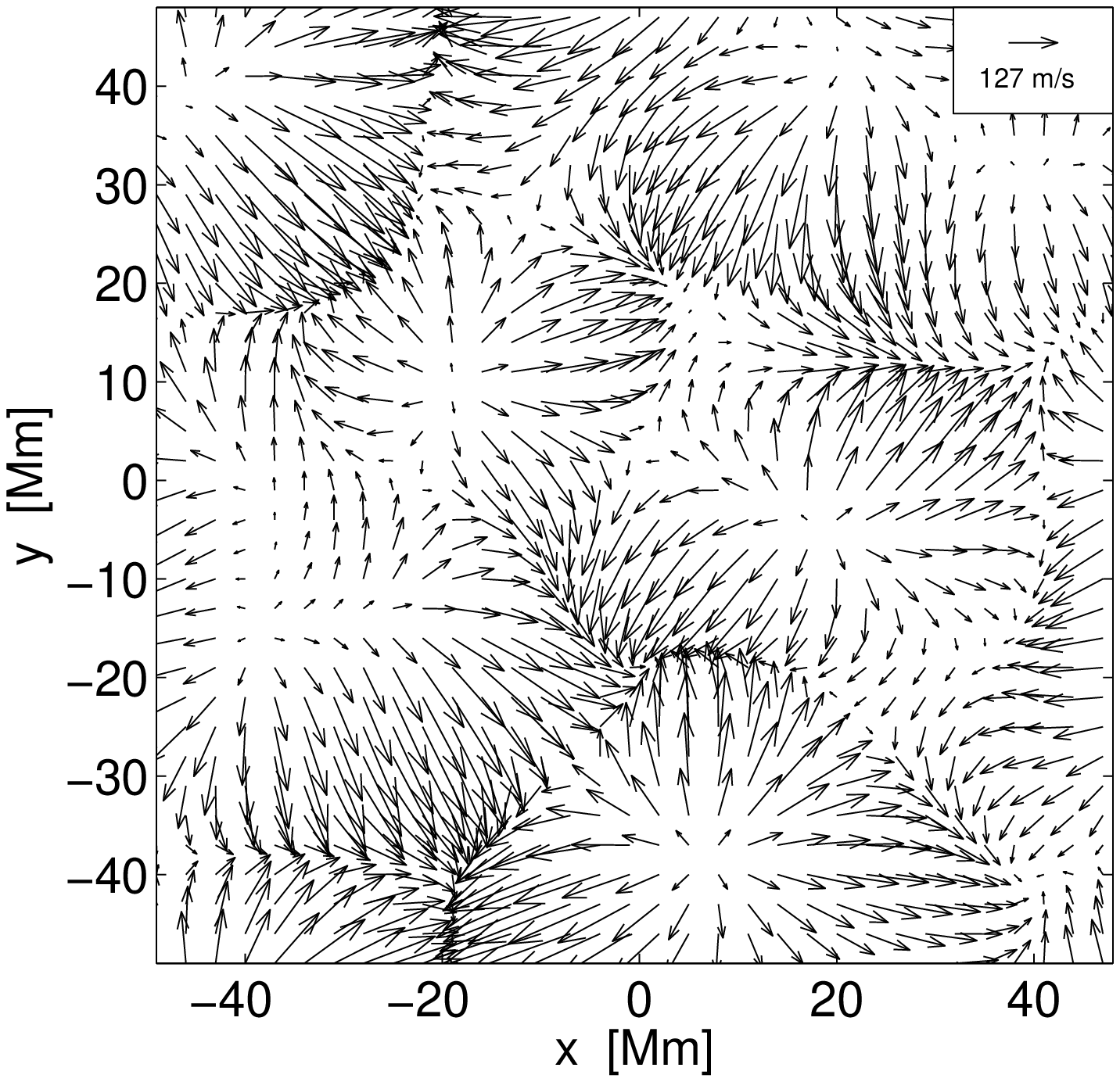}
\end{array}$
\end{center}
\caption{QS2 horizontal $(v_x, v_y)$ inversion flow maps for the ridge (first row), phase-speed (second row), and ridge+phase-speed (third row) travel-time differences for depths (left to right) 1, 3 and 5~Mm. The smoothed simulation flow maps (i.e. $v_{x,y}^{\rm tgt}$) at these depths are shown in the bottom row. The noise for each inversion is $\sim$35~$\rm{ms^{-1}}$ and the reference arrows represent the RMS velocity corresponding to each flow map. The 2D target function at each depth is shown in the upper lefthand corner of the first row figures. The width of the box corresponds to the horizontal FWHM of each target function and represents the approximate spatial resolution of each flow map. All maps in the same column have identical horizontal resolution. Other parameters of each inversion are presented in Table~\ref{tab1}, inversion set~1. Correlation coefficients found between each inversion and the simulation are presented in Table~\ref{tab2}.}
\label{fig:qs2vx}
\end{figure*}

Figure~\ref{fig:qs1vx} shows the resulting $v_{x,y}^{\rm inv}$ horizontal flow maps for QS1 obtained at each depth for the separate ridge and phase-speed filter inversions (rows 1 and 2 respectively), and for the combined ridge+phase-speed filter inversions (row 3). The noise for each inversion is $\sim35\,\rm{m\,s^{-1}}$. For comparison, the simulation flows, $v_{x,y}^{\rm tgt}$ (see Eq.~\ref{valph}), at each corresponding depth are shown in row 4. These represent the best case scenario that we can hope to accomplish with our inversions. The approximate horizontal and vertical resolutions of each inversion (i.e. the horizontal and vertical FWHM of the target function) along with the regularization parameters used are given in Table~\ref{tab1}, inversion set~1. We find that these inversions capture the large-scale supergranule-sized flows present in the simulation, as well as the smaller features throughout the domain at the two shallowest depths. We therefore measure high spatial correlation between $v_{x,y}^{\rm inv}$ and $v_{x,y}^{\rm tgt}$ flow maps down to a depth of 3~Mm for each of the three filtering scenarios (Table~\ref{tab2}). At a depth of 5~Mm, the acoustic wave coverage is small in comparison to the shallower layers, and so it is quite difficult to accurately recover flow structure there. However, the flow correlation at the largest depth is not that small, $\approx 0.6$, and there is visible similarity in the maps. We note here that the $v_x$ correlation is higher than that of $v_y$ in every single case. The same phenomena has been documented previously by \citet{zhao2007}, though it is not clear why there should be a preferential bias of one flow component over the other.

\begin{figure}
 \centering
 \includegraphics[width=1.0\linewidth,clip=]{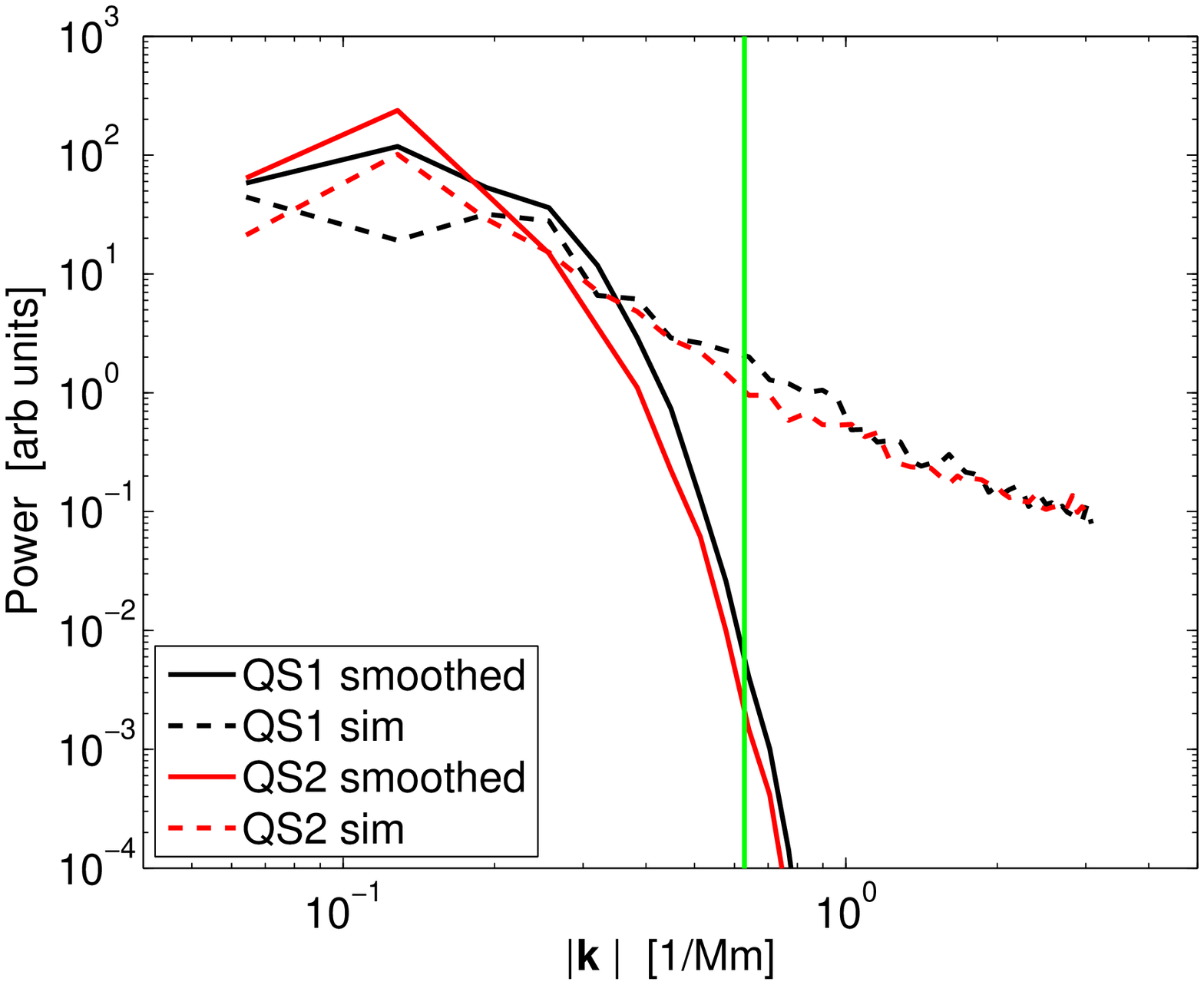}
 \caption{The $v_x$ spatial power averaged over wavevector at a depth of 1~Mm for both time-averaged simulations. The dashed (solid) lines represent the simulation data before (after) convolution with a 3D Gaussian function with a horizontal FWHM=10~Mm and centered at $z_0=1$~Mm. The vertical line marks the wavenumber corresponding to this 10~Mm smoothing value.}
 \label{fig:vpow}
\end{figure}

Figure~\ref{fig:qs2vx} shows the equivalent QS2 $v_{x,y}^{\rm inv}$ horizontal flow maps obtained at each depth for both ridge and phase-speed filter types (rows 1 and 2 respectively), for the combined ridge+phase-speed filters (row 3), and for the simulation at each corresponding depth (row 4). The horizontal and vertical resolutions of each inversion along with the regularization parameters used are identical to the QS1 case and are also given in Table~\ref{tab1}, inversion set~1. Again, each of the three filtering scenarios show strong correlation between $v_{x,y}^{\rm inv}$ and $v_{x,y}^{\rm tgt}$ in the upper 3~Mm, dropping substantially as we reach a depth of 5~Mm (Table~\ref{tab2}). 

The divergence correlation coefficients presented for the QS2 simulation in Table~\ref{tab2} tend to show a slightly higher correlation than for the QS1 case. This could be attributed to the fact that QS1 contains more small-scale structure that the inversions have a hard time resolving, even after smoothing. To support this conjecture, Fig.~\ref{fig:vpow} shows the $v_x$ spatial power averaged over wavevector for both simulations at a depth of 1~Mm. This was computed using the simulation data both before and after convolution with the 1~Mm depth target function. We see in both cases that QS2 has slightly more power over large spatial scales and less at smaller ones. Interestingly, the correlation differences between $v_x$ and $v_y$ found for QS1 are not present in the results for QS2, as seen in Table~\ref{tab2}.

\begin{figure}
 \centerline{
  \includegraphics[width=.5\linewidth,clip=]{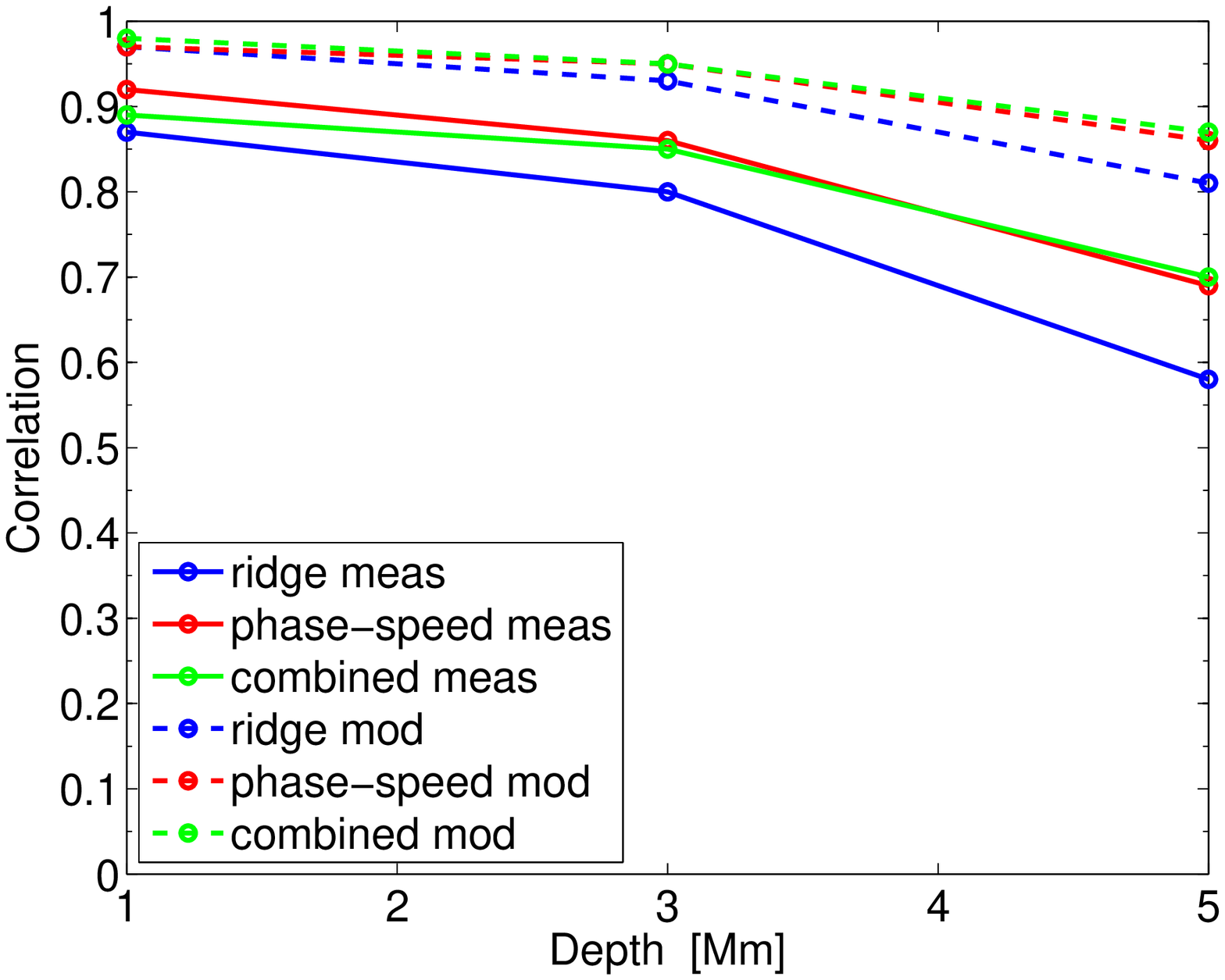} 
  \includegraphics[width=.5\linewidth,clip=]{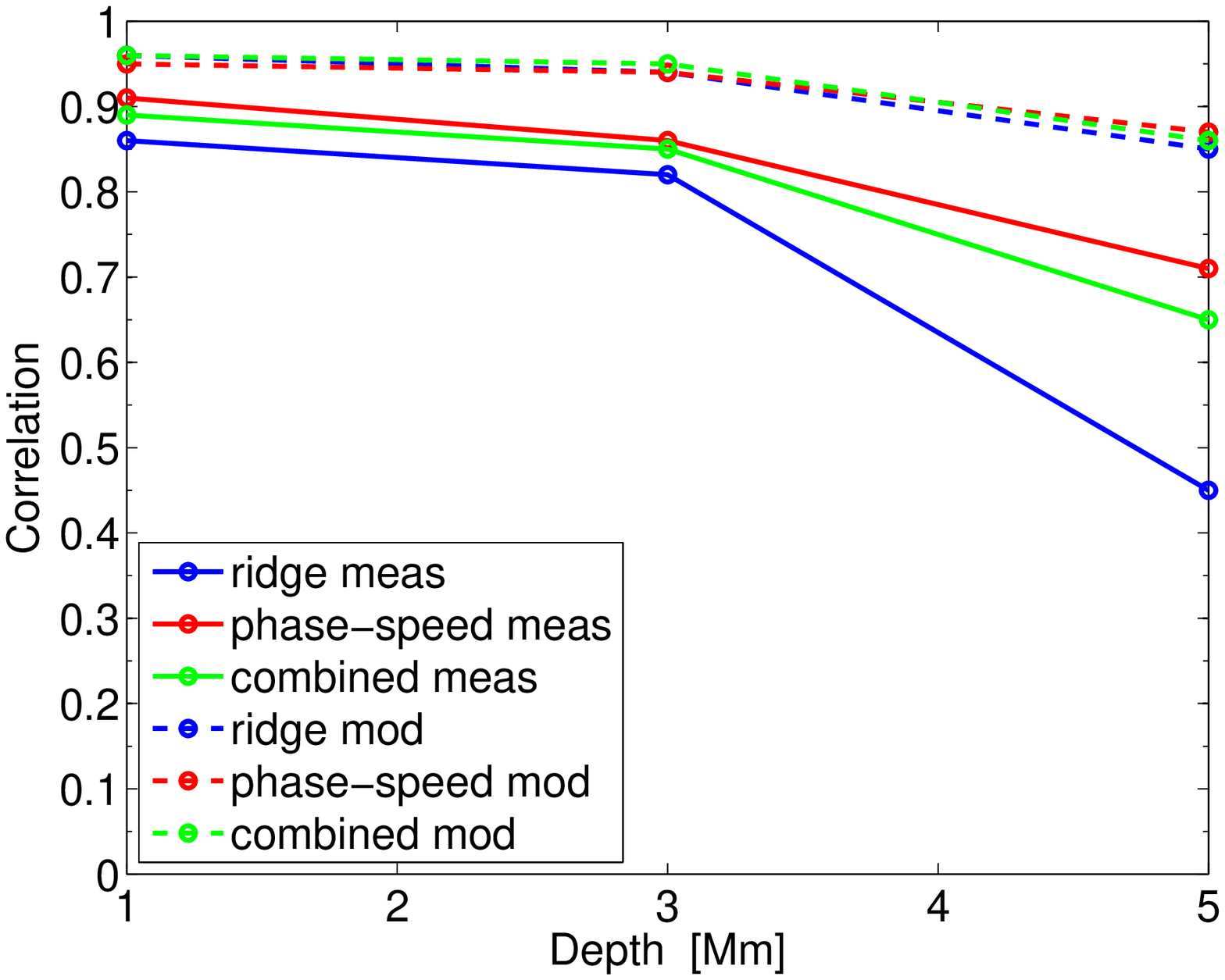}}
 \caption{Correlation between $v_x^{\rm inv}$ and $v_x^{\rm tgt}$ for QS1 (left) and QS2 (right) after inverting measured (``meas,'' solid lines) or forward-modeled travel times (``mod,'' dashed lines).}
 \label{forward_vx}
\end{figure}

Additionally, forward-modeled travel times were inverted to compare with the results from the measured ones to see if there are significant differences in the results. Figure~\ref{forward_vx} shows the correlation between $v_x^{\rm inv}$ and $v_x^{\rm tgt}$ for both QS1 and QS2 at each depth for each filtering scheme when either the measured or modeled travel times were used. In both cases, the $v_x^{\rm inv}$ flows obtained using the forward-modeled travel times are more consistent with $v_x^{\rm tgt}$ at every depth.

\begin{figure}
 \centering
 \includegraphics[width=1.0\linewidth,clip=]{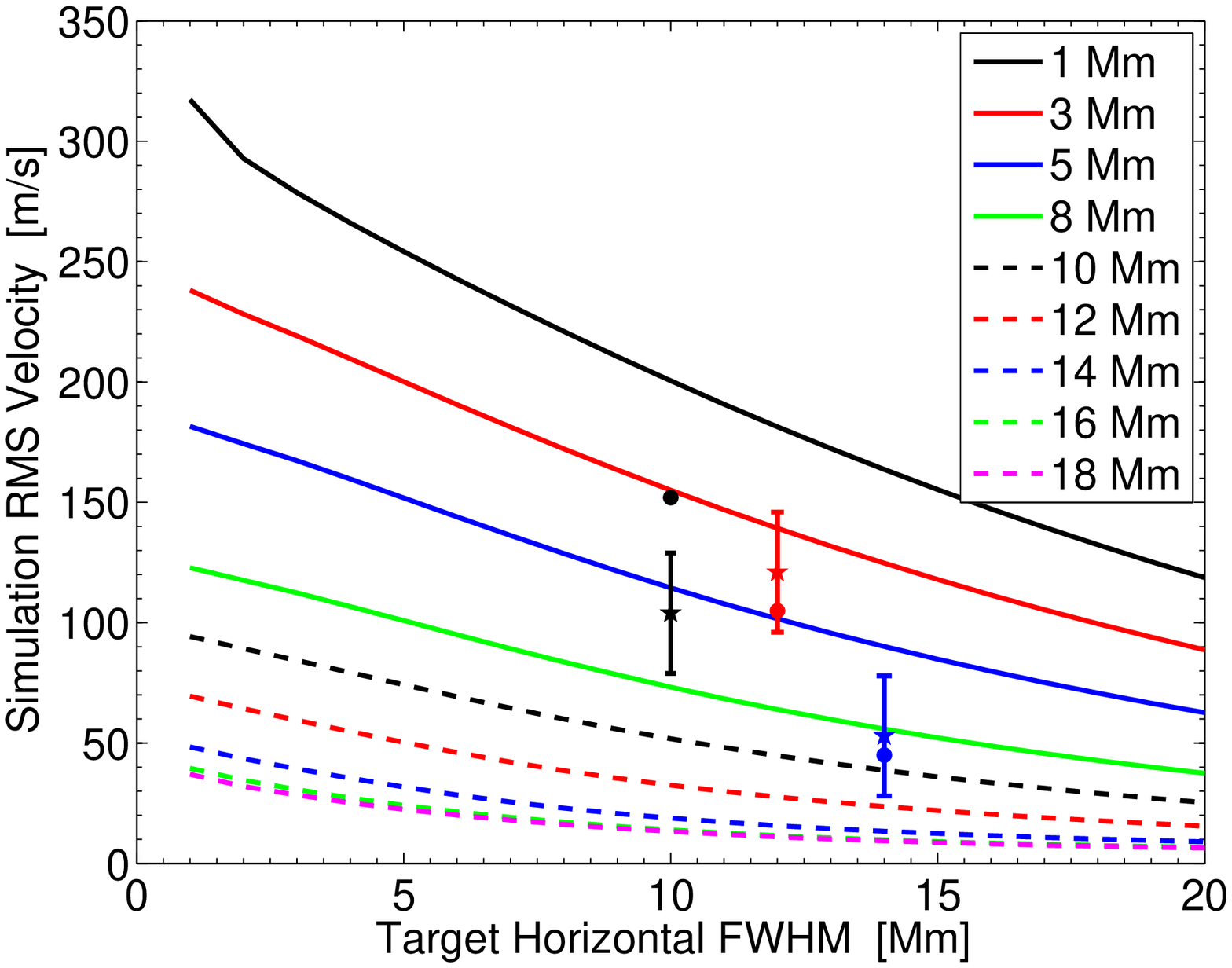}
 \caption{RMS velocity of the QS2 simulation $v_x$ flow component with differing degrees of smoothing at different depths. Each line represents the depth increasing from top to bottom (decreasing flow amplitudes) where more smoothing increases along the $x$ axis. The stars represent the RMS velocity inferred from the three phase-speed inversions for $v_x$ (with color denoting the matching FWHM used for the inversion). The filled circles represent the RMS of $v_x^{(x)}$, i.e, the artificial flows smoothed by the averaging kernel rather than the target function (Eq.~\ref{vbet}). The vertical bars indicate the $\pm 35\,\rm{m\,s^{-1}}$ inversion error at each depth.}
 \label{smoothed}
\end{figure}

\subsubsection{Flow amplitudes}
What is not captured reasonably well by the horizontal inversions, particularly for QS2, is the amplitude of the flows, as is clearly seen in the vector velocity plots as well as in Table~\ref{tab2} where the slopes of the correlation scatter plots are recorded. Here, even the best inversions underestimate those of $v_{x,y}^{\rm tgt}$ by up to 45\% in some cases, well above the standard deviation of the noise. We attribute much of the discrepancy to poor misfit between the inversion averaging kernels and the target function. To explore this further, Figure~\ref{smoothed} shows the raw QS2 simulation $v_x$ flow component at many depths, smoothed with a variety of wider and wider Gaussian target functions. At a given depth and smoothing value, a perfect inversion would fall near to the appropriate curve within the noise level. The starred points are the phase-speed inversion values, clearly well below the expected ones. The filled circles represent the raw simulation data instead smoothed by the averaging kernels ($v_x^{(x)}$ from Eq.~\ref{vbet}) rather than the proper Gaussian functions. As these points fall much closer to the inversion values, especially for the 2 deeper inversions, it is evident that the misfit is the source of the low amplitudes. The shallowest inversion mismatch is farther out of the range of the noise deviation, and is instead likely due to the large near-surface flows in the simulation that we cannot retrieve in a linear inversion. \citet{jackiewicz2007a} first showed a nonlinear response of travel-time differences for steady flows that the first Born approximation does not capture. The break from linearity typically occurs when flows reach $\sim$300-400~$\rm{m\,s^{-1}}$ at nominal travel distances. The depth slices through the simulations presented in Fig.~\ref{fig:QS} show that both simulations clearly possess flows that exceed this linearity threshold near the surface which remain even after smoothing.

\begin{figure*}
\centerline{ \includegraphics[width=1.0\linewidth,clip=]{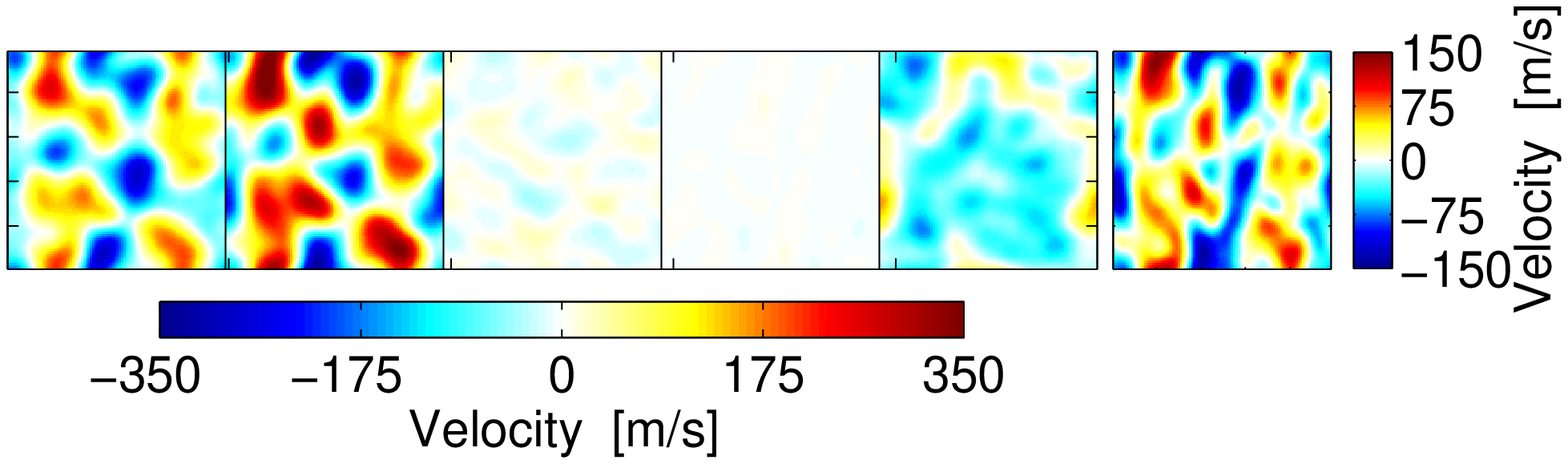}}
 \vspace{-5.5cm}
 \centerline{\hspace{1.2cm}$v_x^{\rm inv}$\hspace{1.6cm}$v_x^{(x)}$\hspace{1.7cm}$v_x^{(y)}$\hspace{1.6cm}$v_x^{(z)}$\hspace{.8cm}$v_x^{\rm inv} - \sum_\beta v_x^{(\beta)}$\hspace{.1cm}$v_x^{\rm tgt} - \sum_\beta v_x^{(\beta)}$ \hfill}
 \vspace{5cm}
 \caption{Contributions of the cross talk terms in the inverted horizontal flows for QS2 at a depth of 1~Mm. From left to right, the first four panels show the $x$-component of the inverted flows, and the individual terms for $v_x^{(x)}$, $v_x^{(y)}$, $v_x^{(z)}$ (according to Eq.~\ref{vbet}), respectively. The last two panels respectively show $v_x^{\rm inv}$ and $v_x^{\rm tgt}$ (defined in Eq.~\ref{valph}) minus the sum of the $v_x^{(x)}$, $v_x^{(y)}$, $v_x^{(z)}$ terms, where $\beta=(x,y,z)$. The first five panels use the horizontal color bar, while the vertical one is for the last panel only.}
 \label{crosstalkx}
\end{figure*}

\begin{figure}
  \centerline{
    \includegraphics[width=.33\textwidth,clip=]{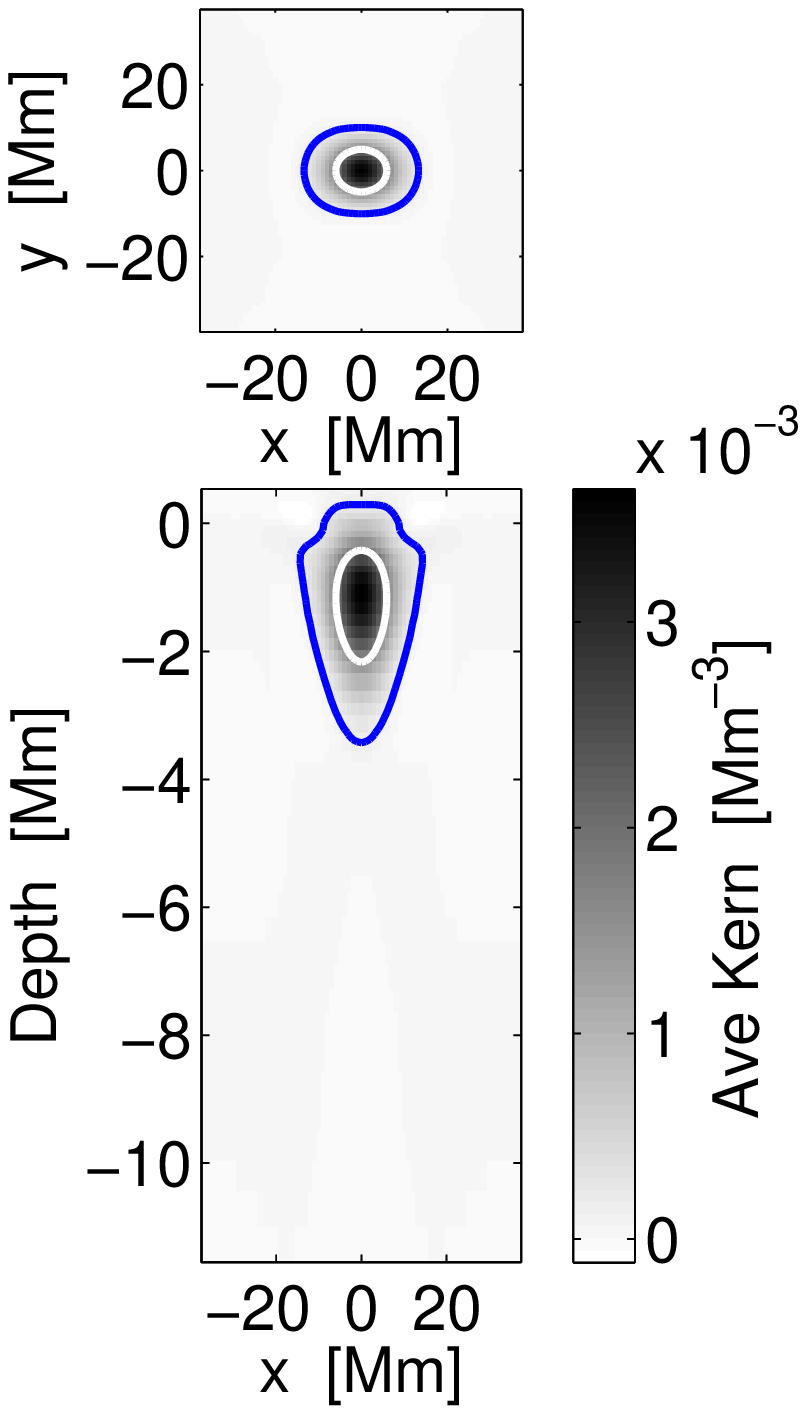}
    \includegraphics[width=.33\textwidth,clip=]{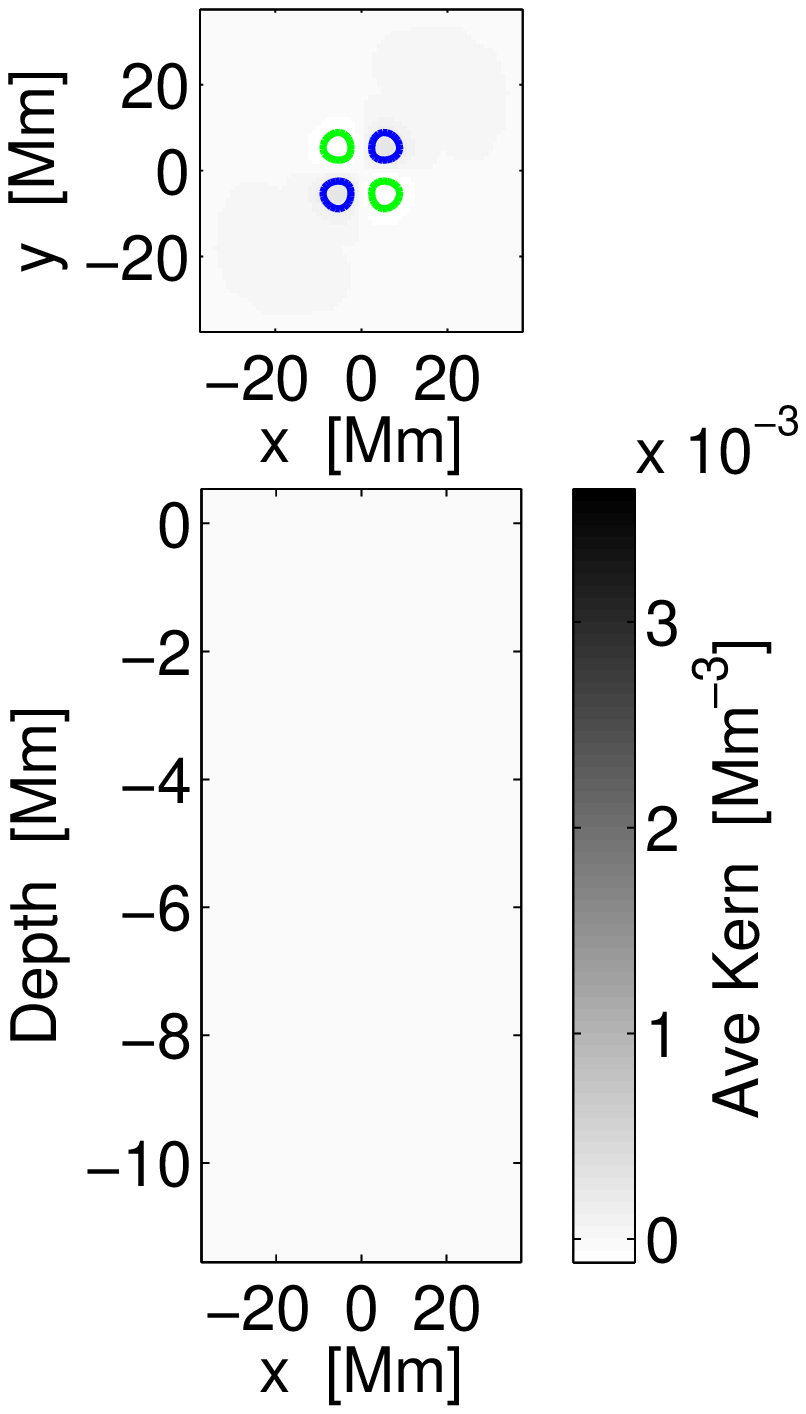}
    \includegraphics[width=.33\textwidth,clip=]{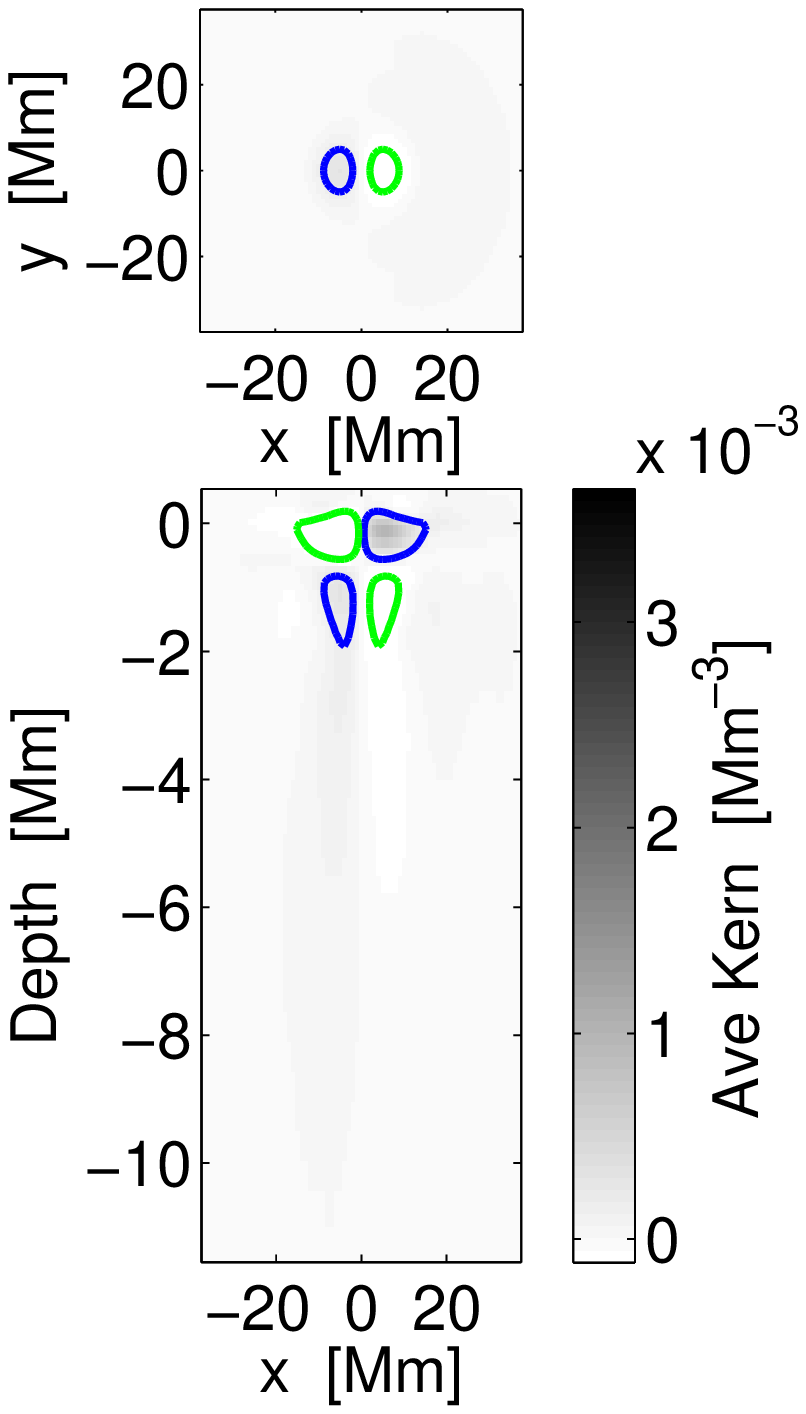}}
  \caption{Averaging kernels for the QS2 horizontal phase-speed flow inversion for $v_x$ at 1Mm depth. The plots show the $x$-component and the crosstalk $y$- and $z$-components of the averaging kernel, from left to right. The top panels are horizontal slices at the target depth of 1Mm, while the lower panels are vertical slices along the $y=0$ line. White contour lines denote 50\% of the maximum kernel value, while the blue and green lines denote the $\pm$5\% contours respectively.}
  \label{fig:avex}
\end{figure}

Further explanation for the poor amplitude inferences could be cross-talk effects from the vertical flow field. Figure~\ref{crosstalkx} shows the components $v_x^{(x)}$, $v_x^{(y)}$, $v_x^{(z)}$ of the near-surface QS2 simulation computed from our $v_x$ inversion averaging kernels. However, we find that little of the off-diagonal terms are ``leaking'' into the targeted component, as inspection of the averaging kernels themselves further verifies in Fig.~\ref{fig:avex} (as a note, averaging kernel plots like Fig.~\ref{fig:avex} have been computed for every inversion listed in Table~\ref{tab1} and can be viewed in the online supplement). The first two panels of Fig.~\ref{crosstalkx} indeed agree rather well. This is expected because of the weaker vertical flows \citep{svanda2011}, and shows that the cross-talk minimization is working well in this case and not contributing to this problem.

The fifth panel in Fig.~\ref{crosstalkx} illustrates a map of the random noise in the inversion whose magnitude is consistent within a factor of 2 with the value predicted from the inversion machinery. The last panel in Fig.~\ref{crosstalkx} reinforces the result that the amplitudes recovered in the horizontal inversions are significantly different (smaller) than those in the inversion, and that this difference is due to the mismatch between the averaging kernels and the target.


\subsubsection{Effects of filtering}
Examining the calculated correlations in Table~\ref{tab2}, there is a slight advantage when using the phase-speed filtering scheme. This is true particularly in the shallow layers of QS1. For QS2 we find that the correlations between phase-speed $v_{x,y}^{\rm inv}$ and $v_{x,y}^{\rm tgt}$ are consistently better than the ridge filtering and combined ridge+phase-speed filtering cases at every depth. 

It is important to note here that no clear overall improvement is found in either simulation when performing inversions in the upper 5~Mm using the combined ridge+phase-speed filtering scheme. In terms of both magnitude and spatial correlation, the combined filtering generally showed no significant advantage over phase-speed filtering alone. However, for the same fixed inversion noise level, the combined filtering can significantly improve the misfit between target and averaging kernel (or conversely, improve the level of inversion noise for the same fixed misfit). This improvement in inversion noise level for the combined filtering scheme was also seen by \citet{svanda2013}, and is likely due to having more kernels. In the case of the near-surface inversions presented here, though, the improved misfit does not seem to translate into any improvement in our inversion results.

We note than in nearly every horizontal inversion performed, for the same noise level, the level of misfit was the worst when ridge filtering alone was used. This is in contrast to \citet{svanda2013} where ridge filtering showed a substantial advantage over phase-speed filtering at a depth of 2.2~Mm when the two methods were compared directly. This was possibly explained by the fact that their ridge filter inversions contained many more independent measurements than the phase-speed filter inversions. For this work, we tried to avoid this potential bias, and the ridge and phase-speed filter inversions were all carried out with nearly identical numbers of independent measurements.

\subsection{A Note on Converging Flow}

\begin{figure}
 \centering
 \includegraphics[width=1.0\linewidth,clip=]{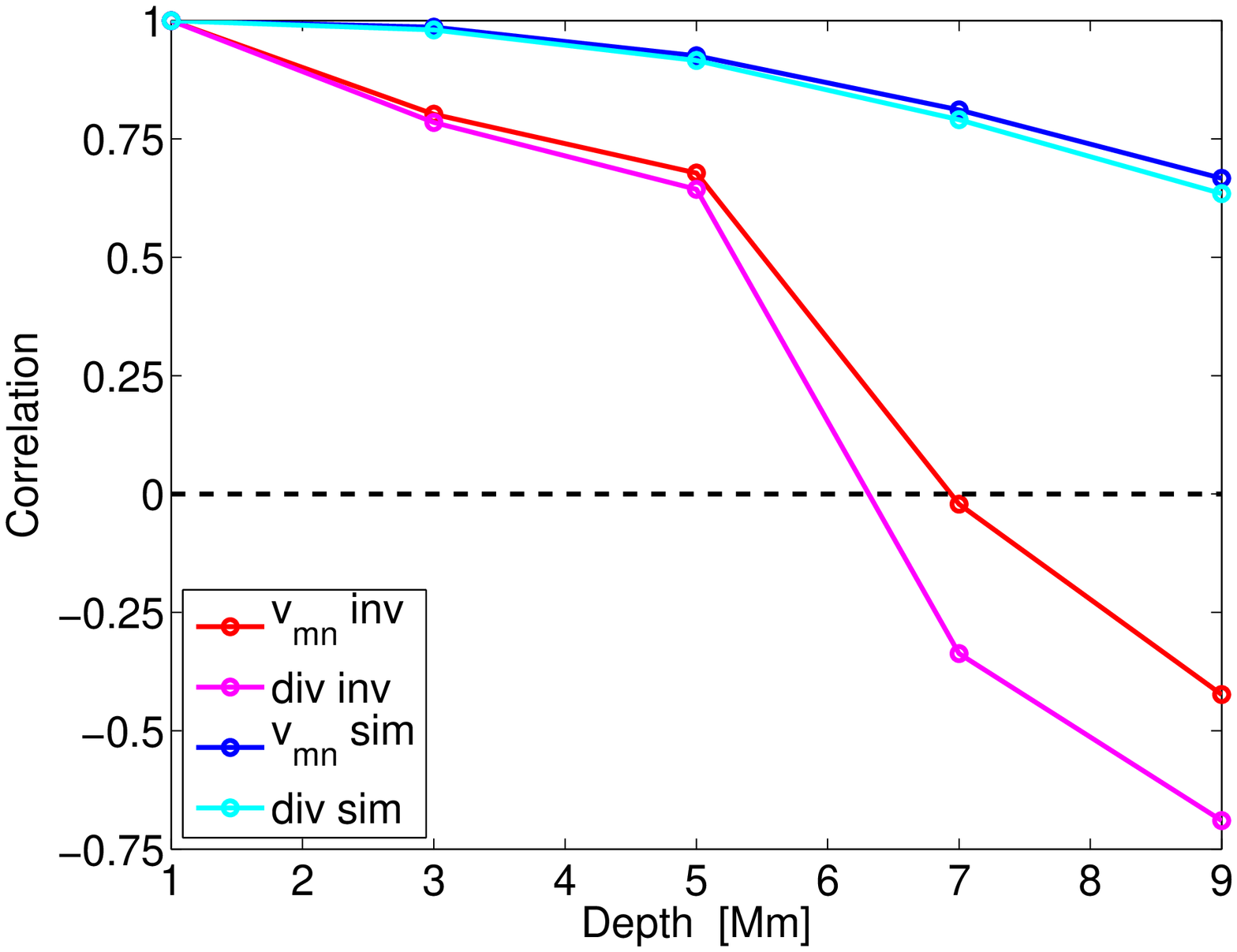}
 \caption{Correlations between the QS2 flows at 1~Mm depth and those at deeper layers. Red and magenta lines are for the inferred flows from the inversions ($v_{x,y}^{\rm inv}$), and blue and cyan are directly from the simulation ($v_{x,y}^{\rm tgt}$). Each case shows the mean correlation found using the individual flow components $v_x$ and $v_y$, as well as the correlation found using the horizontal divergence, $\nabla_{\rm h}\cdot \bm{v}_{\rm h}$.}
 \label{sg}
\end{figure}

There have been several studies in which time-distance helioseismology has been used to examine the flow field around supergranulation within the first few Mm of the upper convection zone \citep{gizon2003, zhao2003, woodard2007, jackiewicz2008, duvall2010, svanda2012}. A major focus of these works is not only to characterize the structure of these features, but also to determine the depth at which supergranules terminate, as this would shed some light on their mechanism of formation. \citet{duvall1998, zhao2003} found evidence of a supergranule return flow below some 10~Mm below the photosphere. These conclusions are drawn based on the correlation of near-surface flow divergence with the divergence of flows recovered at increasing depths. The correlation decreases quickly as one moves deeper through the convection zone and has been shown to eventually reverse sign.

Recently, \citet{svanda2013} suggested that the detection of this return flow could in fact be spurious, simply arising due to the loss of supergranule coherence combined with increasing noise as one probes larger depths. This idea is testable with our simulation data as each domain actually possesses supergranule-sized flows whose large-scale structure stays coherent well throughout the upper half of each simulation domain. Inversions at depths larger than 5~Mm, however, are difficult with our current kernel selection because there is little sensitivity in these regions. The idea here, though, is not to show that we can accurately recover flows at these depths (because we cannot) but to show that a spurious ``return flow'' detection is certainly possible when carrying out inversions where there is little wave signal and where subsurface flows are weak.

Additional horizontal inversions for each of the three filtering schemes were performed at depths of 7 and 9~Mm using QS2. QS2 was chosen simply based on the fact that it contains more well-defined supergranule structure than QS1. Target FWHM values for these inversions are given in Table~\ref{tab1}, inversion set~2. In contrast with the shallow inversion results, we find a marked advantage for the case of ridge+phase-speed filtering over ridge or phase-speed filtering alone. As before, the combined filtering scheme gives a much better misfit between target and averaging kernel for the same amount of noise. This is especially important at large depths when getting a reasonable error generally means having to allow the misfit to decrease drastically. This can become problematic because the averaging kernel becomes increasingly non-localized around the target depth, and begins including unwanted signal from shallower regions. For the same misfit at these depths, the inversion error for the combined filtering case was 20--40~$\rm{m\,s^{-1}}$ less than for the ridge or phase-speed filtering cases alone.

Using the combined filtering scheme, each inversion flow map at depth was correlated with the near-surface one at 1~Mm. Figure~\ref{sg} shows the correlation as a function of depth when both the horizontal divergence, $\nabla_{\rm h}\cdot \bm{v}_{\rm h}^{\rm inv}$, and the separate $v_x^{\rm inv}$ and $v_y^{\rm inv}$ flow component are used as proxies. We see for the inverted quantities that the correlation decreases rapidly at larger depths, where a reversal in sign occurs around 7~Mm. In fact, $v_x^{\rm inv}$ and $v_y^{\rm inv}$ give different crossover depths as indicated by the divergence correlation profile. However, performing the same exercise using the targeted flows (i.e. those of $v_{x,y}^{\rm tgt}$) shows that no reversal exists at any depth. This suggests that the observation of a correlation sign reversal seen in real solar inversions could in fact be spurious, merely arising as a consequence of the current ``standard inversion'' shortcomings.

\subsection{Vertical Flow Inversions}

\begin{figure*}
 \begin{center}$
  \begin{array}{c c c}
   \includegraphics[width=0.33\linewidth,clip=]{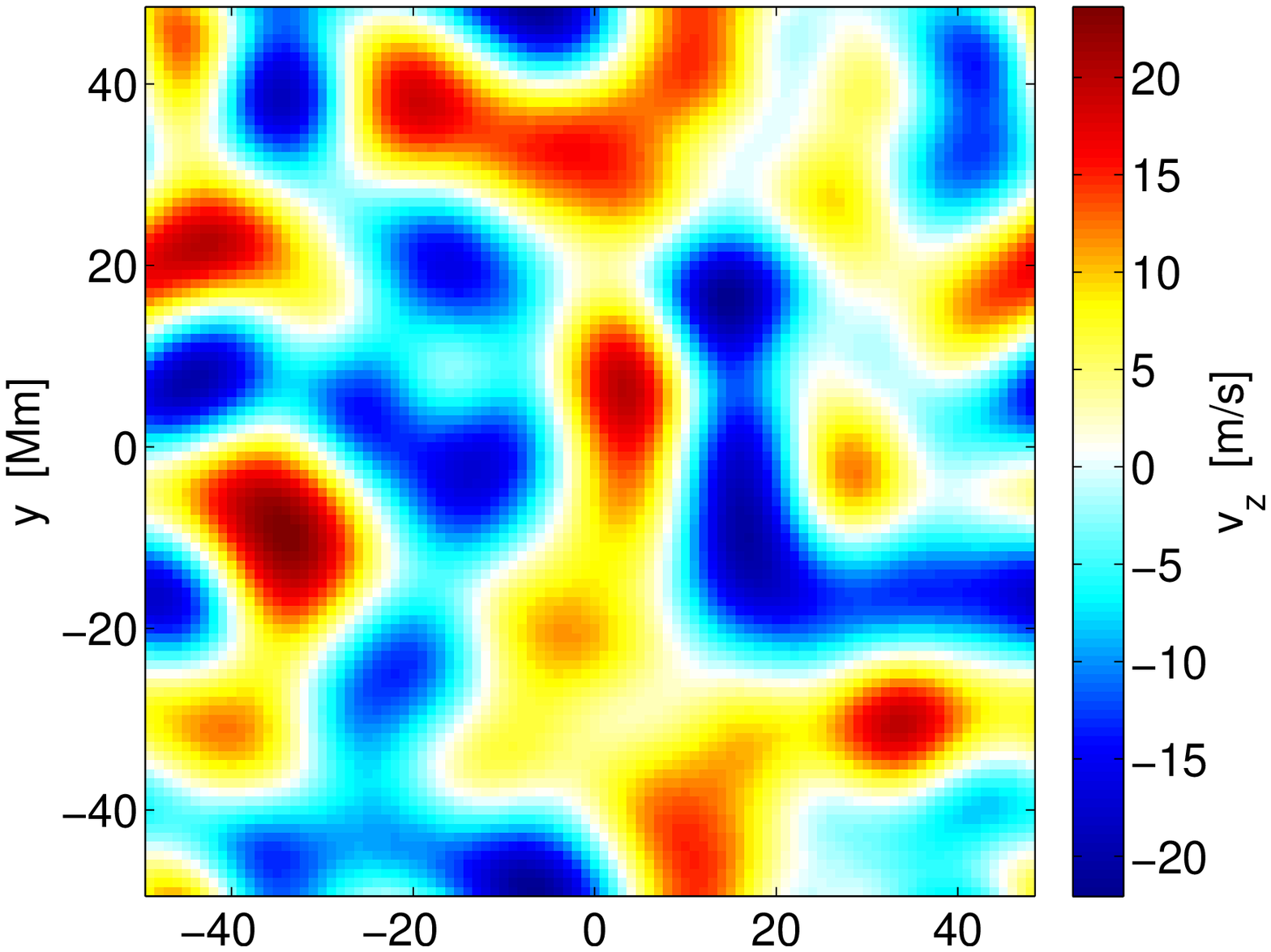} &
   \includegraphics[width=0.33\linewidth,clip=]{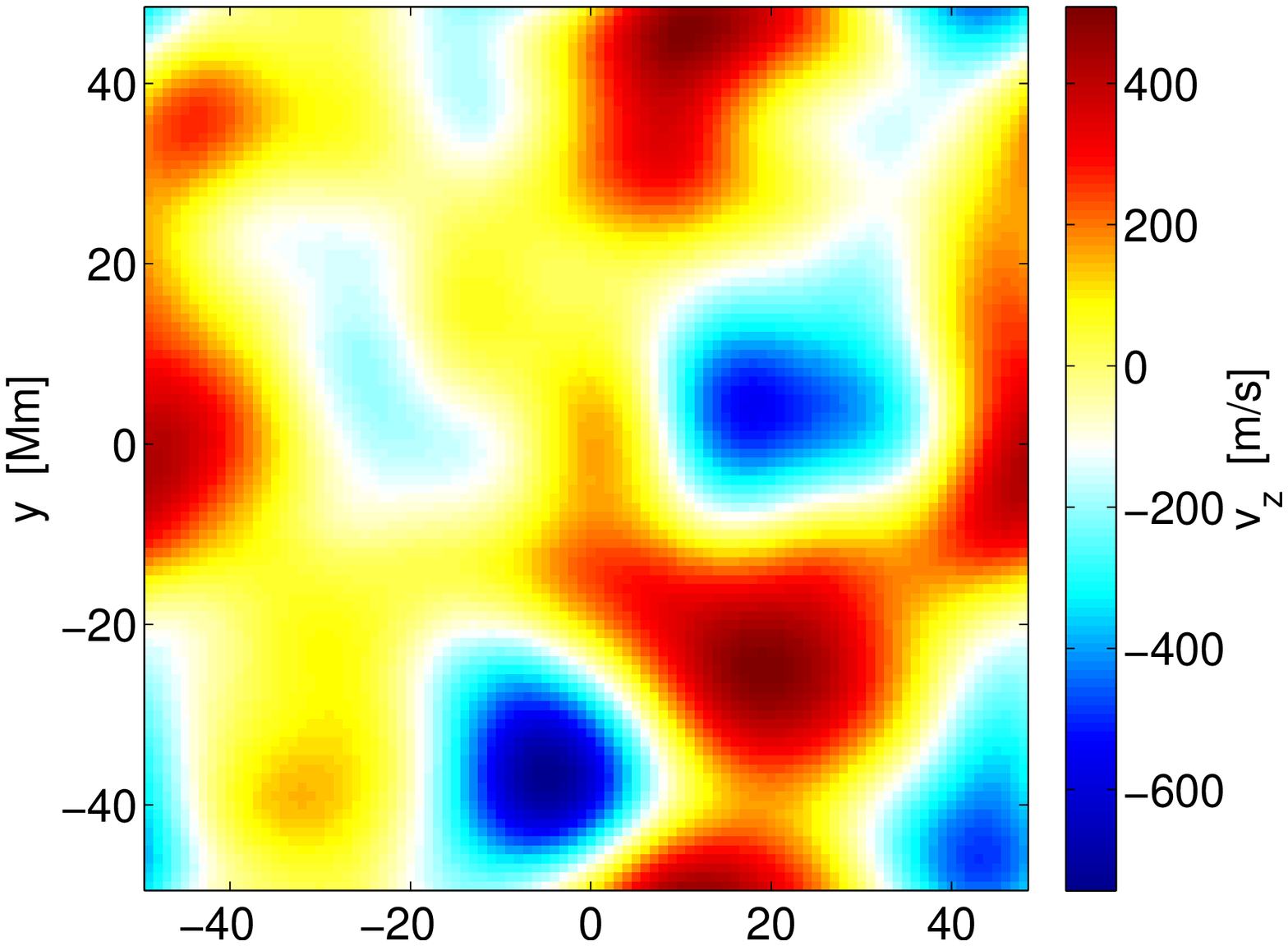} &
   \includegraphics[width=0.33\linewidth,clip=]{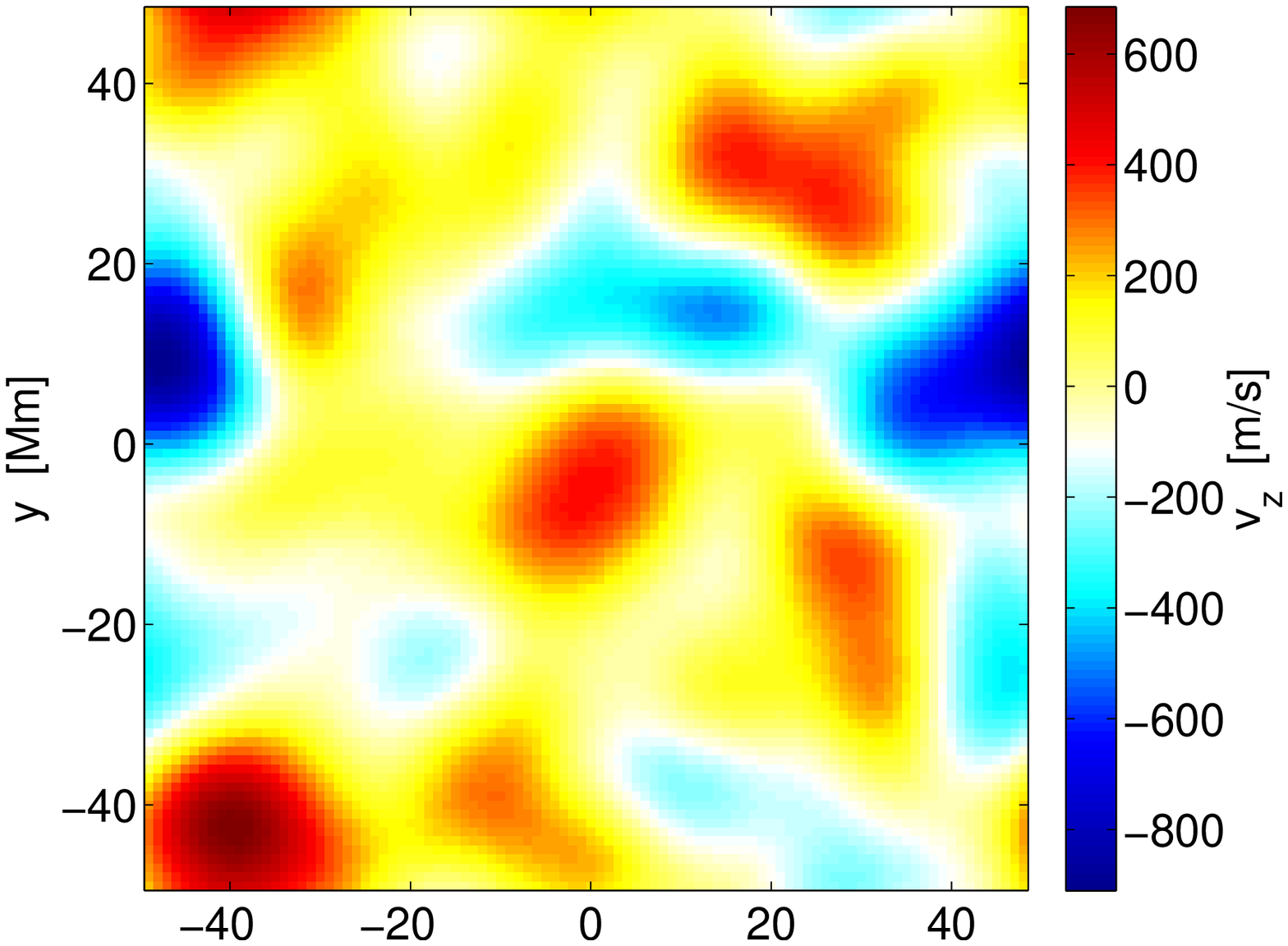}\\
   \includegraphics[width=0.33\linewidth,clip=]{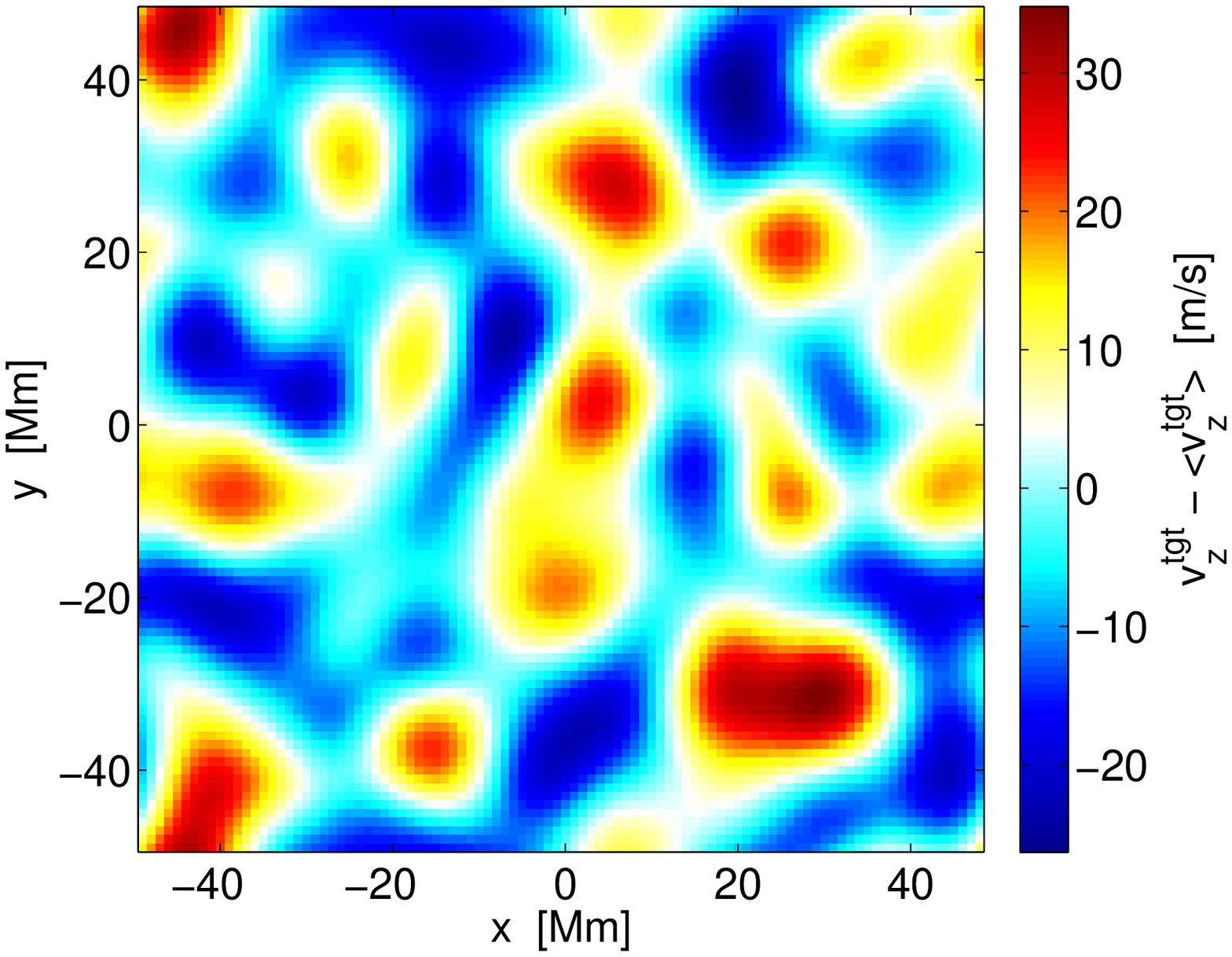} &
   \includegraphics[width=0.33\linewidth,clip=]{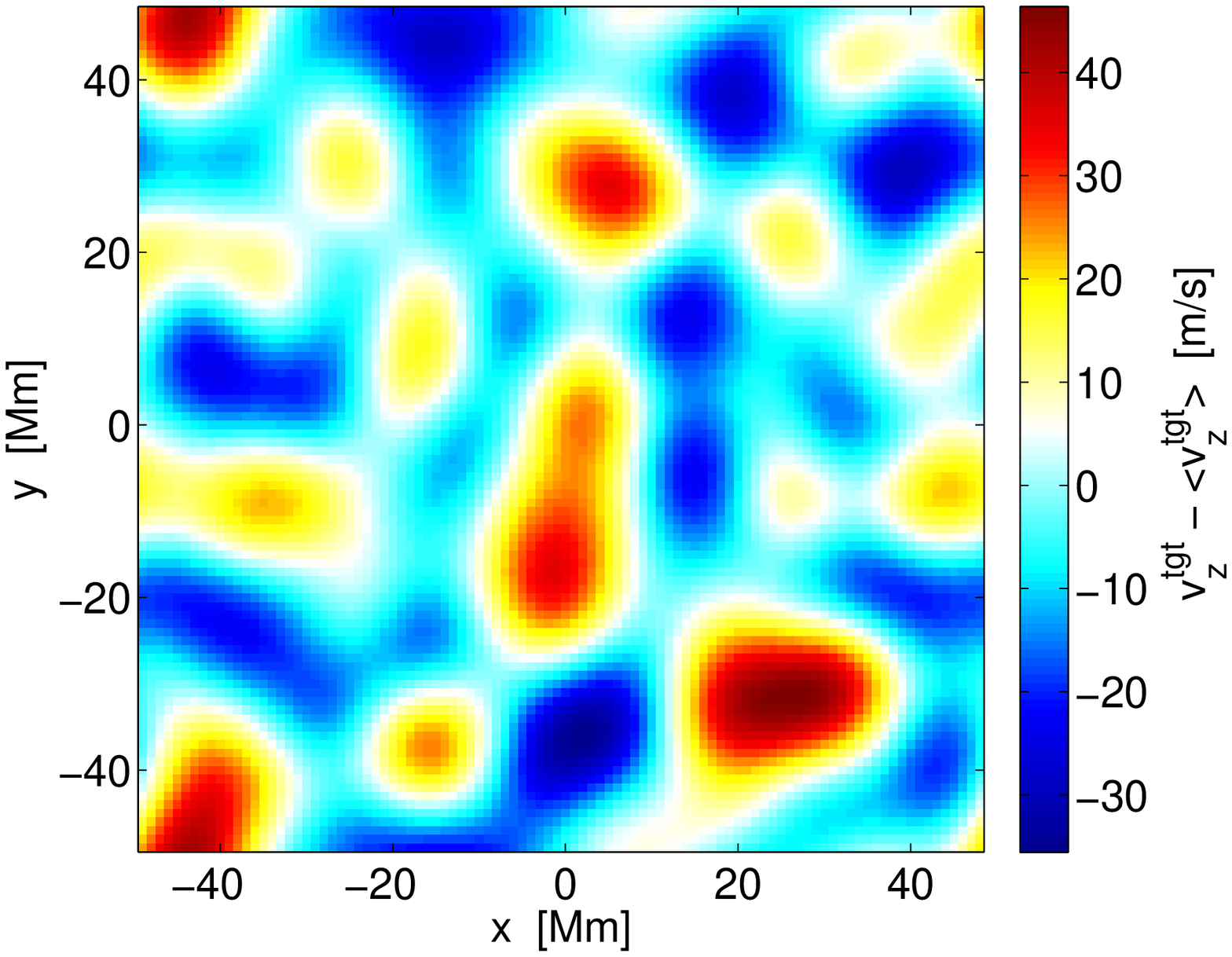} &
   \includegraphics[width=0.33\linewidth,clip=]{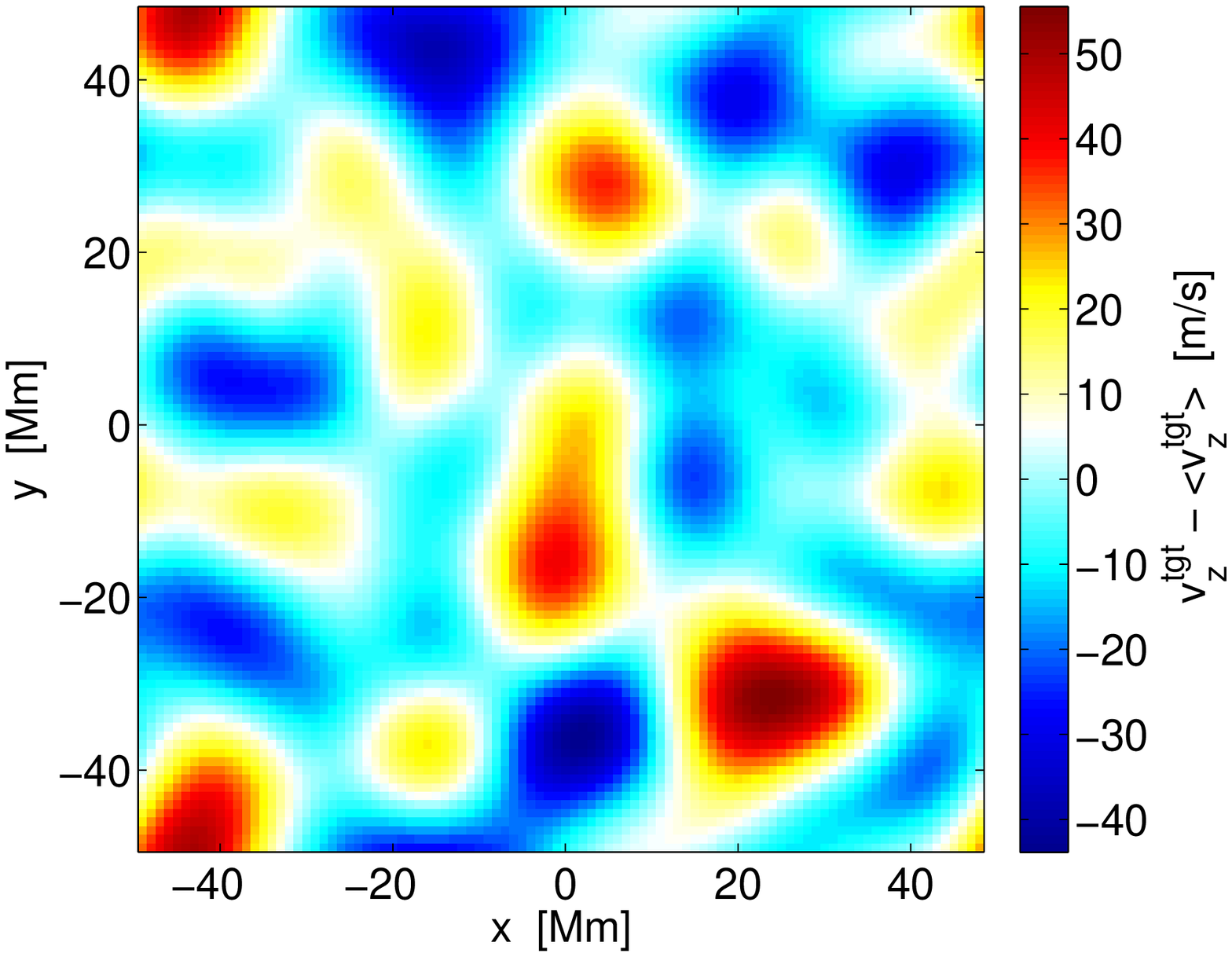}
  \end{array}$
 \end{center}
 \caption{QS1 vertical velocity inversion maps for the phase-speed filtered travel times (top row) for depths (left to right) of 1, 3 and 5~Mm. The smoothed simulation flow maps (i.e. $v_z^{\rm tgt}$) at these depths are shown in the bottom row. The parameters of each inversion (i.e. target FWHM, inversion error, etc.) are presented in Table~\ref{tab1}, inversion set~3. Correlation coefficients found between each inversion and the simulation are presented in Table~\ref{tab2}.}
 \label{fig:qs1vz}
\end{figure*}

\begin{figure*}
 \begin{center}$
  \begin{array}{c c c}
   \includegraphics[width=0.3\linewidth,clip=]{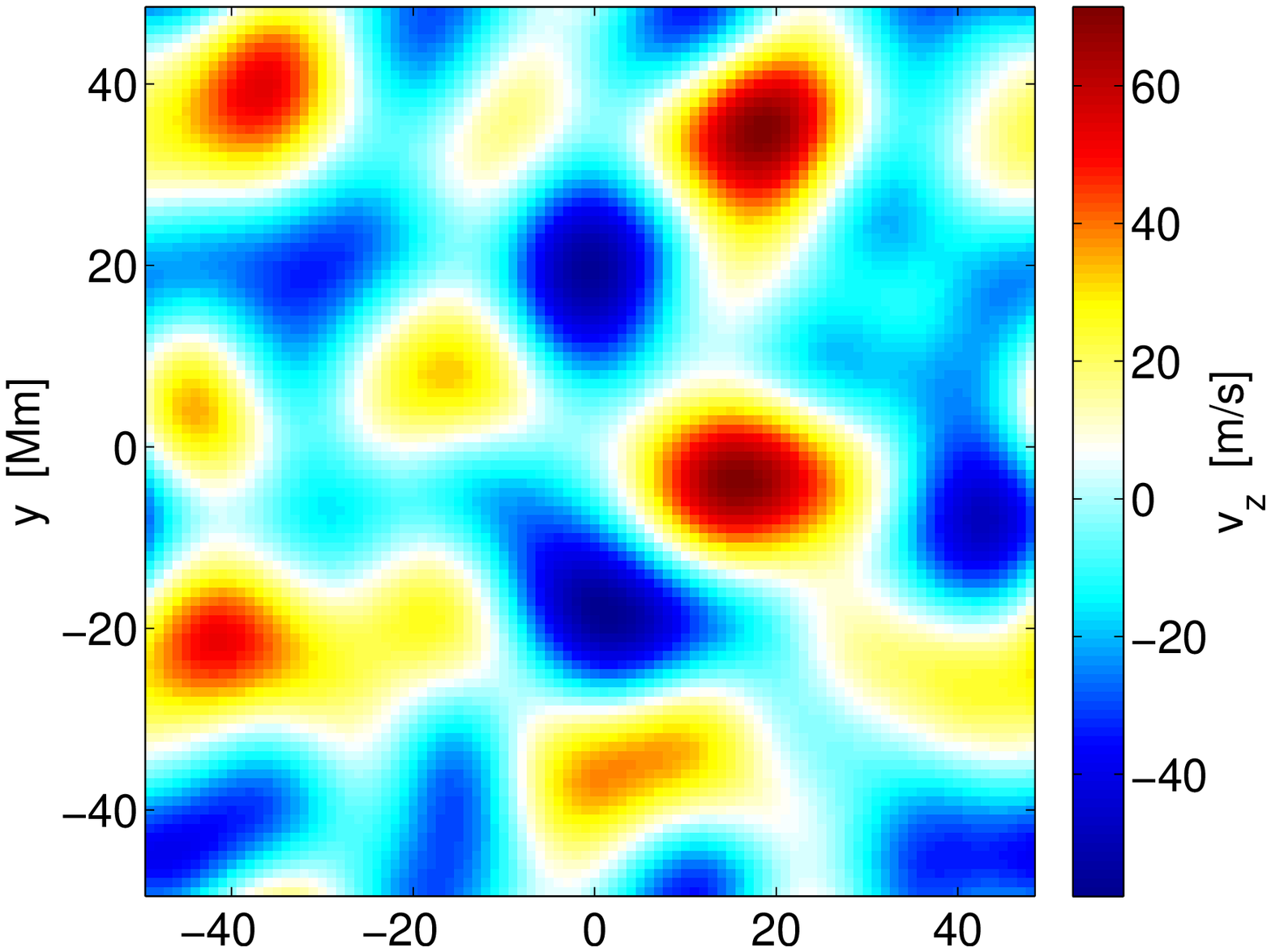} &
   \includegraphics[width=0.3\linewidth,clip=]{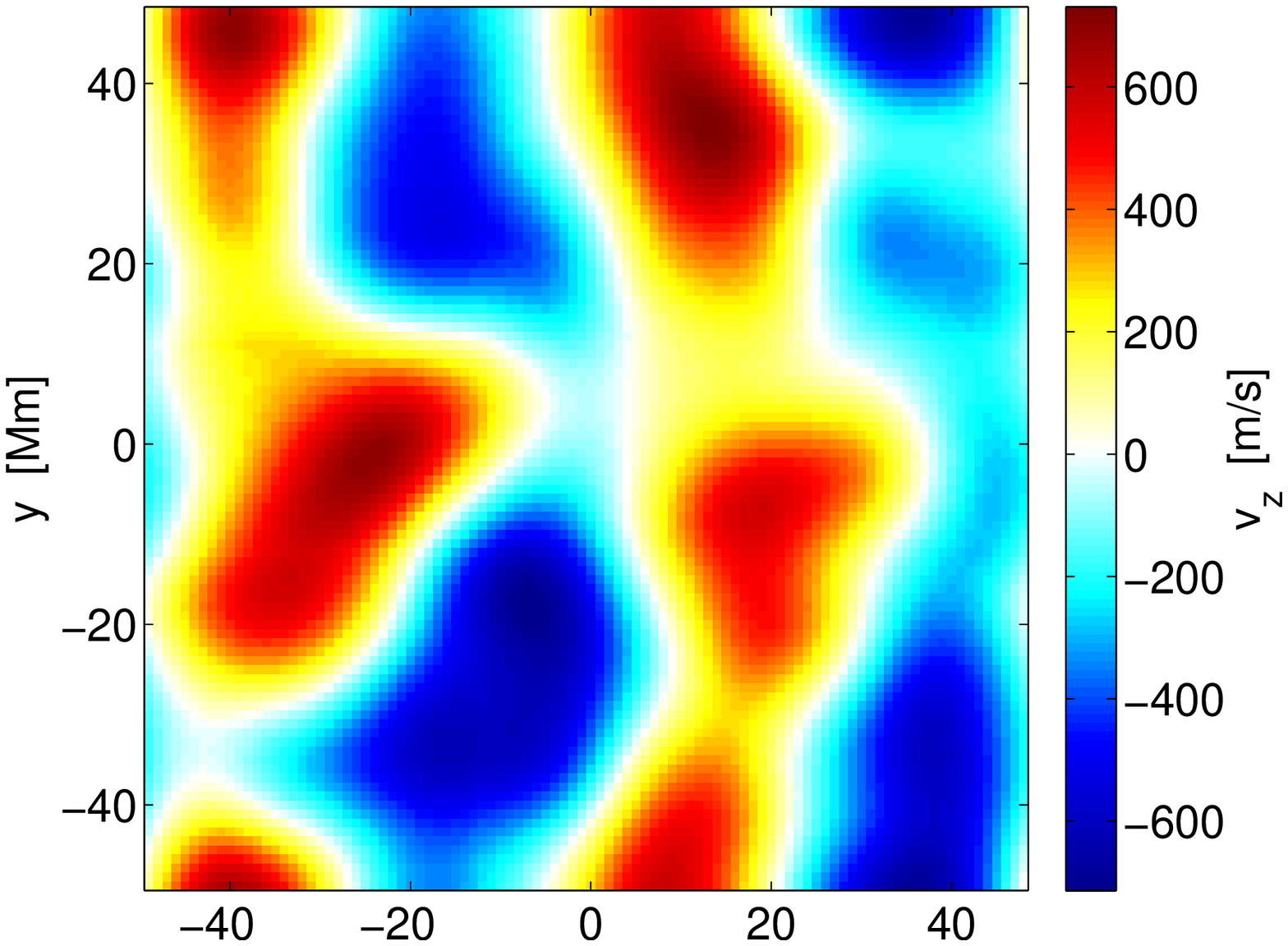} &
   \includegraphics[width=0.3\linewidth,clip=]{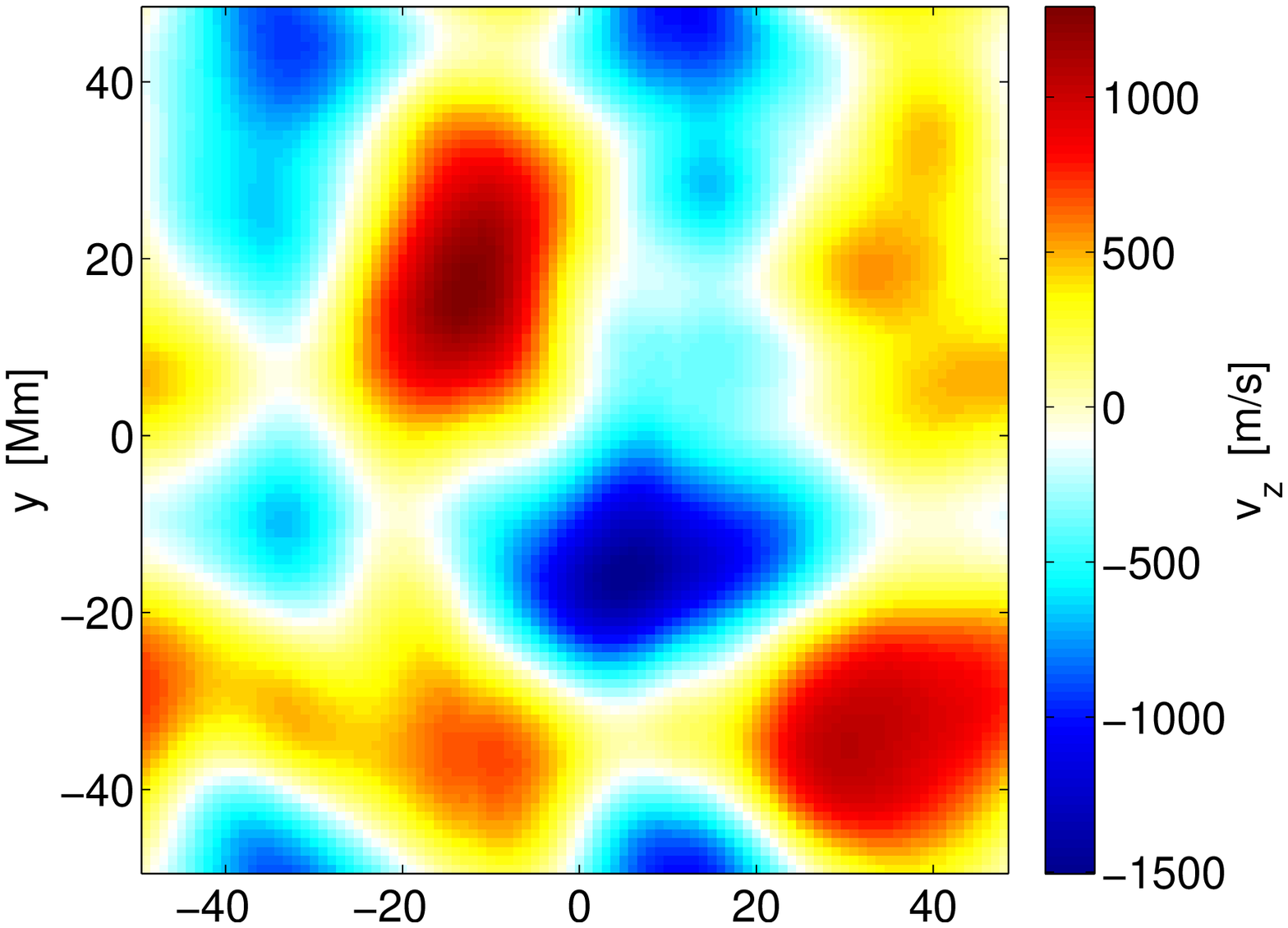}\\
   \includegraphics[width=0.3\linewidth,clip=]{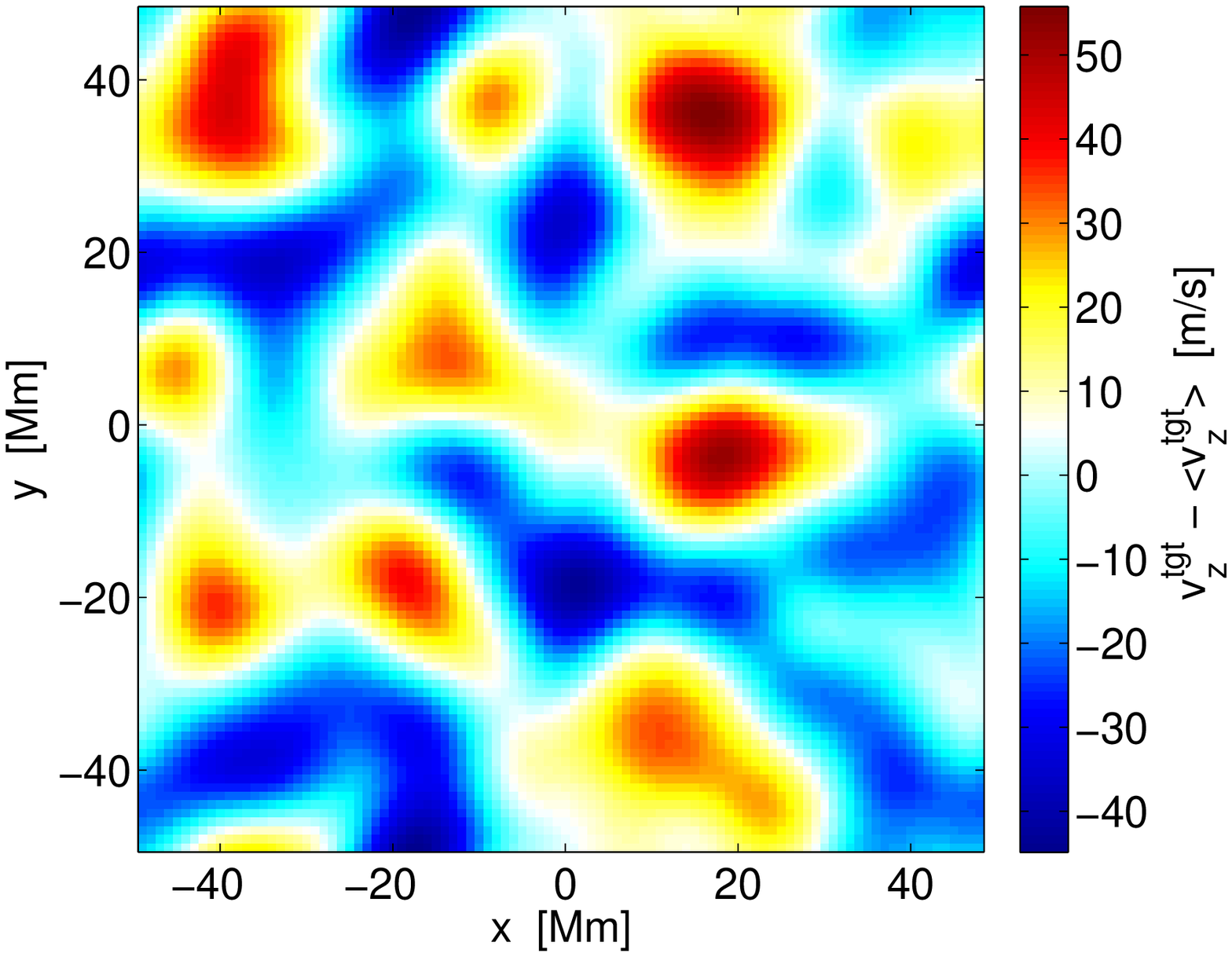} &
   \includegraphics[width=0.3\linewidth,clip=]{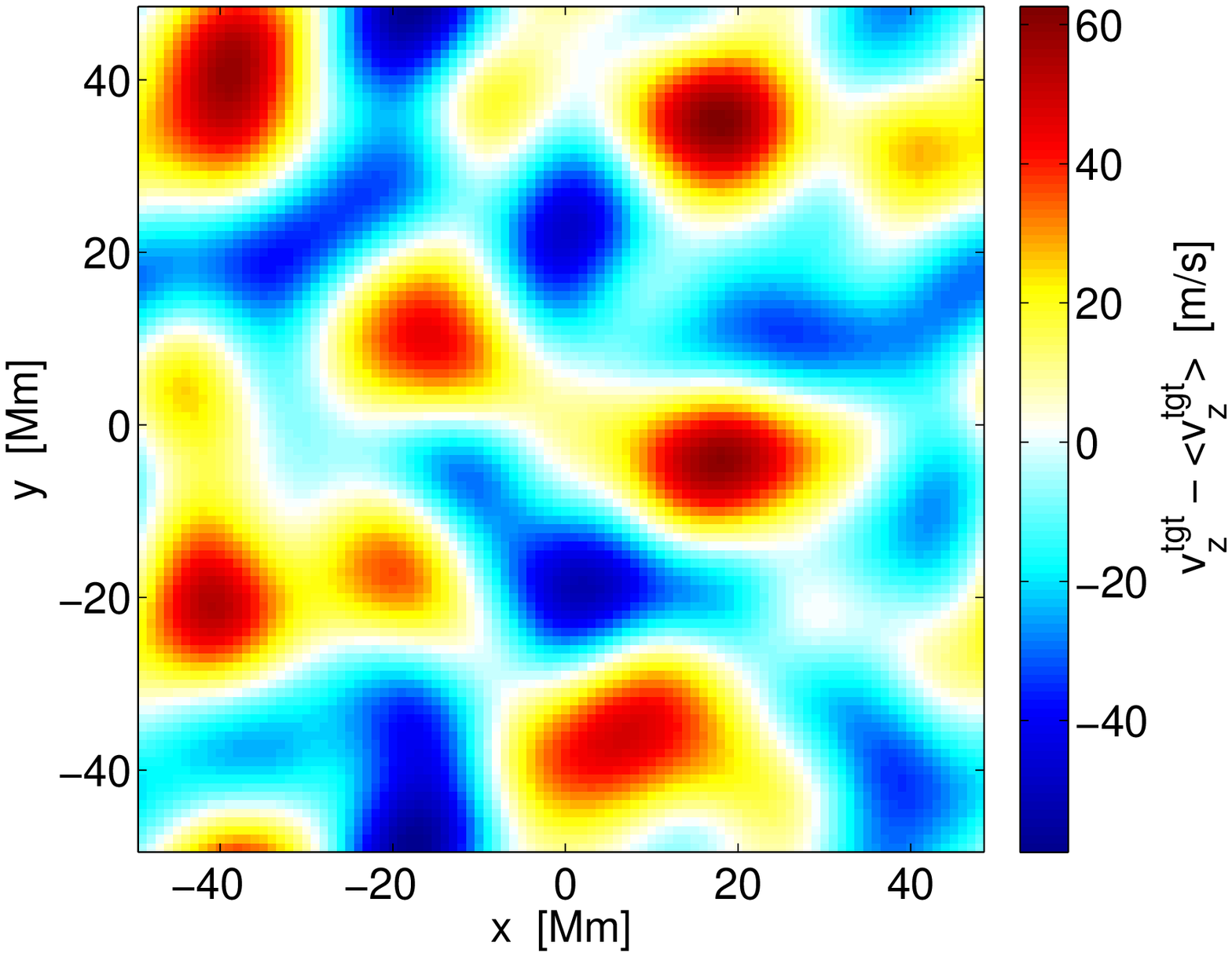} &
   \includegraphics[width=0.3\linewidth,clip=]{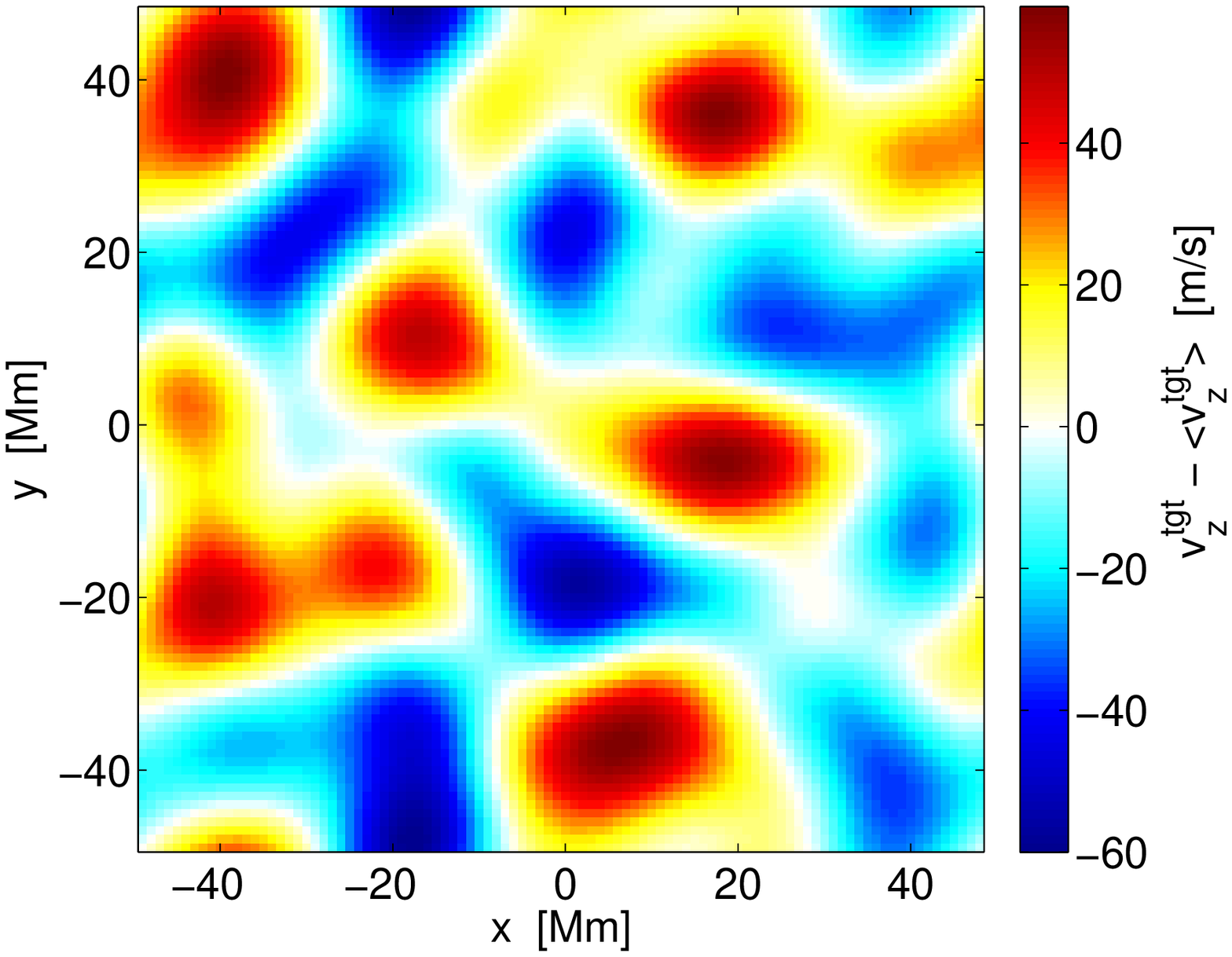}
  \end{array}$
\end{center}
\caption{QS2 vertical velocity inversion maps for the phase-speed filter case (top row) for depths (left to right) of 1, 3 and 5~Mm. The smoothed simulation flow maps (i.e. $v_z^{\rm tgt}$) at these depths are shown in the bottom row. The parameters of each inversion (i.e. target FWHM, inversion error, etc.) are presented in Table~\ref{tab1}, inversion set~3. Correlation coefficients found between each inversion and the simulation are presented in Table~\ref{tab2}.}
\label{fig:qs2vz}
\end{figure*}

Inversions for the vertical velocities in the two simulations were also carried out, the parameters of which are presented in Table~\ref{tab1}, inversion set~3. As pointed out in \citet{svanda2011}, the structure of the vertical velocity sensitivity kernels only allows for the retrieval of the fluctuations of $v_z$ from the mean value $\langle v_z \rangle$. In everything that follows, our comparisons are made between the inverted vertical flows and the mean-subtracted simulation vertical flows. Figure~\ref{fig:qs1vz} shows the resulting QS1 $v_z^{\rm inv}$ flow maps obtained from inverting the phase-speed travel times at each depth (top row). For comparison, the target flows from the simulation, $v_z^{\rm tgt}$, are also shown (bottom row). The strategy for the shallowest inversion was to obtain a noise level in the $10-20~\rm{m\,s^{-1}}$ range. This is motivated by the RMS values of $v_z^{\rm tgt}$ being on the order of 35~$\rm{m\,s^{-1}}$. This gives reasonable averaging kernels here, while at larger depths this constraint on the noise limit has to be considerably relaxed to obtain a reasonable averaging kernel. Consequently, this leads to very inaccurate flow amplitude determinations at larger depth.

We find from Table~\ref{tab2} that the correlation between $v_z^{\rm inv}$ and $v_z^{\rm tgt}$ for QS1 is rather poor in every case. The artificial flows ($v_z^{\rm tgt}$) show a pattern of many small-scale up flows and down flows present over the full range of depths which is not recovered in $v_z^{\rm inv}$. Phase-speed inversions generally resulted in higher correlations than the combined filtering scheme, but the correlation values are very low. The ridge inversions were significantly worse, even showing a relatively strong negative correlation at a depth of 1~Mm. Figure~\ref{fig:qs2vz} shows the equivalent $v_z$ flow maps for QS2. In this case, the near-surface correlations are significantly higher than those of QS1, in the $0.5-0.7$ range at a depth of 1~Mm. Again, the ridge filter correlations are substantially lower than the other filtering schemes, especially in the near-surface layers. At larger depths, the correlations rapidly decrease and the signal is dominated by noise, rendering the recovered amplitudes meaningless.


To ensure that there were no mistakes in the QS1 analysis that led to such poor correlation between $v_z^{\rm inv}$ and $v_z^{\rm tgt}$, a test was carried out by exchanging the inversion weights. While by eye the sets of weights from $v_z^{\rm inv}$ for QS1 and QS2 are very similar, we computed flows according to Eq.~\ref{vinv} with the weights from QS1 and the travel-time measurements from QS2, and vice versa. The results showed no quantitative differences.



\begin{figure*}
 \centerline{\includegraphics[width=1.0\linewidth,clip=]{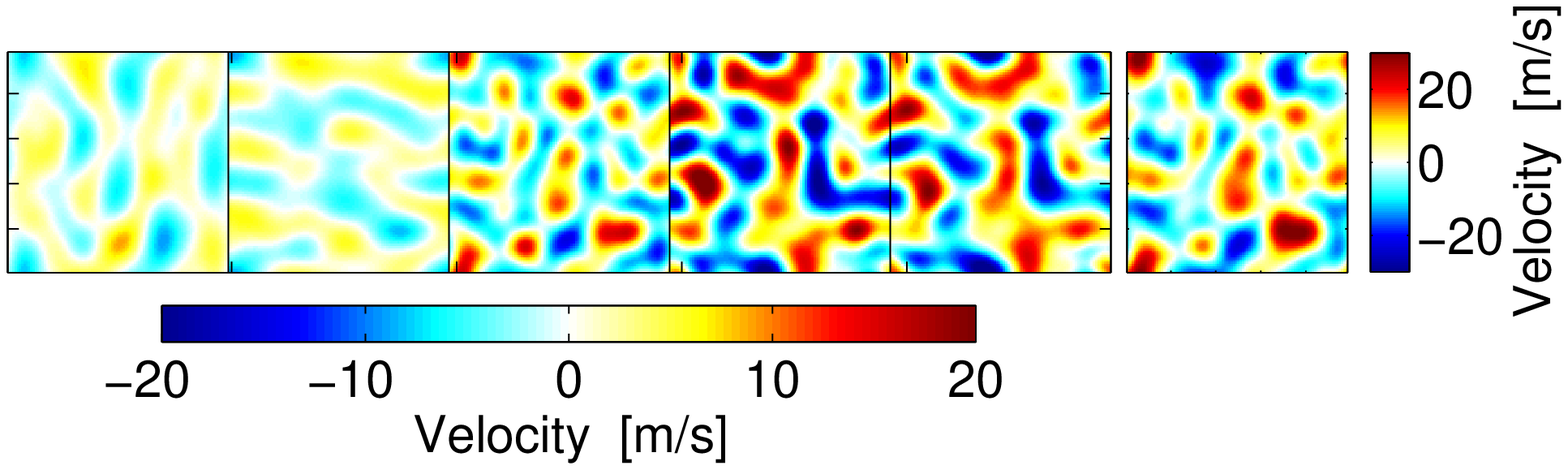}}
  \vspace{-5.5cm}
 \centerline{\hspace{1.1cm}$v_z^{(x)}$\hspace{1.6cm}$v_z^{(y)}$\hspace{1.7cm}$v_z^{(z)}$\hspace{1.6cm}$v_z^{\rm inv}$\hspace{.8cm}$v_z^{\rm inv} - \sum_\beta v_z^{(\beta)}$\hspace{.1cm}$v_z^{\rm tgt} - \sum_\beta v_z^{(\beta)}$\hfill}
 \vspace{5cm}
 \caption{Contributions of the cross talk terms in the inverted vertical flows for QS1 at a depth of 1~Mm. From left to right, the first three panels show the individual terms for $v_z^{(x)}$, $v_z^{(y)}$, $v_z^{(z)}$ (according to Eq.~\ref{vbet}), respectively, while the fourth panel are the inverted flows. The last two panels respectively show $v_z^{\rm inv}$ and $v_z^{\rm tgt}$ (defined in Eq.~\ref{valph}) minus the sum of the $v_z^{(x)}$, $v_z^{(y)}$, $v_z^{(z)}$ terms, where $\beta=(x,y,z)$. The first five panels use the horizontal color bar, while the vertical one is for the last panel only.}
 \label{crosstalkz}
\end{figure*}

\begin{figure}
  \centerline{
    \includegraphics[width=.33\textwidth,clip=]{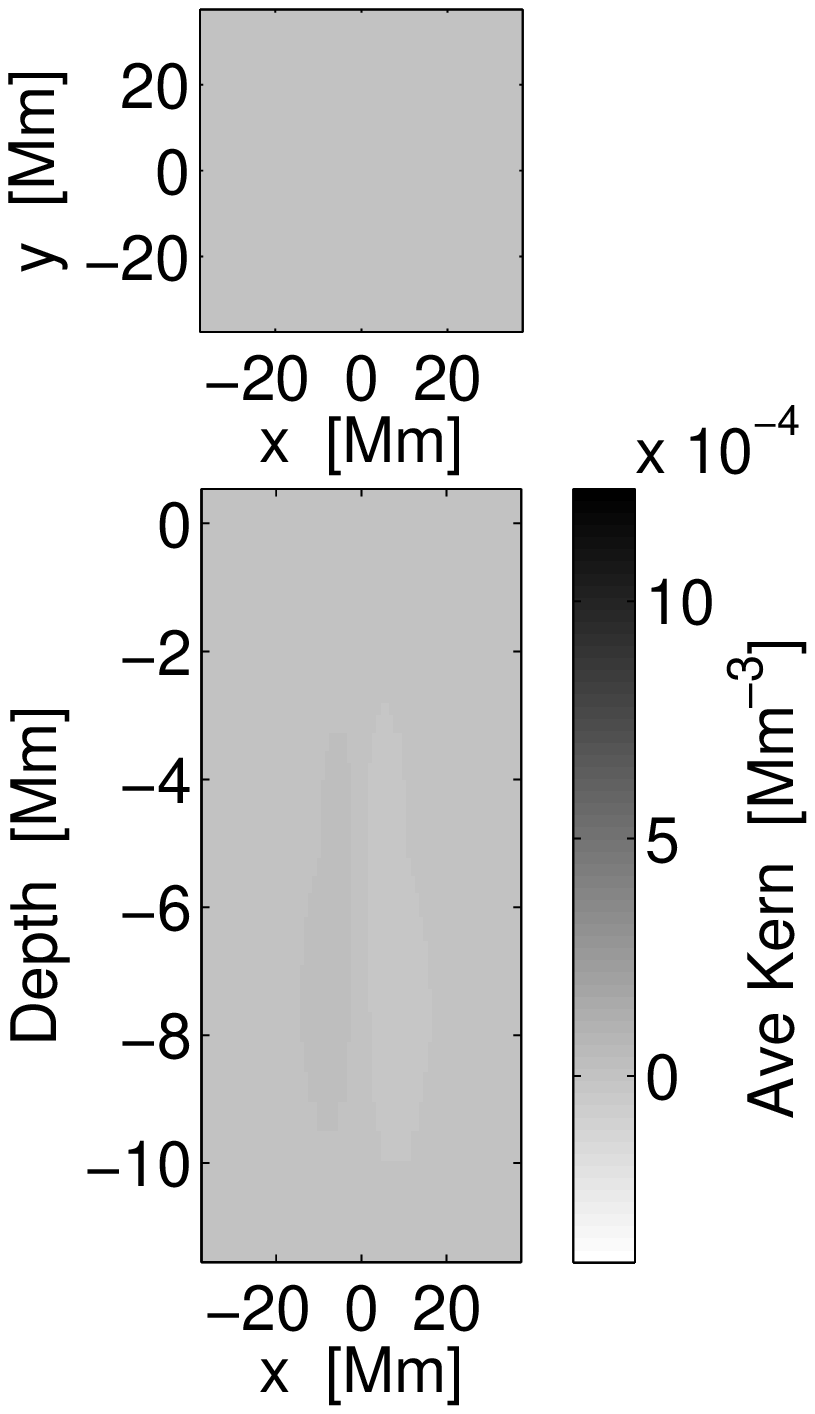}
    \includegraphics[width=.33\textwidth,clip=]{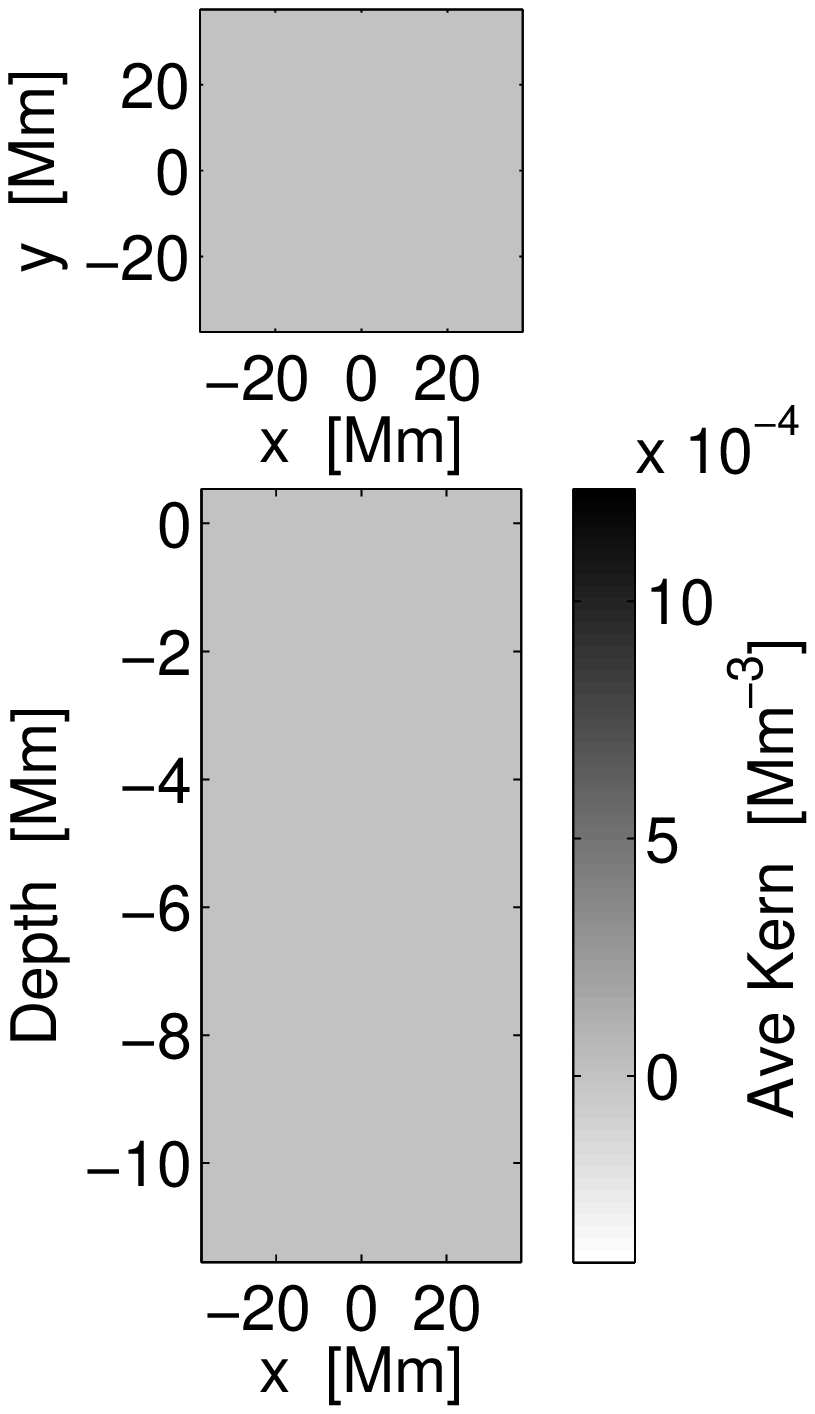}
    \includegraphics[width=.33\textwidth,clip=]{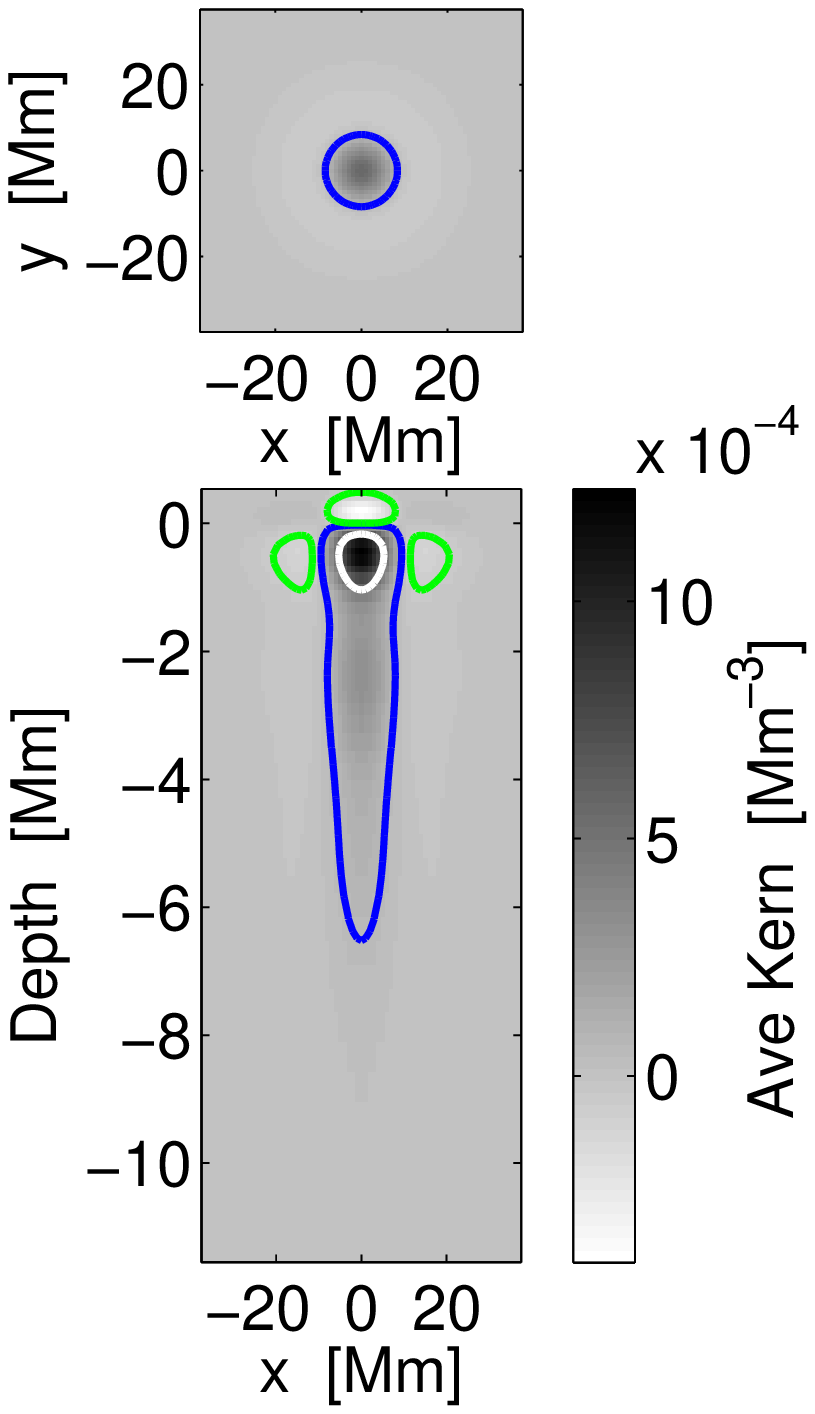}}
  \caption{Averaging kernels for the QS1 vertical phase-speed flow inversion for $v_z$ at 1Mm depth. The plots show the crosstalk $x$- and $y$-components and the $z$-component of the averaging kernel, from left to right. The top panels are horizontal slices at the target depth of 1Mm, while the lower panels are vertical slices along the $y=0$ line. White contour lines denote 50\% of the maximum kernel value, while the blue and green lines denote the $\pm$5\% contours respectively. The color scale is equal in all plots.}
  \label{fig:avez}
\end{figure}

Cross-talk effects in standard helioseismic inversions of numerical simulations can lead to the retrieval of vertical flows with the wrong sign \citep{zhao2007,dombroski2013}. As already mentioned, the inversion procedure here uses a constraint that attempts to limit such effects, while taking into account the associated trade-offs in the misfit and noise. To understand better the poor results for the vertical inversions, we explored the cross-talk contribution.

Representative averaging kernels from the vertical inversions are shown in Fig.~\ref{fig:avez}. We see that the cross-talk minimization in the inversion algorithm appears to be effective, in that the $x, y$ components are relatively small in magnitude, by about a factor of 50 in this example. However, this evidently does not translate into well constrained flows, as Fig.~\ref{crosstalkz} shows the components of the diagonal and off-diagonal flows, $v_z^{(\beta)}$, for QS1. The cross-talk here can reach up to 35\% of the overall maximal flow amplitudes, indicating that the strong horizontal flows play a detrimental role in the retrieval of the vertical velocities because of their larger strength. 

The vertical velocity inversions are impacted strongly by noise. It is therefore challenging to construct averaging kernels that are a close match to the target function. The first statement is demonstrated in Fig.~\ref{crosstalkz} for QS1, whereby the proxy noise component (panel 5) is very similar to the inverted flow component itself (panel 4). The fact that the last panel in Fig.~\ref{crosstalkz} shows a signal that is of the order of the simulated velocities themselves proves the second statement. It should be emphasized that although the vertical-velocity inversions do not perform as well as in the tests of \citet{svanda2011}, the differences should mainly be attributed to the simulation data and overall procedure, as we have not started from a single snapshot with random noise added and tried to recover the input velocities.



\section{Summary and Discussion}\label{sec:dis}
Current time-distance analysis has been tested using two realistic magnetohydrodynamic quiet-Sun simulations. Measurements made using the GB02 and GB04 travel-time definitions are found to correlate very well with one another, varying linearly over a large range of distances. The travel-times computed using GB04 are on average $10\%$ larger than those of GB02 in terms of RMS variation. Correlation between measured and forward-modeled travel-times computed in the first Born approximation is generally high for filters $f, p_1-p_2$ and $\rm{td_1}-\rm{td_4}$, but is found to decrease rapidly for filters $p_3$ and $\rm{td_5}$. We possibly attribute this to inadequate modeling of the simulation power at higher phase-speeds and for higher-order modes.

SOLA inversions were carried out using both travel-time definitions for several filtering schemes, including phase-speed, ridge, and combined phase-speed and ridge measurements to recover flows in the upper layers of both simulations. We find that horizontal flow maps correlate well ($\sim0.8$) with the simulations in the upper 3~Mm of the domains. At a depth of 5~Mm, correlation deteriorates significantly ($\sim0.6$), though some large-scale flow structure is still visible. Simply increasing the number of measurements used in the inversions would likely help to improve wave coverage at larger depths, but this is not a trivial undertaking due to current computational constraints. We find that even for our best inversions, we severely underestimate the flow velocities at every depth, possibly indicating non-linearity of the forward problem caused by the very strong ($>500~\rm{m\,s^{-1}}$) near-surface flows present in both simulation domains.

Inversions employing phase-speed filtering alone seem to show an advantage in the upper 5~Mm when compared to the other filtering schemes. At larger depths, however, the combined ridge+phase-speed filtering produced a better match between averaging kernel and target function for a fixed inversion noise level. Ridge filtering was generally found to give the worst correlation values.

Vertical flow inversions show poor correlation with the simulation over the full range of depths for QS1, but noticeably better results for QS2. Amplitude determination of these vertical velocities fails everywhere but the nearest surface layer, as noise dominates these inferences. While the inversion procedure to minimize the important cross-talk terms appears to work effectively in the inversion procedure itself, we do not accurately retrieve the vertical flows as was the case in \citep{svanda2011}. This is likely due to the differences in the simulation data we used, such as overall flow amplitudes, time-series length, and utilization of forward-modeled travel times.

In summary, the large-scale flows present in these very sophisticated solar-like simulations are not adequately retrievable with current time-distance techniques, and these results cause us to hesitate to invert for real solar features. Improvements to forward and inverse modeling may need to be made for studying individual structures over short time scales. Longer data sets can be used, but this eliminates the possibility of determining individual supergranule vertical flow profiles as their lifetimes are on average only $1-2$~days. However, these findings suggest that perhaps the most promising way to proceed is the statistical averaging scheme developed in \citet{duvall2010} and \citet{svanda2011} for supergranule-type flows. We have quantified the level of discrepancy between the seismic inferences and the known answer for these simulations, and a forthcoming study attempts a similar analysis for the flows in sunspot simulations.

\begin{table*}
 \centering\footnotesize
 \caption{Parameters for each set of inversions presented in the text.}
 \label{tab1}\medskip
 \begin{threeparttable}
\begin{tabular}{c c c c c c c c c c c} 
\hline 
inversion & sim & filter & depth & FWHM$_{\rm{h}}$ & FWHM$_{\rm{z}}$ & $\mu$ & $\nu$ & $\epsilon$ & noise 
\\ & & & & [Mm] & [Mm] & [$\rm{s^2\, m^{-2}}$] & [$\rm{Mm^{3}}$] & [$\rm{s^4 \,m^{-2}}$] & [$\rm{m\,s^{-1}}$] \\ [0.5ex] 

\hline 
set 1 $(v_x, v_y)$ & QS1, QS2 &  all    & 1 Mm  & 10 & 2 & (0.4--1.8)$\times$10$^{-8}$ & 1 & 1.0$\times$10$^{-6}$ & $\sim$35\\
                  &           &        & 3 Mm  & 12 & 2 & (0.4--1.8)$\times$10$^{-8}$ & 1 & 1.0$\times$10$^{-6}$ & $\sim$35\\ \vspace{1ex}
                  &           &        & 5 Mm  & 14 & 2 & (0.4--1.8)$\times$10$^{-8}$ & 1 & 1.0$\times$10$^{-6}$ & $\sim$35\\

set 2 $(v_x, v_y)$  & QS2 & comb & 7 Mm  & 16 & 3 & 4.6$\times$10$^{-10}$ & 1 & 1.0$\times$10$^{-6}$ & 64\\
                  &     &       & 9 Mm  & 18 & 3 & 7.5$\times$10$^{-12}$ & 1 & 1.0$\times$10$^{-6}$ & 86\\
                  
\multicolumn{1}{l}{set 3 $(v_z)$} & QS1 & ridge   & 1 Mm  & 12 & 2 & 7.2$\times$10$^{-9}$ & 80 & 3.2$\times$10$^{-6}$ & 18\\
      &     &        & 3 Mm  & 14 & 2 & 2.3$\times$10$^{-11}$ & 80 & 1.0$\times$10$^{-9}$ & 814\\ \vspace{1ex}
      &     &        & 5 Mm  & 16 & 2 & 1.0$\times$10$^{-11}$ & 80 & 3.2$\times$10$^{-9}$ & 567\\
      &     & phase  & 1 Mm  & 12 & 2 & 3.7$\times$10$^{-8}$  & 80 & 1.0$\times$10$^{-5}$ & 10\\
      &     &        & 3 Mm  & 14 & 2 & 2.7$\times$10$^{-10}$ & 80 & 1.0$\times$10$^{-8}$ & 210\\ \vspace{1ex}
      &     &        & 5 Mm  & 16 & 2 & 2.3$\times$10$^{-11}$ & 80 & 1.0$\times$10$^{-9}$ & 445\\
      &     & comb  & 1 Mm  & 12 & 2 & 3.7$\times$10$^{-8}$  & 80 & 1.0$\times$10$^{-5}$ & 12\\
      &     &        & 3 Mm  & 14 & 2 & 6.1$\times$10$^{-10}$  & 80 & 3.2$\times$10$^{-8}$ & 152\\ \vspace{1ex}
      &     &        & 5 Mm  & 16 & 2 & 2.3$\times$10$^{-11}$ & 80 & 1.0$\times$10$^{-8}$ & 236\\
      & QS2 & ridge   & 1 Mm  & 12 & 2 & 7.2$\times$10$^{-9}$  & 80 & 3.2$\times$10$^{-6}$ & 18\\
      &     &        & 3 Mm  & 14 & 2 & 1.0$\times$10$^{-11}$ & 80 & 1.0$\times$10$^{-9}$ & 814\\ \vspace{1ex}
      &     &        & 5 Mm  & 16 & 2 & 1.0$\times$10$^{-11}$ & 80 & 1.0$\times$10$^{-9}$ & 567\\
      &     & phase  & 1 Mm  & 12 & 2 & 1.9$\times$10$^{-8}$  & 80 & 3.2$\times$10$^{-5}$ & 18\\
      &     &        & 3 Mm  & 14 & 2 & 1.4$\times$10$^{-10}$ & 80 & 1.0$\times$10$^{-8}$ & 210\\ \vspace{1ex}
      &     &        & 5 Mm  & 16 & 2 & 1.2$\times$10$^{-11}$ & 80 & 3.2$\times$10$^{-9}$ & 445\\
      &     & comb  & 1 Mm  & 12 & 2 & 3.7$\times$10$^{-8}$  & 80 & 1.0$\times$10$^{-5}$ & 12\\
      &     &        & 3 Mm  & 14 & 2 & 2.7$\times$10$^{-10}$ & 80 & 3.2$\times$10$^{-8}$ & 152\\ \vspace{1ex}
      &     &        & 5 Mm  & 16 & 2 & 5.2$\times$10$^{-11}$ & 80 & 1.0$\times$10$^{-8}$ & 236\\
[1ex] 
\hline 
\end{tabular}
 \begin{tablenotes}[para,flushleft]
Columns from left to right are the inversion set number, the specific simulation, the type of filtering, target inversion depth, the horizontal target width, the vertical target size, the noise, cross-talk, and weight spread trade-off parameters, respectively, and the inversion noise level.
 \end{tablenotes}
 \end{threeparttable}
\end{table*}

\begin{table*}
 \centering\footnotesize
 \caption{Descriptive correlation statistics between inverted flows and corresponding simulation data.}
 \label{tab2}\medskip
 \begin{threeparttable}
  \begin{tabular}{c c c c c c c c c c c c c} 
\hline 
sim & fiter & depth & $v_x$ & $v_y$ & $v_z$ & $\rm{mag_h}$ & div & s$_{v_x}$ & s$_{v_y}$ & s$_{v_z}$ & s$_{\rm{mag_h}}$ & s$_{\rm{div}}$\\ [0.5ex] 
\hline 

QS1  &  ridge    &  1  &  0.87  &  0.74  &-0.37&  0.57  &  0.87  &  0.90  &  0.93  &-0.29&  0.64  &  1.20\\
          &  ridge    &  3  &  0.80  &  0.66  &0.01&  0.45  &  0.78  &  0.63  &  0.67  &0.00&  0.41  &  1.08\\\vspace{1ex}
          &  ridge    &  5  &  0.58  &  0.53  &-0.03&  0.04  &  0.57  &  0.40  &  0.59  &0.00&  0.03  &  1.04\\
          &  phase  &  1  &  0.92  &  0.70  &0.36&  0.58  &  0.86  &  0.97  &  0.82  &0.46&  0.68  &  0.95\\
          &  phase  &  3  &  0.86  &  0.74  &0.34&  0.62  &  0.85  &  0.65  &  0.76  &0.02&  0.53  &  0.89\\\vspace{1ex}
          &  phase  &  5  &  0.69  &  0.62  &0.21&  0.23  &  0.69  &  0.73  &  0.88  &0.01&  0.32  &  1.46\\
          &  comb   &  1  &  0.89  &  0.70  &0.28&  0.56  &  0.85  &  1.05  &  0.95  &0.33&  0.77  &  1.14\\
          &  comb   &  3  &  0.85  &  0.72  &0.25&  0.57  &  0.83  &  0.57  &  0.67  &0.04&  0.44  &  0.80\\\vspace{1ex}
          &  comb   &  5  &  0.70  &  0.59  &0.12&  0.21  &  0.65  &  0.46  &  0.55  &0.01&  0.17  &  0.80\\

QS2  &  ridge    &  1  &  0.86  &  0.90  &0.50&  0.63  &  0.92  &  1.07  &  1.21  &0.65&  0.85  &  1.39\\
          &  ridge    &  3  &  0.82  &  0.75  &0.39&  0.63  &  0.73  &  1.31  &  1.27  &0.01&  1.09  &  1.78\\\vspace{1ex}
          &  ridge    &  5  &  0.45  &  0.50  &-0.02&  0.26  &  0.33  &  0.99  &  0.98  &0.00&  0.51  &  1.30\\
          &  phase  &  1  &  0.91  &  0.89  &0.81&  0.68  &  0.93  &  1.57  &  1.55  &0.67&  1.27  &  1.84\\
          &  phase  &  3  &  0.86  &  0.84  &0.53&  0.65  &  0.88  &  0.92  &  0.97  &0.04&  0.77  &  1.27\\\vspace{1ex}
          &  phase  &  5  &  0.71  &  0.78  & -0.11&  0.54  &  0.81  &  0.97  &  1.09  &-0.01&  0.76  &  1.71\\
          &  comb   &  1  &  0.89  &  0.90  &0.79&  0.65  &  0.92  &  1.25  &  1.20  &0.58&  0.92  &  1.30\\
          &  comb   &  3  &  0.85  &  0.77  &0.49&  0.63  &  0.82  &  0.95  &  0.90  &0.06&  0.75  &  1.30\\\vspace{1ex}
          &  comb   &  5  &  0.65  &  0.71  &0.04&  0.43  &  0.74  &  0.83  &  0.83  &0.00&  0.53  &  1.53

\\[1ex] 
\hline 
\end{tabular}
 \begin{tablenotes}[para,flushleft]
Columns 4-6 are the correlation coefficients for the given flow component. `${\rm mag_h}$' denotes $\sqrt{v_x^2+v_y^2}\equiv |\bm{v}_h|$, and `div' the horizontal divergence $\nabla_{\rm h}\cdot\bm{v}_{\rm h}$. The last 5 columns give the slope of the best-fit lines through the correlation plots for the same quantities.
 \end{tablenotes}
 \end{threeparttable}
\end{table*}

\acknowledgements
The authors gratefully acknowledge support by the NASA SDO Science Center through contract NNH09CE41C awarded to NWRA and very fruitful discussions with Doug Braun. K.D. and J.J. also acknowledge funding from a NASA EPSCoR award to NMSU under contract NNX09AP76A. M. Rempel is partially supported through NASA contracts NNH09AK02I, NNH12CF68C and NASA grant NNX12AB35G. The National Center for Atmospheric Research is sponsored by the National Science Foundation. Resources supporting this work were provided by the NASA High-End Computing (HEC) Program through the NASA Advanced Supercomputing (NAS) Division at Ames Research Center under project s0925.

\bibliography{mybib}

\end{document}